\documentclass[12pt,eqsecnum,amsfonts,amssymb,aps]{revtex4-2}

\input epsf

\usepackage[usenames]{color}
\usepackage{graphicx}
\usepackage{graphics}
\usepackage{bm}
\usepackage{rotating}

\newcommand{\beq}{\begin{equation}}
\newcommand{\eeq}{\end{equation}}
\newcommand{\beqs}{\begin{eqnarray}}
\newcommand{\eeqs}{\end{eqnarray}}

\begin{document}

\baselineskip 6.0mm

\title{Chromatic Zeros on the Limit $G^{(p,\ell)}_\infty$ of the Family 
$G^{(p,\ell)}_m$ of Hierarchical Graphs}

\author{Shu-Chiuan$^a$ Chang and Robert Shrock$^b$}

\affiliation{(a) \ Department of Physics \\
National Cheng Kung University, Tainan 70101, Taiwan}

\affiliation{(b) \ C. N. Yang Institute for Theoretical Physics and \\
  Department of Physics and Astronomy \\
  Stony Brook University, Stony Brook, NY 11794}

\begin{abstract}

  We calculate the continuous accumulation set ${\cal B}_q(p,\ell)$ of zeros of
  the chromatic polynomial $P(G^{(p,\ell)}_m,q)$ in the limit $m \to \infty$,
  on a family of graphs $G^{(p,\ell)}_m$ defined such that $G^{(p,\ell)}_m$ is
  obtained from $G^{(p,\ell)}_{m-1}$ by replacing each edge (i.e., bond) on
  $G^{(p,\ell)}_m$ by $p$ paths each of length $\ell$ edges, starting with the
  tree graph $T_2$. Our method uses the property that the chromatic polynomial
  $P(G,q)$ of a graph $G$ is equal to the $v=-1$ evaluation of the partition
  function of the $q$-state Potts model, together with (i) the property that
  $Z(G^{(p,\ell)}_m,q,v)$ can be expressed via an exact closed-form real-space
  renormalization (RG) group transformation in terms of
  $Z(G^{(p,\ell)}_{m-1},q,v')$, where $v'=F_{(p,\ell),q}(v)$ is a rational
  function of $v$ and $q$ and (ii) ${\cal B}_q(p,\ell)(v)$ is the locus in the
  complex $q$-plane that separates regions of different asymptotic behavior of
  the $m$-fold iterated RG transformation $F_{(p,\ell),q}(v)$ in the $m \to
  \infty$ limit. Thus, our results involve calculations of region diagrams in
  the complex $q$-plane showing the type of behavior that occurs in the $m \to
  \infty$ limit of the $m$-fold iterated RG transformation mapping
  $F_{(p,\ell),q}(v)$ starting with the initial value $v=v_0=-1$.  Calculations
  are presented of the maximal point $q_c(G^{(p,\ell)}_\infty)$ at which the
  locus ${\cal B}_q$ crosses the real-$q$ axis, as well as several other points
  at which, depending on $p$ and $\ell$, the locus ${\cal B}_q$ crosses this
  axis. We give explicit results for a variety of $(p,\ell)$ cases and observe
  a number of interesting features.  Calculations of the ground-state
  degeneracy of the Potts antiferromagnet at $q_c(G^{(p,\ell)}_\infty)$ are
  presented.  This work extends a previous study with R. Roeder of the
  $(p,\ell)=(2,2)$ case to higher $p$ and $\ell$ values.

\end{abstract}

\maketitle


\section{Introduction} 
\label{intro_section}

For a given graph $G$, the chromatic polynomial $P(G,q)$ counts the number of
ways of assigning $q$ colors to the vertices of $G$ subject to the condition
that adjacent vertices have different colors. This is called a proper
$q$-coloring of (the vertices of) $G$.  Chromatic polynomials have long been of
interest in mathematical graph theory \cite{biggs,bollobas,dong} and are also
closely related to statistical mechanics, since $P(G,q)$ is equal to the
zero-temperature partition function of the $q$-state Potts antiferromagnet on
$G$ \cite{wurev}.  In turn, the Potts model has been of interest as a model of
phase transitions and critical phenomena.  On a graph $G$, the partition
function of the $q$-state Potts model, denoted $Z(G,q,v)$, is a polynomial in
$q$ and a temperature-dependent Boltzmann variable, $v$, where $v \in [-1,0]$
for the antiferromagnet and $v \ge 0$ for the ferromagnet. In particular, for
an arbitrary graph $G$, the special case $v=-1$, i.e., the zero-temperature
Potts antiferromagnet, is of particular interest, since this evaluation of the
Potts model partition function yields the chromatic polynomial;
$Z(G,q,-1)=P(G,q)$.  In the original statistical physics formulation, $q$ is a
positive integer specifying the number of possible values of a classical spin
defined at a given site of a lattice, $\sigma_i \in \{1,...,q\}$, but, via a
graph-theoretic expression (Eq. (\ref{cluster}), below), $q$ can be generalized
to a real, or, indeed, complex quantity.  On a family of $n$-vertex graphs, as
$n \to \infty$, an infinite subset of the zeros of $Z(G,q,v)$ merge to form a
continuous accumulation set.  In this $n \to \infty$ limit, using the formal
symbol $G_\infty \equiv \lim_{n \to \infty} G_n$, we denote the continuous
accumulation set of locus of zeros of $Z(G_n,q,v)$ in the limit $n \to \infty$
(i) in the complex $q$-plane, for a given $v$, as ${\cal B}_q(G_\infty,v)$ and
(ii) in the complex $v$-plane, for a given $q$, as ${\cal B}_v(G_\infty,q)$.
Since we focus on the continuous accumulation set of chromatic zeros here,
i.e., ${\cal B}_q(G_\infty,v)$ with $v=-1$, we will use the simplified notation
${\cal B}_q(G_\infty) \equiv {\cal B}_q(G_\infty,v=-1)$.  For a generic
$n$-vertex graph, $G$, the calculation of $Z(G,q,v)$ for arbitrary $q$ and $v$,
and similarly, the calculation of $P(G,q)$ for arbitrary $q$, become
exponentially difficult as $n$ grows sufficiently large.  Families of graphs
where the calculation of $Z(G,q,v)$ is tractable for general $q$ and $v$ are
thus of great value. An example is provided by hierarchical families of
graphs. A hierarchical family of graphs $G_m$ is defined by starting with a
given graph $G_0$ and applying a transformation to $G_0$ obtain $G_1$,
iterating this RG transformation to obtain $G_2$, and so forth, to obtain
$G_{m+1}$ from $G_m$.  The formal limit $\lim_{m \to \infty} G_m \equiv
G_\infty$ then defines a hierarchical lattice $G_\infty$, which is generically
a self-similar, fractal object.  By performing a sum over spins at each
iterative step, one can construct an exact functional transformation relating
$Z(G_{m+1},q,v)$ to $Z(G_m,q,v')$, where $v'$ is related to $v$ according to a
rational function $v'=F_q(v)$. The properties of this model can then be
determined in the $m \to \infty$ limit. This is an exact real-space
renormalization-group (RG) transformation. The properties of iterated functions
and fractals have been of considerable importance in mathematics and physics
(some reviews include \cite{beardon}-\cite{devaney}).

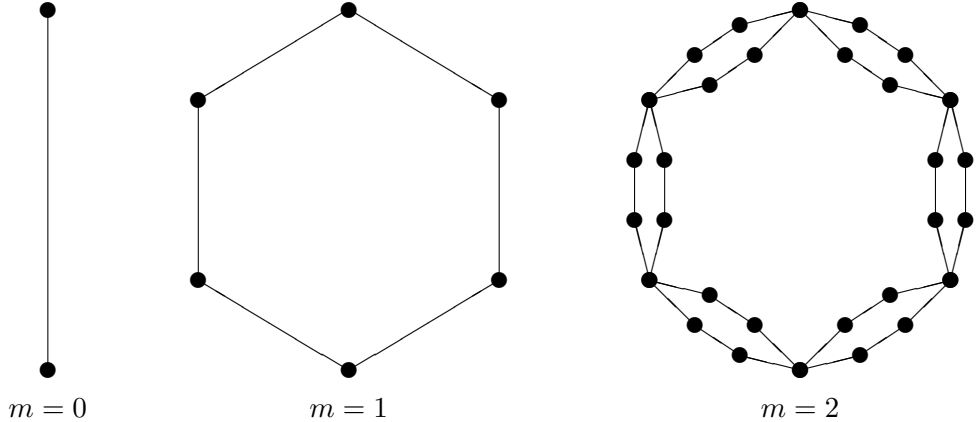
\begin{figure}[htbp]
\unitlength 1mm 
\begin{picture}(120,48)
\put(0,0){\line(0,1){48}}
\multiput(0,0)(0,48){2}{\circle*{2}}
\put(0,-5){\makebox(0,0){$m=0$}}
\put(40,0){\line(5,3){20}}
\put(40,0){\line(-5,3){20}}
\multiput(20,12)(40,0){2}{\line(0,1){24}}
\put(40,48){\line(5,-3){20}}
\put(40,48){\line(-5,-3){20}}
\multiput(40,0)(0,48){2}{\circle*{2}}
\multiput(20,12)(0,24){2}{\circle*{2}}
\multiput(60,12)(0,24){2}{\circle*{2}}
\put(40,-5){\makebox(0,0){$m=1$}}
\multiput(100,0)(12,10){2}{\line(4,1){8}}
\multiput(108,2)(-2,4){2}{\line(3,2){6}}
\multiput(100,0)(14,6){2}{\line(1,1){6}}
\multiput(100,0)(20,12){2}{\circle*{2}}
\multiput(108,2)(6,4){2}{\circle*{2}}
\multiput(106,6)(6,4){2}{\circle*{2}}
\multiput(120,12)(-2,16){2}{\line(1,4){2}}
\multiput(122,20)(-4,0){2}{\line(0,1){8}}
\multiput(120,12)(2,16){2}{\line(-1,4){2}}
\multiput(120,12)(0,24){2}{\circle*{2}}
\multiput(122,20)(0,8){2}{\circle*{2}}
\multiput(118,20)(0,8){2}{\circle*{2}}
\multiput(100,48)(12,-10){2}{\line(4,-1){8}}
\multiput(108,46)(-2,-4){2}{\line(3,-2){6}}
\multiput(100,48)(14,-6){2}{\line(1,-1){6}}
\multiput(100,48)(20,-12){2}{\circle*{2}}
\multiput(108,46)(6,-4){2}{\circle*{2}}
\multiput(106,42)(6,-4){2}{\circle*{2}}
\multiput(100,0)(-12,10){2}{\line(-4,1){8}}
\multiput(92,2)(2,4){2}{\line(-3,2){6}}
\multiput(100,0)(-14,6){2}{\line(-1,1){6}}
\multiput(100,0)(-20,12){2}{\circle*{2}}
\multiput(92,2)(-6,4){2}{\circle*{2}}
\multiput(94,6)(-6,4){2}{\circle*{2}}
\multiput(80,12)(-2,16){2}{\line(1,4){2}}
\multiput(82,20)(-4,0){2}{\line(0,1){8}}
\multiput(80,12)(2,16){2}{\line(-1,4){2}}
\multiput(80,12)(0,24){2}{\circle*{2}}
\multiput(82,20)(0,8){2}{\circle*{2}}
\multiput(78,20)(0,8){2}{\circle*{2}}
\multiput(80,36)(12,10){2}{\line(4,1){8}}
\multiput(88,38)(-2,4){2}{\line(3,2){6}}
\multiput(80,36)(14,6){2}{\line(1,1){6}}
\multiput(80,36)(20,12){2}{\circle*{2}}
\multiput(88,38)(6,4){2}{\circle*{2}}
\multiput(86,42)(6,4){2}{\circle*{2}}
\put(100,-5){\makebox(0,0){$m=2$}}
\end{picture}
\vspace*{5mm}
\caption{\footnotesize{$G^{(p,\ell)}_m$ graphs with $(p,\ell)=(2,3)$ and 
$m=0, \ 1, \ 2$.}}
\label{GP2L3_fig}  
\end{figure}

\begin{figure}[htbp]
\unitlength 1mm 
\begin{picture}(120,48)
\put(0,0){\line(0,1){48}}
\multiput(0,0)(0,48){2}{\circle*{2}}
\put(0,-5){\makebox(0,0){$m=0$}}
\put(40,0){\line(0,1){48}}
\put(40,0){\line(1,1){24}}
\put(40,0){\line(-1,1){24}}
\put(40,48){\line(1,-1){24}}
\put(40,48){\line(-1,-1){24}}
\multiput(40,0)(0,48){2}{\circle*{2}}
\multiput(16,24)(24,0){3}{\circle*{2}}
\put(40,-5){\makebox(0,0){$m=1$}}
\put(100,0){\line(0,1){48}}
\put(100,0){\line(1,1){24}}
\put(100,0){\line(-1,1){24}}
\put(100,48){\line(1,-1){24}}
\put(100,48){\line(-1,-1){24}}
\multiput(100,0)(-3,12){2}{\line(1,4){3}}
\multiput(100,0)(3,12){2}{\line(-1,4){3}}
\multiput(100,48)(-3,-12){2}{\line(1,-4){3}}
\multiput(100,48)(3,-12){2}{\line(-1,-4){3}}
\multiput(100,0)(9,15){2}{\line(5,3){15}}
\multiput(100,0)(15,9){2}{\line(3,5){9}}
\multiput(100,0)(-9,15){2}{\line(-5,3){15}}
\multiput(100,0)(-15,9){2}{\line(-3,5){9}}
\multiput(100,48)(9,-15){2}{\line(5,-3){15}}
\multiput(100,48)(15,-9){2}{\line(3,-5){9}}
\multiput(100,48)(-9,-15){2}{\line(-5,-3){15}}
\multiput(100,48)(-15,-9){2}{\line(-3,-5){9}}
\multiput(100,0)(0,48){2}{\circle*{2}}
\multiput(76,24)(24,0){3}{\circle*{2}}
\multiput(97,12)(3,0){3}{\circle*{2}}
\multiput(97,36)(3,0){3}{\circle*{2}}
\multiput(109,15)(3,-3){3}{\circle*{2}}
\multiput(91,15)(-3,-3){3}{\circle*{2}}
\multiput(109,33)(3,3){3}{\circle*{2}}
\multiput(91,33)(-3,3){3}{\circle*{2}}
\put(100,-5){\makebox(0,0){$m=2$}}
\end{picture}
\vspace*{5mm}
\caption{\footnotesize{$G^{(p,\ell)}_m$ graphs with $(p,\ell)=(3,2)$ and 
$m=0, \ 1, \ 2$.}} 
\label{GP3L2_fig}
\end{figure}
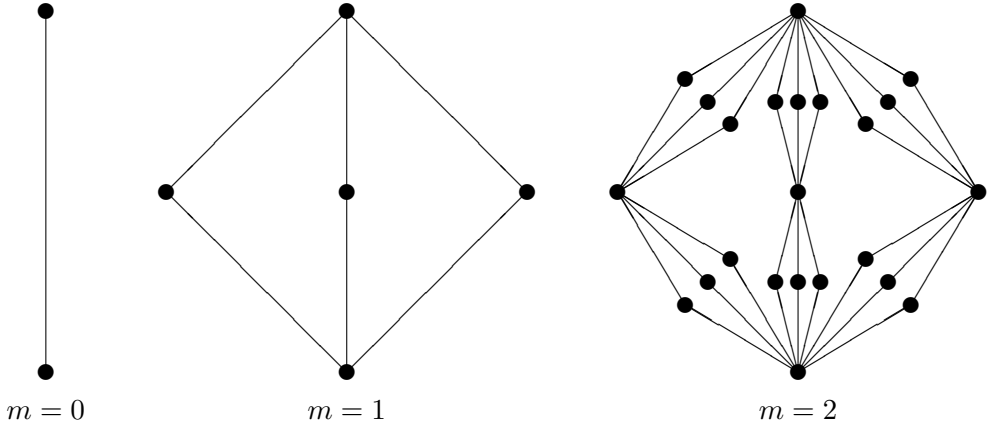

An interesting family of hierarchical graphs is defined as follows. One starts
with an initial graph $G_0$ consisting of two vertices (i.e., sites) and a bond
(denoted as edge in mathematical graph theory) joining them. The iterative
graphical transformation replaces this single edge by $p$ paths, each
consisting of $\ell$ links. This procedure is repeated iteratively. The graph
resulting from the $m$'th iteration of this procedure is denoted
$G^{(p,\ell)}_m$. In Figs. \ref{GP2L3_fig} and \ref{GP3L2_fig} we show
illustrative examples of these graphs, namely $G^{(p,\ell)}_m$ for
$(p,\ell)=(2,3)$ and $(3,2)$, with $m=0, \ 1, \ 2$.  We define the formal limit
$\lim_{m \to \infty} G^{(p,\ell)}_m \equiv G^{(p,\ell)}_\infty$.  We restrict
to the nontrivial range $p \ge 2$ and $\ell \ge 2$. In this range, the limit
$G^{(p,\ell)}_\infty$ is a self-similar object.  The lowest member of this
doubly infinite family of hierarchical graphs $G^{(p,\ell)}_m$ is the family
$G^{(2,2)}_m \equiv D_m$, called the Diamond Hierarchical Lattice (DHL). In
\cite{dhl} with R. Roeder, we studied ${\cal B}_q(G^{(2,2)}_\infty,v)$ and
${\cal B}_v(G^{(2,2)}_\infty,q)$ for the Potts model (see also
\cite{chio_roeder,rig}).


In this paper we report new results on the continuous accumulation set of zeros
of the chromatic polynomial $P(G^{(p,\ell)}_m,q)$ in the limit $m \to \infty$,
denoted ${\cal B}_q(G^{(p,\ell)}_\infty)$, with higher values of $(p,\ell)$
going beyond the case $(p,\ell)=(2,2)$. The resulting locus depends on the
values of $p$ and $\ell$, and is thus denoted in full as ${\cal
  B}_q(G^{(p,\ell)}_\infty)$. We will often use the simplified notation ${\cal
  B}_q(p,\ell) \equiv {\cal B}_q(G^{(p,\ell)}_\infty)$.  For completeness, some
review of the $(p,\ell)=(2,2)$ case is also included. Our method uses the fact
that the chromatic polynomial $P(G,q)$ of a graph $G$ is equal to the $v=-1$
evaluation of the partition function of the $q$-state Potts model together with
(i) the properties that $Z(G^{(p,\ell)}_m,q,v)$ can be expressed via an exact
closed-form real-space renormalization group transformation in terms of
$Z(G^{(p,\ell)}_{m-1},q,v')$, where $v'=F_{(p,\ell),q}(v)$ is a rational
function of $v$ and $q$ and (ii) ${\cal B}_q(p,\ell)(v)$ is the locus in the
complex $q$-plane that separates regions of different asymptotic behavior of
the $m$-fold iterated RG transformation $F_{(p,\ell),q}(v)$ in the $m \to
\infty$ limit, starting from the initial value $v=v_0=-1$.  Thus, for each
$(p,\ell)$ family, our results involve calculations of region diagram in the
complex $q$-plane showing the type of behavior that occurs in the $m \to
\infty$ limit of the above-mentioned $m$-fold iterated RG transformation.  We
will refer to this as the chromatic region diagram for the $(p,\ell)$ case,
i.e., the continuous accumulation set of the chromatic polynomial
$P(G^{(p,\ell)}_m,q)$ in the limit $m \to \infty$.  We have also calculated
${\cal B}_q(G^{(p,\ell)}_\infty,v)$ for an initial value $v=v_0$ in the
nonzero-temperature range $-1 < v_0 \le 0$ for the Potts antiferromagnet and
also the range $v_0 \ge 0$ for the Potts ferromagnet; the results will be
reported elsewhere.  There have been numerous studies of spin models on
hierarchical lattices, most of which have analyzed the zeros of $Z(D_m,q,v)$ in
the complex plane of the temperature-like Boltzmann variable $v$, e.g.,
\cite{gk}-\cite{blr2} (see \cite{dhl} for further references). A number of
these works studied the Julia sets of various RG transformations
\cite{julia}. Some related studies of the Potts model partition function
$Z(G^{(p,\ell)}_m,q,v)$ for $p,\ell \ge 3$ and/or $\ell \ge 3$ include
\cite{qiao01}, \cite{yang88}-\cite{qiao_memoir}.  Studies of zeros of the Potts
model partition function on other hierarchical graph families, including
Sierpinski and Hanoi graphs, include, e.g., \cite{desimoi}-\cite{han}.

This paper is organized as follows. In Sections
\ref{background_section}-\ref{transformation_section} we review some relevant
background on the Potts model, the hierarchical family $G^{(p,\ell)}_m$, and
the iterative RG transformation that relates $Z(G^{(p,\ell)}_{m+1},q,v)$ to
$Z(G^{(p,\ell)}_m,q,v')$ and $v'$ to $v$. In Section
\ref{general_locus_section} we present plots of the loci ${\cal
  B}_q(G^{(p,\ell)}_\infty)$ and associated region diagrams in the complex
$q$-plane for a variety of $(p,\ell)$ families.  Sections
\ref{qc_L_even_section} and \ref{qc_L_odd_section} contain further results on
these loci for even and odd $\ell$, respectively, including calculations of
various special points $q_c$, $q_\infty$, $q_{\rm int}$, $q_x$, and $q_L$
(dependent on $(p,\ell)$) where ${\cal B}_q(G^{(p,\ell)}_\infty)$ crosses the
real-$q$ axis. Some further properties of the loci ${\cal B}$ are presented in 
Section \ref{further_section}, and our conclusions are given in Section
\ref{conclusions_section}. Some ancillary information is contained in two
appendices.


\section{Background} 
\label{background_section}

In this section we discuss some relevant background from graph theory
and statistical physics.  For further details, see, e.g., \cite{dhl}. 
A graph $G=(V,E)$ is defined by its set
$V$ of vertices (= sites) and its set $E$ of edges (= bonds).  We denote
$n=n(G)=|V|$ and $e(G)=|E|$ as the number of vertices and edges of
$G$. At temperature $T$, the partition function of the $q$-state Potts
model is given by 
$Z = \sum_{ \{ \sigma_i \} } e^{-\beta {\cal H}}$, with the Hamiltonian
\beq
{\cal H} = -J \sum_{e_{ij}} \delta_{\sigma_i, \sigma_j} \ ,
\label{ham}
\eeq
where $i$ and $j$ label adjacent vertices of $G$; $\sigma_i$ are
classical spin variables on these vertices, taking values in the set
$I_q = \{1,...,q\}$; $\delta_{rs}$ is the Kronecker delta function,;
$\beta = (k_BT)^{-1}$ with 
$k_B$ the Boltzmann constant; and $e_{ij}$ is the edge joining
the vertices $i$ and $j$ in $G$ \cite{wurev}. We define the notation  
\beq
K = \beta J \ , \quad y = e^K \ , \ v = y-1 \ . 
\label{kdef}
\eeq
The signs of the spin-spin interaction constant $J$ favoring
ferromagnetic (FM) and antiferromagnetic (AFM) spin configurations are
$J > 0$ and $J < 0$, respectively.  Hence, the physical ranges of $v$
are $v \ge 0$ for the Potts ferromagnet (FM) and $-1 \le v \le 0$ for
the Potts antiferromagnet (AFM).  The value $v=0$, i.e., $K=0$,
corresponds to infinite temperature, while the zero-temperature values of
$v$ are $v=-1$ for the antiferromagnet and $v=\infty$ for the ferromagnet. 

Using the identity
$e^{K \delta_{\sigma_i \sigma_j}}= 1+v\delta_{\sigma_i \sigma_j}$,
one can reexpress the partition function for the $q$-state Potts model
in the form
\beq
Z = \sum_{\{ \sigma_i \}} \prod_{e_{ij}} (1 + v\delta_{\sigma_i \sigma_j}) \ . 
\label{zv}
\eeq
This partition function is invariant under a global symmetry that acts
on the spins, namely the mapping $\sigma_i \to \pi_q(\sigma_i)$, where
$\pi_q$ is an element of the permutation group on $q$ objects, denoted
$S_q$. At high temperatures, this symmetry is realized explicitly in
the physical states, while in the $n \to \infty$ (thermodynamic) limit
on a regular lattice graph with dimensionality greater than a lower critical
dimensionality, it can be broken spontaneously with the presence of a
nonzero long-range ordering of the spins.  This ordering is
ferromagnetic or antiferromagnetic, depending on where $J > 0$ or $J < 0$,
respectively.

A spanning subgraph of $G$ is $G' = (V,E')$ with $E' \subseteq E$. The
number of connected components of $G'$ is denoted $k(G')$.  The
partition function of the Potts model can equivalently be expressed in
a purely graph-theoretic manner as the sum over spanning subgraphs
\cite{fk}
\beq
Z(G,q,v) = \sum_{G' \subseteq G} q^{k(G')} \, v^{e(G')}  \ .
\label{cluster}
\eeq
Eq. (\ref{cluster}) shows that the partition function $Z(G,q,v)$ is a
polynomial in $q$ and $v$ with positive integer coefficients for each
nonzero term. As is evident from
Eq. (\ref{cluster}), $Z(G,q,v)$ has degree $n(G)$ in $q$ and degree $e(G)$ in
$v$, or equivalently, in $y$.  Since $k(G') \ge 1$ for all $G'$, $Z(G,q,v)$
always contains an overall factor of $q$, so one can define a reduced
partition function 
\beq
Z_r(G,q,v) \equiv q^{-1} Z(G,q,v) \ , 
\label{zr}
\eeq
which is also a polynomial in $q$ and $v$

The expression in Eq. (\ref{cluster}) allows one to
generalize both $q$ and $v$ from their physical ranges to complex
values, as is necessary in order to analyze the zeros of $Z(G,q,v)$ in
$q$ for fixed $v$ and the zeros of $Z(G,q,v)$ in $v$ for fixed $q$.
Since the coefficients in $Z(G,q,v)$ are real (actually in ${\mathbb
  Z}_+$, but all we use here is the reality), it follows that for real
$v$, the zeros of $Z(G,q,v)$ in the $q$-plane and the accumulation
locus ${\cal B}_q(v)$ are invariant under complex conjugation $q \to
q^*$, and for real $q$, the zeros of $Z(G,q,v)$ in the $v$-plane and
the accumulation locus ${\cal B}_v(q)$ are invariant under complex
conjugation $v \to v^*$, i.e., 
\beq 
v \in {\mathbb R} \ \Rightarrow \ {\cal B}_q(v) \ {\rm is \ invariant \ under}
 \ q \to q^*
\label{Bq_invariance}
\eeq
and
\beq
q \in {\mathbb R} \ \Rightarrow \ {\cal B}_v(q) \ {\rm is \ invariant \ under}
 \ v \to v^* \ . 
\label{Bv_invariance}
\eeq

As noted above, the $T \to 0$ limit for the Potts antiferromagnet means $K \to
-\infty$ and thus $v \to -1$. In this limit (see Eq. (\ref{zv})), the only spin
configurations that contribute to $Z(G,q,v)$ are those for which the spins on
adjacent vertices are different.  Hence,
\beq
P(G,q) = Z(G,q,-1) \ , 
\label{pz}
\eeq
where $P(G,q)$ is the chromatic polynomial, counting the number of proper
$q$-colorings of (the vertices of) $G$.  From Eqs. (\ref{zr}) and (\ref{pz}),
it follows that $P(G,q)$ always contains an overall factor of $q$.  Since
$G^{(p,\ell)}_m$ always contain at least one edge, $P(G^{(p,\ell)}_m,q)$ also
contains an overall factor of $q-1$.  The minimum integer number $q$ that
allows a proper $q$-coloring of $G$ is the chromatic number, $\chi(G)$.  An
important property of $G^{(p,\ell)}_m$ is that it is bipartite, and hence
\beq
\chi(G^{(p,\ell)}_m)=2 
\label{chig}
\eeq
and
\beq
P(G^{(p,\ell)}_m, 2)=2 \ . 
\label{pgq2}
\eeq

Part of the interest in chromatic polynomials from a statistical physics point
of view is their connection with ground-state entropy in a Potts
antiferromagnet.  On a given $n$-vertex graph $G_n$, the ground-state (i.e.,
zero-temperature) degeneracy per vertex (i.e., site) of the Potts
antiferromagnet is
\beq
W(G_n,q) = [P(G,q)]^{1/n(G)} \ .
\label{wfinite}
\eeq
In the $n(G) \to \infty$ limit of a given
family of $n$-vertex graphs $G$, denoted $G_\infty$,
the ground-state degeneracy per vertex of the Potts antiferromagnet is
\beq
W(G_\infty,q) = \lim_{n(G) \to \infty} [P(G,q)]^{1/n(G)} \ ,
\label{wdef}
\eeq
and the corresponding ground-state entropy per vertex is
\beq
S_0(G_\infty,q) = k_B \ln [W(G_\infty,q)] \ .
\label{s0}
\eeq
In normal physical systems, including the Potts model, the entropy is 
non-negative, so $W(G,q) \ge 1$ and $W(G_\infty,q)$. 
For real $q < \chi(G)$, $P(G,q)$ can be negative; in this case, since there is
no obvious choice for which of the $n$ roots of $(-1)$ to pick in
Eq. (\ref{wfinite}) or Eq. (\ref{wdef}), one can only determine the magnitudes
$|W(G,q)|$ and $|W(G_\infty)|$ \cite{w,a}.
As discussed in \cite{w,a}, for certain values of $q$, denoted $q_s$, one must
take account of the noncommutativity
\beq
\lim_{n(G) \to \infty} \lim_{q \to q_s} [P(G,q)]^{1/n(G)} \neq
\lim_{q \to q_s}\lim_{n(G) \to \infty} [P(G,q)]^{1/n(G)} \ .
\label{wnoncom}
\eeq
This noncommutativity will not be relevant for our calculations of
$W(G^{(p,\ell)}_\infty,q)$ evaluated at $q=q_c(G^{(p,\ell}_\infty)$, in Section
\ref{wc_subsection}, since $\chi(G^{(p,\ell)}_m)=2$, and we find that
$q_c(G^{(p,\ell}_\infty) > 2$ (see Eq. (\ref{qcgt2})).  For calculations of
other special points, including $q_\infty(G^{(p,\ell)}_\infty)$, etc., where
${\cal B}_q(p,\ell)$ crosses the real axis, we take the order of limits to be
$m \to \infty$ first, since this is inherent in the definition of ${\cal
  B}_q(p,\ell)$, and then the limit of $q$ approaching the respective crossing
point.

For a given graph $G$, the zeros of $P(G,q)$ are called the chromatic zeros.
In analyzing these chromatic zeros and their limiting behavior as $n(G) \to
\infty$, it is useful to recall some rigorous results concerning zero-free
regions on the real-$q$ axis. Since the signs of descending powers of $q$ in
$P(G,q)$ alternate, $P(G,q)$ has no zeros in the interval $(-\infty,0)$.  For
an arbitrary graph $G$, there are also no chromatic zeros in the interval (0,1)
\cite{woodall} and none in the interval $(1,32/27]$ \cite{jackson,thomassen}.
Thus, although it was shown in
\cite{dhl} that ${\cal B}_q$ crosses the real $q$ axis at the point $q=32/27$,
this point itself is not a chromatic zero.  Since $P(G,q)$ always has a factor
of $q$, it always vanishes at $q=0$, and if, as is the case here, $G$ has at
least one edge, then $P(G,q)$ also vanishes at $q=1$.

The Potts model partition function is equivalent to a function of 
considerable interest in graph theory, namely the Tutte polynomial
\cite{biggs,bollobas,tutte1,tutte2}. 
The Tutte polynomial, denoted $T(G,x,y)$, of a graph $G$ is defined by
\beq
T(G,x,y) = \sum_{G' \subseteq G} (x-1)^{k(G')-k(G)} (y-1)^{c(G')} \ ,
\label{t}
\eeq
where, as above, $k(G')$ denotes the number of connected components of the
spanning subgraph $G'$, and $c(G')$ denotes the number of linearly independent
circuits on $G'$, given by $c(G')=e(G')+k(G')-n(G')$ 
(note that $n(G')=n(G)$).  With
$y=e^K=v+1$, as defined in Eq. (\ref{kdef}), and
\beq
x = 1 + \frac{q}{v} \ ,
\label{x}
\eeq
it follows that 
\beq
Z(G,q,v) = (x-1)^{k(G)}(y-1)^{n(G)}T(G,x,y) \ .
\label{zt}
\eeq
Thus, the partition function of the Potts model is equivalent, up to the
indicated prefactor, to the Tutte polynomial on a given graph $G$, with
the correspondences (\ref{x}) and (\ref{kdef}) relating the Potts variables
$q$ and $v$ to the Tutte variables $x$ and $y$.

Zeros of $Z(G,q,v)$ in $q$ for a given $v$ and zeros of $Z(G,q,v)$ in $v$ for a
given $q$ are of interest partly because for many families of graphs, such as
strips of regular lattices of given width and arbitrary length $m$, denoted
$\Lambda_m$, in the $m \to \infty$ limit, an infinite subset of these
respective zeros typically merge to form certain continuous loci.  As stated
above, for a one-parameter family of graphs $G_m$, we define the locus ${\cal
  B}_q(G_\infty,v)$ as the continuous accumulation set of zeros of $Z(G_m,q,v)$
in the complex $q$-plane as $m \to \infty$. (There may also be discrete zeros
that do not lie on this locus.) For infinite-length, finite-width strips of
regular lattices, and also chain graphs, ${\cal B}_q$ is generically comprised
of algebraic curves, including possible line segments
\cite{bkw_chrom}-\cite{sokal_dense}.  The underlying reason for this is that
$P(G,q)$, and, more generally, $Z(G,q,v)$, for these classes of graphs consist
of a sum of $m$'th powers of certain algebraic functions, denoted generically
as $\lambda_j$, where $m$ is the length of the strip, and the loci ${\cal
  B}_q(G_\infty,v)$ occur at values of $q$ where there are two or more
$\lambda_j$ functions that are largest in magnitude and degenerate in
magnitude. An early mathematical analysis of this sort of behavior was given in
\cite{bkw}. Thus, one calculates the locus ${\cal B}_q$ for the $m \to \infty$
limit of a family of graphs $G_m$ of this type by computing the various
$\lambda$ functions and mapping out the loci where there are degeneracies in
magnitude between two (or more) dominant $\lambda$ functions.  The loci ${\cal
  B}_q$ for the $m \to \infty$ limits of various families of graphs may be
connected or disconnected. For example, the locus ${\cal B}_q$ for the
infinite-length limit of the cyclic square-lattice strip of width $L_y=2$
vertices (i.e., the ladder graph with periodic longitudinal boundary
conditions) is a connected set of curves separating the $q$-plane into four
regions (shown in Fig. 3 of \cite{w}), while the locus ${\cal B}_q$ for the
infinite-length limit of the square-lattice strip of width $L_y=3$ with free
longitudinal boundary conditions is comprised of three disconnected arcs (shown
in Fig. 3a of \cite{strip}).  The infinite-length limit of a cyclic chain of
polygon subgraphs connected via $e_g$ edges betweeen each polygon exhibits
connected loci ${\cal B}_q$ when $e_g=0$ and disconnected loci ${\cal B}_q$
when $e_g \ge 1$ (e.g., Fig. 2 in \cite{nec}). Thus, a considerable variety of
behavior is found concerning the connectedness of the loci ${\cal B}_q$ for
these families of graphs.

The method that we used with R. Roeder in \cite{dhl} for calculating ${\cal
  B}_q(G^{(2,2)}_\infty,v=-1)$ and that we use here for calculating ${\cal
  B}_q(G^{(p,\ell)}_\infty,v=-1)$ with higher $p$ and $\ell$ is quite different
from the procedure described above.  Rather than determining the set of
relevant $\lambda$ functions and then computing the locus where there is a
degeneracy in magnitude of the dominant $\lambda$ functions, we use the
properties that (i) $Z(G^{(p,\ell)}_m,q,v)$ can be expressed via an exact
closed-form real-space renormalization group transformation in terms of
$Z(G^{(p,\ell)}_{m-1},q,v')$, where $v'=F_{(p,\ell),q}(v)$ is a rational
function of $v$ and $q$ and (ii) ${\cal B}_q(G^{(p,\ell)}_\infty)$ is the
locus, in the complex $q$-plane, that separates regions of different asymptotic
behavior of the $m$-fold iterated RG transformation $F_{(p,\ell),q}(v)$ in the
$m \to \infty$ limit. This will be discussed in detail below in
Sections \ref{transformation_section} and \ref{general_locus_section}.  
We focus on the determination of special points where ${\cal
B}_q(G^{(p,\ell)}_\infty)$ crosses the real-$q$ axis.
One of the results of our study is a determination of the points where ${\cal
  B}_q(G^{(p,\ell)}_\infty)$ crosses the real-$q$ axis.  This locus crosses
this axis at a maximal (i.e. rightmost) point denoted $q_c(G^{(p,\ell)}_\infty)
\equiv q_c(p,\ell)$ and at a leftmost point denoted $q_L(G^{(p,\ell)}_\infty)
\equiv q_L(p,\ell)$. 

We note that the property that ${\cal B}_q$ crosses the real $q$ axis at a
point $q_0$ does not imply that $P(G,q)$ vanishes at this point.  The precise
meaning of the property that ${\cal B}_q$ crosses the real $q$ axis at a point
$q_0$ is that in the limit $n(G) \to \infty$, the zeros of $P(G,q)$ approach
arbitrarily close to $q_0$.  This type of behavior is familiar from statistical
physics.  For example, for integral $q \ge 2$ on the (infinite) square lattice
$\Lambda_{sq}$, the continuous locus of zeros ${\cal B}_v(q)$ of
$Z(\Lambda_{sq},q,v)$ in the $v$-plane for the $q$-state Potts ferromagnet
crosses the real $v$ axis at $v_c=\sqrt{q}$ (see, e.g., \cite{wurev}), but, as
is evident from Eq. (\ref{cluster}), for the finite-temperature $q$-state Potts
ferromagnet, since $v > 0$, all terms contributing to $Z(\Lambda,q,v)$ for any
finite square lattice are positive, so that $Z(\Lambda,q,v)$ does not vanish at
$v_c$. This crossing of ${\cal B}_v(q)$ separates the paramagnetic phase with
$0 \le v \le v_c$ with explicit $S_q$ symmetry from the ferromagnetically
ordered phase with $v > v_c$, in which the $S_q$ symmetry is spontaneously
broken.


\section{Hierarchical Graphs $G^{(p,\ell)}_m$ and Limit 
  $G^{(p,\ell)}_\infty$} 
\label{GPL_section}

In this section we discuss further details concerning the hierarchical family
of $m$'th iterate graphs $G^{(p,\ell)}_m$ studied here. Unless otherwise
stated, we assume that $p$ and $\ell$ are integers in the nontrivial ranges $p
\ge 2$ and $\ell \ge 2$. We have mentioned above how one defines
$G^{(p,\ell)}_m$ iteratively, starting with the $m=0$ initial graph,
$G^{(p,\ell)}_0=T_2$, the tree graph with two vertices. As stated above, the
formal limit of these iterations, $m \to \infty$, is denoted as
$G^{(p,\ell)}_\infty$.  We now discuss the number of edges and vertices on the
$m$'th iterate graph, $G^{(p,\ell)}_m$.  In the first iteration, the single
edge of the initial $T_2$ graph is replaced by $p$ paths, each of length $\ell$
edges, thereby producing $p \ell$ edges in $G^{(p,\ell)}_1$.  In the second
iteration, each of the $p \ell$ edges in $G^{(p,\ell)}_1$ is again replaced by
$p$ paths, each of length $\ell$ edges, so that $G^{(p,\ell)}_2$ has $(p
\ell)^2$ edges.  Continuing this process, one obtains the result
\beq
e(G^{(p,\ell)}_m) = (p \ell)^m \ . 
\label{eGPLm}
\eeq
We next derive the formula for $n(G^{(p,\ell)}_m)$.  Starting from the
initial graph $T_2$, the first iteration retains the two end vertices
and adds $(\ell-1)$ vertices on each of the $p$ paths, so that
$n(G^{(p,\ell)}_1)=2+p(\ell-1)$. In the second iteration, one has the
two original end vertices plus the $p(\ell-1)$ vertices produced by
the first iteration and, in addition, since one replaces each of the
$(p \ell)$ edges in $G^{(p,\ell)}_1$ by $(p \ell)$ new edges, this
adds $p \ell(\ell-1)$ new vertices, so
$n(G^{(p,\ell)}_2) = 2 + p(\ell-1)+(p \ell)[p(\ell-1)]$.
Proceeding to higher $m$ in this
manner, one finds that the number of vertices has the form
$n(G^{(p,\ell)}_m) = a(p \ell)^m+b$, where $a$ and $b$ depend on $p$ snd
$\ell$. One can determine $a$ and $b$ by evaluating this general form for
$m=0$ and $m=1$ and setting the respective expressions equal to the
explicitly derived results for these $m$ values.  Thus, one gets the
equations
\beq
m=0: \quad a+b=2
\label{n_m0}
\eeq
\beq
m=1: \quad a(p \ell)+b = 2+p(\ell-1) \ . 
\label{n_m1}
\eeq
These are two linear equations for the two quantities $a$ and $b$;
solving them, we obtain
\beq
a = \frac{p(\ell-1)}{p \ell-1}
\label{asol}
\eeq
and
\beq
b = \frac{p(\ell+1)-2}{p \ell-1} \ .
\label{bsol}
\eeq
Thus, we derive  the general result
\beq
n(G^{(p,\ell)}_m) = \frac{p(\ell-1)(p \ell)^m+p(\ell+1)-2}{p \ell-1} \ . 
\label{nGPLm}
\eeq
Evidently, in the nontrivial range $p \ge 2$ and $\ell \ge 2$ that we consider
here, both $e(G^{(p,\ell)}_m)$ and $n(G^{(p,\ell)}_m)$ are exponentially
increasing functions of the iteration index $m$.

We denote the number of edges connecting to a vertex $v_i$ in a graph $G$ as
the degree of this vertex, $\Delta(v_i)$.  If all of the vertices in a graph 
have the same degree, this graph is said to be $\Delta$-regular. 
Except for the initial $m=0$ graph $T_2$ (for which $\Delta=1$) and the
$m=1$ graph $G^{(p,\ell)}_1$ with $p=2$, for which $\Delta=2$,
$G^{(p,\ell)}_m$ is not a $\Delta$-regular graph. However, as in
earlier work \cite{wn}, for an arbitrary graph $G=G(V,E)$,  one can
define an effective ( = average) vertex degree
\beq
\Delta_{\rm eff}(G) = \frac{2e(G)}{n(G)} \ .
\label{Delta_eff}
\eeq
Using Eqs. (\ref{eGPLm}) and (\ref{nGPLm}), we calculate 
\beq
\Delta_{\rm eff}(G^{(p,\ell)}_m) =
\frac{2(p \ell-1)}{p(\ell-1)+[p(\ell+1)-2](p \ell)^{-m}} \ .
\label{Delta_eff_GPLm}
\eeq
Hence, in the $m \to \infty$ limit, we have 
\beq
\Delta_{\rm eff}(G^{(p,\ell)}_\infty) = \frac{2(p \ell-1)}{p(\ell-1)} \ .
\label{Delta_eff_GPL}
\eeq
In the nontrivial range $p \ge 2$ and $\ell \ge 2$, $\Delta_{\rm
  eff}(G^{(p,\ell)}_\infty)$ is a monotonically increasing function of $p$ for
fixed $\ell$ and a monotonically decreasing function of $\ell$ for fixed $p$.
These properties are evident from the derivatives
\beq
\frac{\partial \Delta_{\rm eff}(G^{(p,\ell)}_\infty)}{\partial p}
= \frac{2}{p^2(\ell-1)} 
\label{dDeltadP}
\eeq
and
\beq
\frac{\partial \Delta_{\rm eff}(G^{(p,\ell)}_\infty)}{\partial \ell}
= -\frac{2(p-1)}{p(\ell-1)^2} \ , 
\label{dDeltadL}
\eeq
which are, respectively, positive-definite and negative-definite in this
range $p \ge 2$ and $\ell \ge 2$.
In the diagonal case $p=\ell \equiv s$,
we have
\beq
\Delta_{\rm eff}(G^{(s,s)}_\infty) = 2\bigg (1 + \frac{1}{s} \bigg ) \ . 
\label{Delta_eff_Gss}
\eeq
As is evident, $\Delta_{\rm eff}(G^{(s,s)}_\infty)$ decreases monotonically
from the value 3 for $s=2$ to the limiting value 2 as $s \to \infty$.

We recall the procedure for calculating the Hausdorff
dimension $d_H$ of a hierarchical lattice $G_\infty$, which will be applied
to $G^{(p,\ell)}_\infty$. (For rigorous mathematical discussions
of Hausdorff dimensions of fractal objects, see, e.g., 
\cite{beardon}-\cite{peitgen}.)  If the RG
transformation replaces each edge by $\ell$ edges
and gives rise to $N$ copies of the original graph, then $N =
\ell^{d_H}$, $d_H = \ln(N)/\ln(\ell)$.  In the case of the
iteration procedure for $G^{(p,\ell)}_m$, one has $N=p \ell$, yielding the
result that
\beq
d_{H,G^{(p,\ell)}_\infty}= \frac{\ln(p \ell)}{\ln \ell }
= 1 + \frac{\ln p}{\ln \ell} \ .
\label{dim_hausdorff}
\eeq
Early studies showed that properties of statistical mechanical systems such as
the Ising and general Potts model on $n \to \infty$ limits of hierarchical
lattice graphs $G_\infty$ are different from those on regular lattices (e.g.,
\cite{gk,gam,saclay,bambihu85,bzl}. Nevertheless, hierarchical lattices with
closed-form exact RG transformations provide a valuable theoretical framework
in which one can investigate these properties.  


\section{Iterative Transformation on $Z(G^{(p,\ell)}_m,q,v)$} 
\label{transformation_section}

By carrying out the summation over the spins at intermediate vertices
at each stage, one finds the following iterative
transformation for the partition function of the Potts model on
hierarchical family of graphs $\{ G^{(p,\ell)}_m \}$ \cite{qin_yang91} 
\beq
Z(G^{(p,\ell)}_{m+1},q,v)=Z(G^{(p,\ell)}_m,q,v') \, [A_\ell(q,v)]^{p(p \ell)^m}
 \ , 
\label{rg}
\eeq
where
\beq
v' = F_{(p,\ell),q}(v) = \bigg [ \frac{(q+v)^\ell+(q-1)v^\ell}
                                        {(q+v)^\ell-v^\ell} \ \bigg ]^p  - 1
\label{vprime}
\eeq
and
\beq
A_\ell(q,v) = \frac{1}{q} \Big [ (q+v)^\ell - v^\ell \Big ]
= \frac{v^\ell}{q} \Big [ x^\ell - 1 \Big ] \ ,  
\label{A_L}
\eeq
where $x$ is the Tutte variable defined in Eq. (\ref{x}). 
Note that the numerator in the square brackets
in Eq. (\ref{vprime}) is the Potts partition function of the circuit graph
with $\ell$ vertices (and thus also $\ell$ edges), $C_\ell$,
\beq
Z(C_\ell,q,v) = (q+v)^\ell + (q-1)v^\ell = v^\ell(x^\ell+q-1) \ .
\label{zcn}
\eeq
The iterative transformation (\ref{vprime}) embodies the action of the
real-space renormalization group action here.  The RG fixed point (RGFP) is
determined by the condition that this transformation leaves $v$ unchanged,
which we denote as :
\beq
F_{(p,\ell),q}(v)=v \ ,  
\label{RGFPeq}
\eeq
i.e., $F_{(p,\ell),q}(v)-v=0$. Since the left-hand side of (\ref{RGFPeq}) is a
rational function, this equation is equivalent to the equation in which the
numerator of this rational function is set equal to zero.  Some illustrative
examples are given in Appendix \ref{RGFP_appendix}.  Although we do not add a
subscript to $v'$ or $y'$, it is understood that these quantities are
transformed at each iteration.  We use the same notation as in \cite{dhl} to
denote multifold functional composition, namely $f^2(z) \equiv (f \circ f)(z)
\equiv f(f(z))$, $f^3(z) \equiv f(f(f(z)))$, etc.; explicitly, for our case,
\beq
F^2_{(p,\ell),q}(v) \equiv F_{(p,\ell),q}\Big (F_{(p,\ell),q}(v) \Big ) \ ,
\label{composition}
\eeq
and so forth for higher values of the iteration index $m$. This notation is
commonly used in complex dynamics literature (e.g., \cite{beardon,devaney}),
but the reader is cautioned not to confuse this with the common notation
$f^n(z) \equiv [f(z)]^n$.

It is often convenient to use an equivalent RG transformation defined
as a function of $y$, which thus is a mapping from $y=v+1$ to  $y'=v'+1$,
namely 
\beqs
y'=r_{(p,\ell),q}(y) &=& \bigg [ \frac{(q+y-1)^\ell+(q-1)(y-1)^\ell}
  {(q+y-1)^\ell-(y-1)^\ell} \ \bigg ]^p \cr\cr
&=& \bigg [ \frac{x^\ell+(q-1)}{x^\ell-1}\bigg ]^p =
\bigg [ 1 + \frac{q}{x^\ell-1} \bigg ]^p \ , 
\label{yprime}
\eeqs
The transformation $r_{(p,\ell),q}(y)$  can be expressed as a twofold
composition \cite{qiao_memoir}.  For this purpose, let us define 
\beq
t_{a,q}(y) = \Big (1 + \frac{q}{y-1} \Big )^a = x^a \ . 
\label{taq}
\eeq
Then
\beq
r_{(p,\ell),q}(y) = t_{p,q}\Big ( t_{\ell,q}(y) \Big ) \ . 
\label{taqtaq}
\eeq

One can equivalently express $T(G^{(p,\ell)}_{m+1},x,y)$ in terms of
$T(G^{(p,\ell)}_m,x',y')$, but the transformation is more complicated because
both $x$ and $y$ change, to $x'$ and $y'$.  In contrast, in the transformation
(\ref{rg})-(\ref{A_L}) relating $Z(G^{(p,\ell)}_{m+1},q,v)$ to
$Z(G^{(p,\ell)}_m,q,v')$, there is a change in only one of the variables,
namely $v \to v'$ in Eq. (\ref{vprime}), but no change in $q$. 

The limit as $m \to \infty$ of this RG map is of particular interest.
For compact notation we define
\beqs
F^\infty_{(p,\ell),q}(v) &\equiv& \lim_{m \to \infty} F^m_{(p,\ell),q}(v) \ , 
\cr\cr
r^\infty_{(p,\ell),q}(y) &\equiv& \lim_{m \to \infty} r^m_{(p,\ell),q}(y) \ . 
\label{Fqv_inf}
\eeqs

Previous studies of the continuous accumulation set of chromatic zeros ${\cal
  B}_q$ for infinite-length limits of sections of regular lattices or chain
graphs with periodic (or twisted periodic) longitudinal boundary conditions
(which minimize finite-size effects) have shown that these are boundaries
(whence the symbol ${\cal B}$) that separate regions in the complex $q$-plane
where $W(G_\infty,q)$ takes on different analytic forms
\cite{baxter87}-\cite{a}.  This is true more generally for ${\cal B}_q$ for a
given value of $v$ different from $v=-1$ \cite{a}. The analogue in the complex
$v$-plane is that ${\cal B}_v$, for a given $q$, separates regions where the
(reduced) free energy of the Potts model takes on different analytic forms.
For the case of a hierarchical lattice, one can determine the region diagram as
a function of $v$ by calculating the asymptotic behavior of the $m$-fold
composition of the iteration transformation $F_q^m(v)$, or equivalently
$r_q^m(y)$, as $m \to \infty$. In particular, for the present case of
$G^{(p,\ell)}_m$ graphs, the continuous accumulation set of chromatic zeros is
determined by the behavior of $F_q^m(v)$ (resp. $r_q^m(y)$), starting with the
initial value $v=v_0=-1$ (resp., $y=y_0=0$), as $m \to \infty$. Henceforth, we
will refer to this limit simply as $F_q^\infty(-1)$ (resp. $r_q^\infty(0)$).
Corresponding to each of these types of behavior there are regions in the
complex $q$-plane.  The boundaries separating any two of these regions comprise
the locus ${\cal B}_q$ for the given value of $v$.  Insofar as we restrict our
consideration to the initial value of $v$ being $v=-1$, i.e., the
zero-temperature Potts antiferromagnet, we simplify the notation ${\cal
  B}_q(-1) \equiv {\cal B}(v=-1)$ to ${\cal B}_q$, with it being understood
that this symbol refers to the case $v=-1$.

When presenting our new results for ${\cal B}_q$ and associated region
diagrams here we will use the same color coding that we used in
Ref. \cite{dhl} with R. Roeder, namely:
\beq
  F^\infty_{(p,\ell),q}(-1) = 0, \ i.e., \
    r^\infty_{(p,\ell),q}(0) = 1: \quad {\rm white} 
\label{white}
\eeq
\beq
  F^\infty_{(p,\ell),q}(-1) = r^\infty_{(p,\ell),q}(0) = \infty: \quad 
{\rm blue} 
\label{blue}
\eeq
\beq
F^\infty_{(p,\ell),q}(-1) \neq 0, \ \infty, \ i.e., \ 
r^\infty_{(p,\ell),q}(0) \neq 1, \ \infty: \quad {\rm black}. 
\label{black}
\eeq
Physically (in the nontrivial case $J \ne 0$), $v=0$ means $\beta=0$, i.e.,
infinite temperature $T$ with either sign of $J$, while $v=\infty$ means $J >
0$ and $T=0$, the zero-temperature ferromagnet. The zero-temperature
antiferromagnetic case is not included among these options because the RG
transformation does not, in general, preserve a negative sign of the spin-spin
coupling, $J$, whereas, in contrast, it does preserve a positive sign of $J$.
We use the term ``region diagram'' (in the complex $q$-plane) to refer
to a plot of ${\cal B}_q(v_0)$ for general $v_0$ and ``chromatic
region diagram'' for the case in which the initial value $v_0=-1$,
the chromatic polynomial case. In both cases, the color
coding for white, blue, and black regions was given above.

We describe some details of our calculation of the chromatic region diagram 
and locus ${\cal B}_q(-1)$.  Starting with the initial value $v=v_0=-1$, we
choose a given point $q$ on the negative real axis and 
compute the $m$-fold composition of iterations $F^m_{(p,\ell),q}(-1)$ up to 
$m_{\rm max}=100$. Then 

\begin{enumerate}

\item 

When the absolute value $|v|$ decreases below 
$10^{-8}$ after a certain iteration stage $m \le m_{\rm max}$, 
the point $q$ is assigned to the white region. 

\item 

When the value of $|y|=|v+1|$ exceeds $10^8$ after a certain iteration stage 
$m \le m_{\rm max}$, the point $q$ is assigned to the blue region. 

\item 

If, after $m_{\rm max}=100$ iterations, the resultant value of $|v|$ is not 
less than $10^{-8}$ and the value of $|y|=|v+1|$ is not larger than 
   $10^8$, then the point $q$ is assigned to the black region. 

\end{enumerate}

After this assignment has been made for the point $q$, the same
procedure is carried out for the next value of $q$, chosen a small
distance $\epsilon$ to the right, i.e., equal to $[{\rm Re}(q) +
  \epsilon] + i \,{\rm Im}(q)$. The value of $\epsilon$ is typically
$3 \times 10^{-3}$ for many of the global plots and is reduced to
commensurately smaller values for plots showing detailed, magnified
views of relevant portions of the region diagram.  In effect, we make
a horizontal left-to-right scan in this manner. Then we increase ${\rm
  Im}(q)$ by $\epsilon$, starting with the initial point ${\rm Re}(q)
+ i\,[{\rm Im}(q)+\epsilon]$ and perform the corresponding horizontal
scan.  Since ${\cal B}_q(v)$ is invariant under complex conjugation
(recall Eq. (\ref{Bq_invariance})), the part of the region diagram
with ${\rm Im}(q) < 0$ is just the reflection about the real-$q$ axis
of the part with ${\rm Im}(q) > 0$ and does not require additional
calculation.  The iteration transformation $v' = F_{(p,\ell),q}(v)$ in
Eq. (\ref{vprime}) (or equivalently, $y' = r_{(p,\ell),q}(y)$) is a
rational function and hence in the procedure above, for a given
$(p,\ell)$, we calculate the values of $q$ where poles occur and avoid
them. For a given $(p,\ell)$, these poles are automatically in the
blue areas of the respective region diagrams.  We have performed a
number of checks on these calculations of chromatic region diagrams
and associated continuous accumulation sets of chromatic zeros ${\cal
  B}_q$ for the various $(p,\ell)$ cases that we consider.  In
particular, we have checked that our calculations of the crossing
points $q_c$, $q_\infty$, $q_{\rm int}$, $q_x$, and $q_L$ agree with
the respective points on the region diagrams computed for the various
$(p,\ell)$ cases.


\section{Locus of Chromatic Zeros of $G^{(p,\ell)}_\infty$ }
\label{general_locus_section}

\subsection{General Properties}
\label{locus_general}

In this section we analyze the continuous accumulation locus of the zeros of
the chromatic polynomial $P(G^{(p,\ell)}_m,q)$ in the complex $q$-plane in the
limit $m \to \infty$, denoted ${\cal B}_q(G^{(p,\ell)}_\infty)(v=-1)$, and
abbreviated as ${\cal B}_q(p,\ell)$. As mentioned above, to calculate ${\cal
  B}_q(p,\ell)$, we use the property that the locus ${\cal B}_q(p,\ell)$
separates regions of different asymptotic behavior of the $m$-fold iterated RG
function $F^m_{(p,\ell),q}(v)$ in the complex $q$-plane in the $m \to \infty$
limit, starting from the initial value $v=v_0=-1$.  Our current study extends
our previous work with R. Roeder in \cite{dhl} for the lowest nontrivial case,
$(p,\ell)=(2,2)$ to higher values of $p$ and $\ell$ (see also
\cite{chio_roeder} \cite{rig}).  Thus, we map out this locus by analyzing this
iterated transformation.  We scan over real and complex values of $q$ and, for
each value, we apply the iterated RG transformation sufficiently many times to
decide on which type of behavior occurs, as described in the previous section,
with the corresponding color coding.  As noted, to simplify the notation,
unless otherwise indicated, we will keep the the argument $v=-1$ for the locus
${\cal B}_q(-1)$ implicit and simply refer to this locus as ${\cal
  B}_q(p,\ell)$ or just ${\cal B}_q$.

We find the following general properties of ${\cal B}_q(p,\ell)$
that apply for all $p$ and $\ell$ under consideration: 

\begin{enumerate}

\item
  \label{finite}
  
  For given fixed, finite values of $p$ and $\ell$, ${\cal
    B}_q(p,\ell)$ extends over only a finite region of the complex
  $q$-plane.

\item
  \label{crossing_qc}
  
  The locus ${\cal B}_q(p,\ell)$ crosses the real-$q$ axis at a maximal point
  denoted $q_c(G^{(p,\ell)}_\infty)$, which depends on $p$ and $\ell$.  For
  compact notation, we will usually refer to $G^{(p,\ell)}_\infty$ simply as
  the $(p,\ell)$ case and will use the shorthand 
\beq
q_c(G^{(p,\ell)}_\infty) \equiv q_c(p,\ell) \ . 
\label{qcshorthand}
\eeq
We find that $q_c(p,\ell) > 2$ and observe several monotonicity relations for
$q_c(G^{p,\ell)}_\infty)$ as a function of $p$ and $\ell$, as will be discussed
below.

\item 
  \label{q_left}

  The locus ${\cal B}_q(p,\ell)$ crosses the real-$q$ axis at a leftmost ($L$)
  point $q_L(p,\ell)$.  For all of the cases that we have studied, we find
\beq
q_L(p,\ell) = 0 \quad {\rm if} \ p \le \ell 
\label{qLzero}
\eeq
and
\beq
q_L(p,\ell) < 0 \quad {\rm if} \ p > \ell \ . 
\label{qLneg}
\eeq
In our previous studies of loci ${\cal B}_q(v)$ for many families of graphs, we
showed that ${\cal B}_q(v=-1)$ crossed the real axis at a leftmost point
$q_L(G_\infty)=0$, but we also showed that there are self-dual families of
graphs where this leftmost crossing is shifted to the right, to
$q_L(G_\infty)=1$ \cite{sdg}.  Interestingly, as stated above as property
(\ref{qLneg}), in our current work, we have discovered the first cases, to our
knowledge, where $q_L$ is negative, i.e., the continuous accumulation set of
chromatic zeros locus ${\cal B}_q$ (formally, ${\cal B}_q(v=-1)$) crosses the
negative real-$q$ axis for a family of nonrandom graphs.  We are aware of only
one previous example where ${\cal B}_q$ crosses the negative real-$q$ axis,
namely for (the $n \to \infty$ limit of) a certain family of random graphs
\cite{dolan}.  The property that ${\cal B}_q$ crosses the negative-$q$ axis
implies that an infinite set of chromatic zeros approach arbitrarily close to
the negative real-$q$ axis. In this context, we recall that a general result in
graph theory is that for an arbitrary graph $G$, the chromatic polynomial
$P(G,q)$ never has a zero on the negative real axis.

\item
  \label{left_cusp}
  
  For all $(p,\ell)$ families the outer part of ${\cal B}_q(p,\ell)$ in the
  neighborhood of $q_L$ has the form of a cusp opening to the left.

\item
\label{white_outermost}
  
In the region of the complex $q$-plane outside of the outermost part of ${\cal
  B}_q(p,\ell)$ and extending infinitely far away from the origin, the limit
$F^\infty_{(p,\ell),q}(-1)=0$, color-coded white. Hence, along the real axis,
the semi-infinite white real intervals $q > q_c(p,\ell)$ and $q < q_L(p,\ell)$
are analytically connected via routes in the complex $q$-plane outside of the
outermost part of ${\cal B}_q$.

\item
  \label{black_right_of_zero}
  
  In the interval of the real-$q$ axis immediately to the right of the point
  $q_L$, the limit $F^\infty_{(p,\ell),q}(-1)$ is neither zero nor infinite,
  and is thus color-coded black. This also applies to the region in the
  complex-$q$ region in immediately adjacent to this real interval, i.e., 
  not separated from this real interval by a component of ${\cal B}_q(p,\ell)$.

\end{enumerate}

There are also a number of other properties of the loci ${\cal B}_q(p,\ell)$
that depend on $p$ and $\ell$.  We proceed to present and analyze figures
showing these loci and the associated region diagrams for a number of different
values of $p$ and $\ell$.  In later sections, we will present calculations of
specific points where ${\cal B}_q(p,\ell)$ intersects the real-$q$ axis,
including $q_c(p,\ell)$, and the reader may wish to consult those results while
viewing the plots to be given below. 

The loci ${\cal B}_q(p,\ell)$ exhibit certain common properties 
depending on whether $p$ and $\ell$ are even or odd.  Therefore, 
we divide our discussion according to these classes, namely 

\begin{itemize}

\item 
  $p$ even and $\ell$ even, denoted as $(p_{\rm even}, \ell_{\rm even})$ 

\item
  
  $p$ odd and $\ell$ even: \quad $(p_{\rm odd}, \ell_{\rm even})$ 

\item 
  
  $p$ odd and $\ell$ odd: \quad $(p_{\rm odd}, \ell_{\rm odd})$ 

\item

  $p$ even and $\ell$ odd : \quad $(p_{\rm even}, \ell_{\rm odd})$. 

\end{itemize}
For the figures showing region diagrams and loci ${\cal B}_q$ in each of these
classes, we order the presentation of figures by increasing values of $p$, and,
for each $p$, increasing values of $\ell$.


\subsection{ Chromatic Region Diagrams and Loci ${\cal B}_q$ 
for $p$ Even and $\ell$ Even}
\label{P_even_L_even}

\begin{figure}
  \begin{center}
\includegraphics[height=10cm]{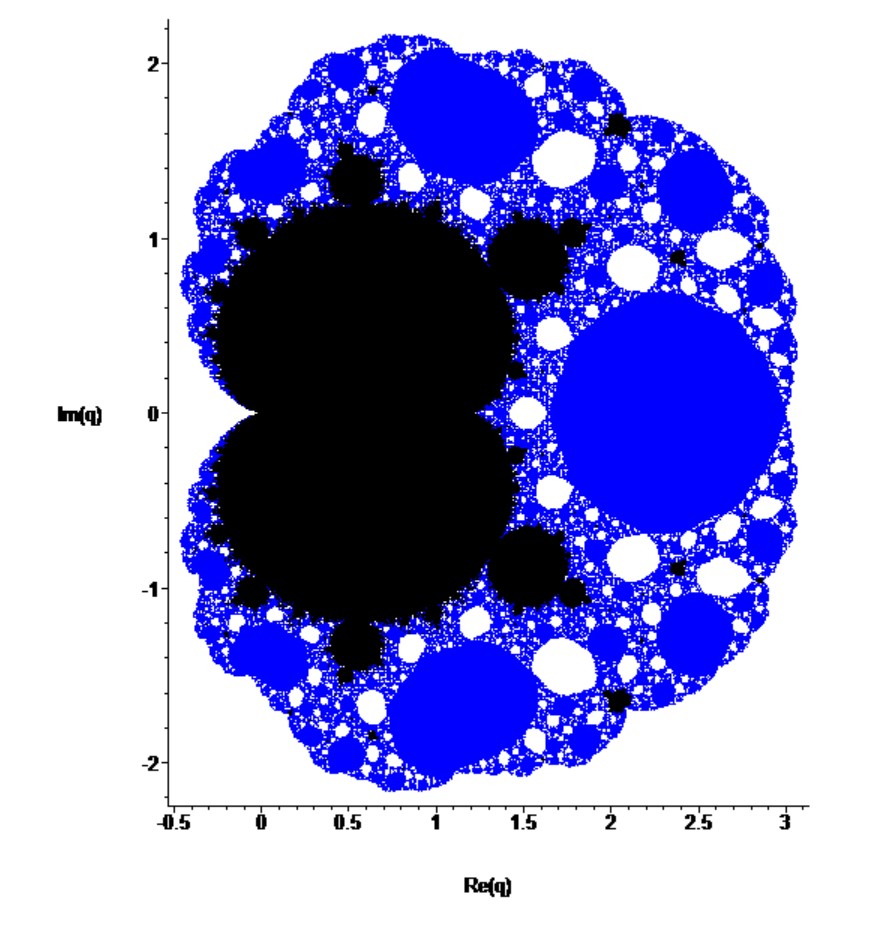}
    \end{center} 
  \caption{\footnotesize{Chromatic region diagram and locus ${\cal B}_q$ 
for $(p,\ell)=(2,2)$.}}
  \label{P2L2y0plot_fig}
\end{figure}

\begin{figure}
  \begin{center}
\includegraphics[height=10cm]{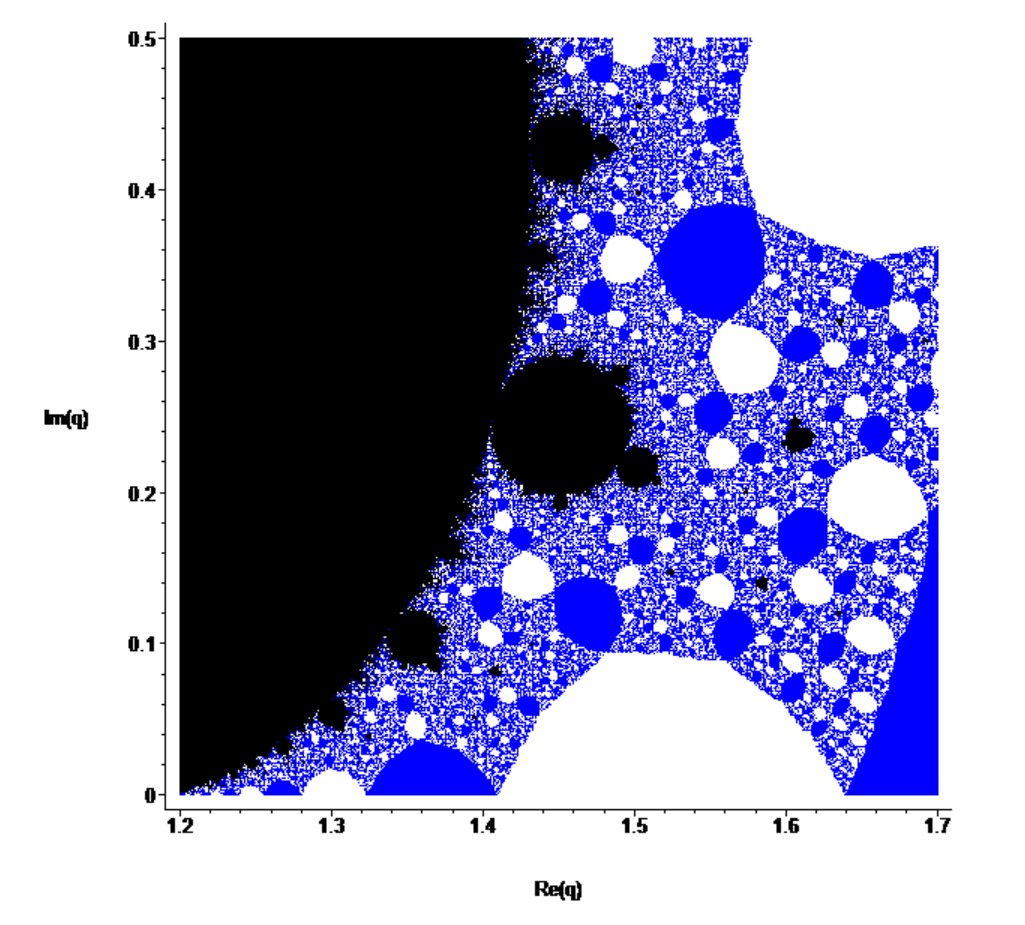}
    \end{center} 
    \caption{\footnotesize{Chromatic region diagram and locus ${\cal B}_q$ for
        $(p,\ell)=(2,2)$, showing detailed structure for the real interval $1.2
        < q < 1.7$ and associated area of the complex $q$-plane with ${\rm
          Im}(q) > 0$. This depicts part of the infinite sequence $S_\infty$ of
        crossings of the locus ${\cal B}_q$ on the real-$q$ axis. In this and
        similar detailed figures below, the corresponding area with ${\rm
          Im}(q) < 0$ is just the complex-conjugate and hence is not shown.}}
  \label{P2L2y0plotq1.2_1.7_fig}
\end{figure}

\begin{figure}
  \begin{center}
\includegraphics[height=10cm]{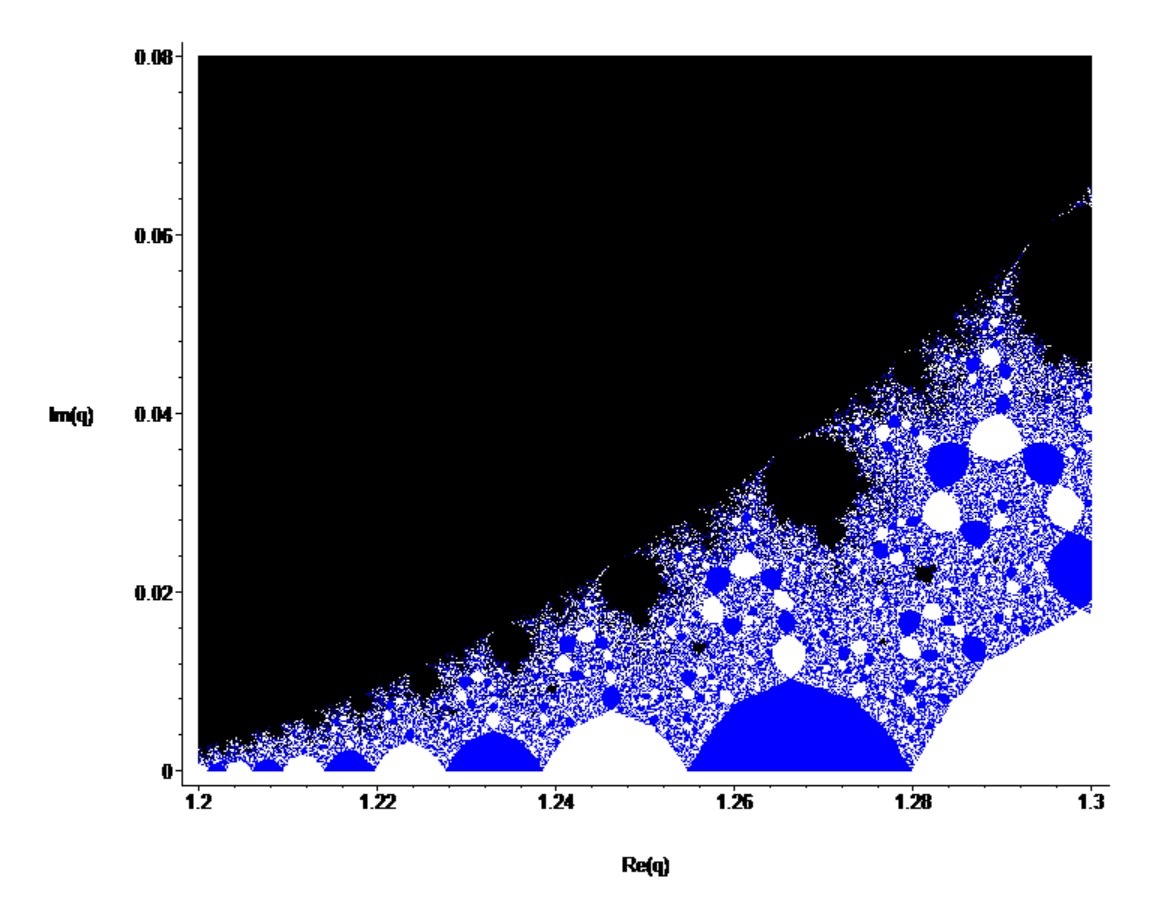}
    \end{center} 
  \caption{\footnotesize{Chromatic region diagram and locus ${\cal B}_q$
    for $(p,\ell)=(2,2)$, showing detailed structure for 
the real interval $1.2 < q < 1.3$ and associated area of the complex $q$-plane
with ${\rm Im}(q) > 0$.}}
  \label{P2L2y0plotq1.2_1.3_fig}
\end{figure}

\begin{figure}
  \begin{center}
\includegraphics[height=6cm]{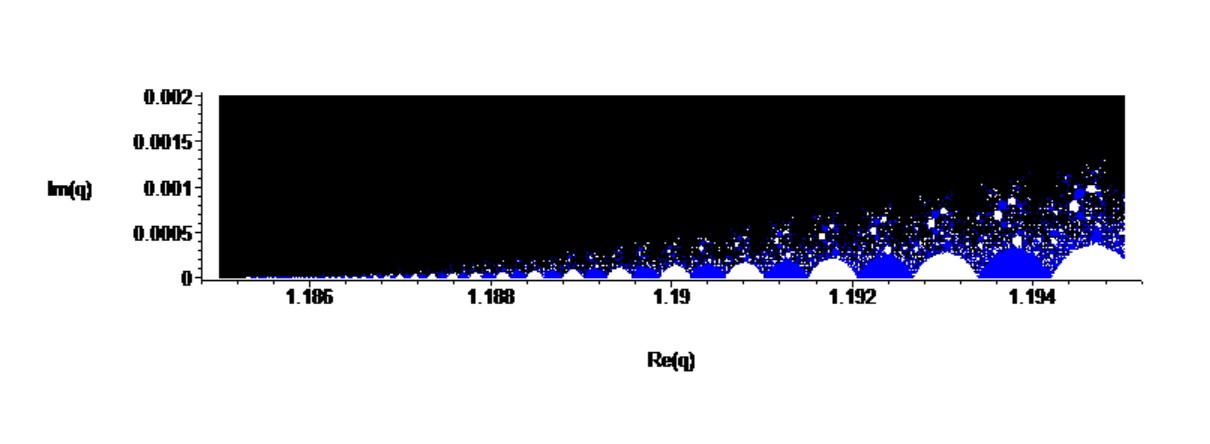}
    \end{center} 
  \caption{\footnotesize{Chromatic region diagram and locus ${\cal B}_q$
    for $(p,\ell)=(2,2)$, showing detailed structure for 
the real interval $1.185 < q < 1.195$ and associated area of the complex 
$q$-plane with ${\rm Im}(q) > 0$.}}
  \label{P2L2y0plotq1.185_1.198_fig}
\end{figure}

In this section we present and analyze the loci ${\cal B}_q$ and associated
region diagrams in the complex $q$ plane that we have calculated for the limits
$G^{(p,\ell)}_\infty$ with $(p_{\rm even},\ell_{\rm even})$.  We recall that
the term ``chromatic region diagram'' means that the initial value of $v=v_0$
for the RG iterations is $v_0=-1$, i.e., the zero-temperature Potts
antiferromagnet, for which the identity in Eq. (\ref{pz}) holds. 
We begin by giving some further details on the lowest nontrivial case,
$(p,\ell)=(2,2)$, which we analyzed previously with R. Roeder in \cite{dhl}.
In Fig. \ref{P2L2y0plot_fig} we show the general region diagram for this
case. This plot augments Fig. 3 in \cite{dhl} by the addition of
numerically marked coordinate axes for ${\rm Re}(q)$ and ${\rm Im}(q)$.  It was
shown in \cite{dhl} that ${\cal B}_q$ crosses the positive real-$q$ axis at an
infinite sequence of points that were denoted $q_r$, $r=1,2,...$.
In our current framework with general $p$ and $\ell$, it is necessary for 
clarity to explicitly indicate their dependence on $(p_{\rm even},\ell_{\rm
  even})$, writing $q_r(2,2)$.  As one moves
leftward from $q_c(2,2)=3$, this sequence starts with $q_1(2,2)=1.638897$ and
continues with $q_2(2,2)=1.4097005$, etc. \cite{dhl}. 
Figures \ref{P2L2y0plotq1.2_1.7_fig}, \ref{P2L2y0plotq1.2_1.3_fig},
and \ref{P2L2y0plotq1.185_1.198_fig} display successive portions of this
infinite sequence, moving from right to left. It was also shown in \cite{dhl}
that as one moves from right to left, this infinite sequence approaches the
limiting point
\beq
q_\infty(2,2)=\frac{32}{27} =1.185185...
\label{qinf_P2L2}
\eeq 
from above. Note that the point $q=33/27$, itself, is not a chromatic
zero; indeed, it is the upper boundary included in one of the zero-free regions
on the real-$q$ axis of a chromatic polynomial for an arbitrary graph, namely
the interval $(1,32/27]$ \cite{jackson}.

One of our important findings in the present work, generalizing the result for
$(p,\ell)=(2,2)$ in \cite{dhl}, is that the region diagrams in the $(p,\ell)$
families with even $p$ and even $\ell$, denoted $(p_{\rm even},\ell_{\rm
  even})$, include similar infinite sequences of points where ${\cal B}_q$
crosses the real-$q$ axis. We denote these points as $q_r(p_{\rm
  even},\ell_{\rm even})$, $r=1,2,...$, showing the dependence on the family
$(p_{\rm even},\ell_{\rm even})$, and we denote the overall infinite sequence
as $S_\infty(p_{\rm even},\ell_{\rm even})$. For each $(p_{\rm even}, \ell_{\rm
  even})$, as $r \to \infty$, this infinite sequence of crossing points on
${\cal B}_q$ converges from above to a limit point that we denote as
$q_\infty(p_{\rm even},\ell_{\rm even})$.  Summarizing the nature of this
sequence (with the dependence of $q_r$ and $q_\infty$ on $(p_{\rm
  even},\ell_{\rm even})$ understood implicitly), the infinite sequence
$S_\infty$ has the structure
\beq
S_\infty = \{ q_\infty < ... < q_4 < q_3 < q_2 < q_1 \ \} \ . 
\label{Sinf}
\eeq
For a given family $(p_{\rm even},\ell_{\rm even})$, as $r$ increases and one
moves to the left along the real-$q$ axis, the widths of each interval,
\beq
w_{r,r+1} \equiv q_r - q_{r+1} \ {\rm for \ a \ given} \ 
(p_{\rm even},\ell_{\rm even})
\label{width}
\eeq
become smaller, and approach zero as $r \to \infty$:
\beq
\lim_{r \to \infty} w_{r,r+1}(p_{\rm even},\ell_{\rm even}) = 0 \ . 
\label{width_to_zero}
\eeq
This property is, of course, necessary for the existence of the limit point 
$q_\infty(p_{\rm even},\ell_{\rm even})$.  Associated with each interval
$q_{r+1} < q < q_r$ in this sequence there is a ``bubble'' region in the 
complex-$q$ plane.  

\begin{figure}
  \begin{center}
\includegraphics[height=10cm]{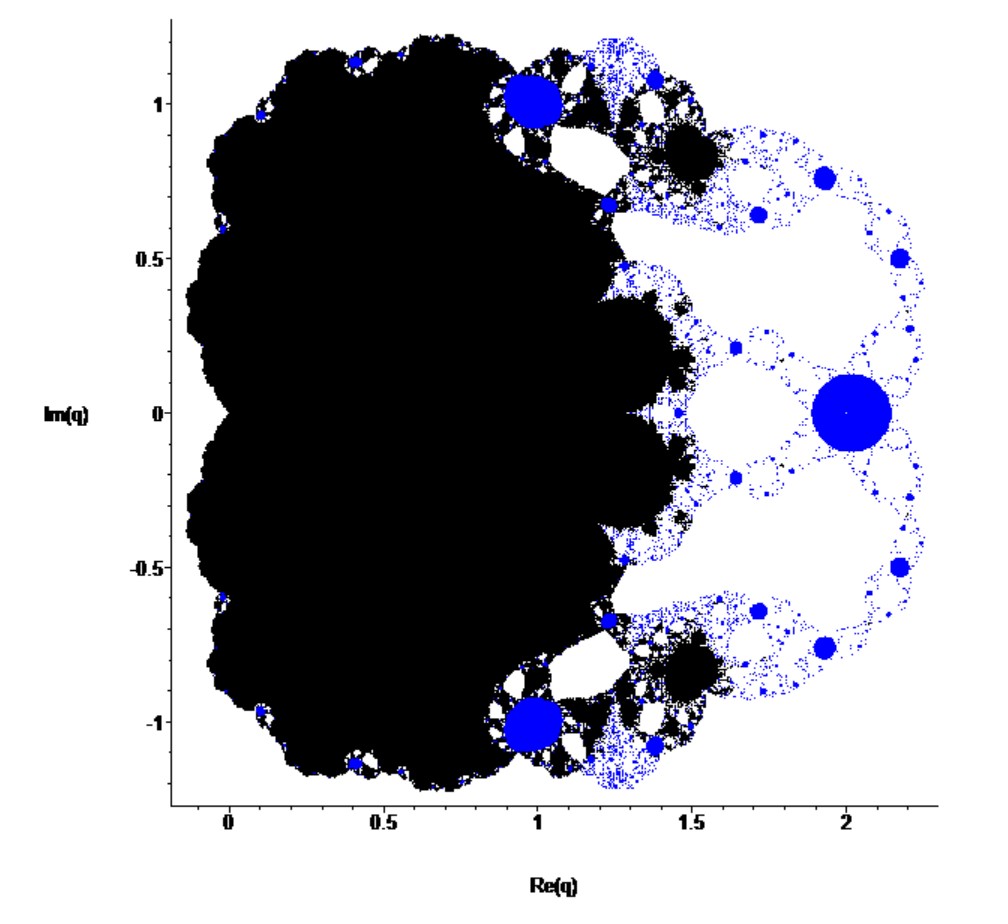}
    \end{center} 
  \caption{\footnotesize{Chromatic region diagram and locus ${\cal B}_q$ 
for $(p,\ell)=(2,4)$.}}
  \label{P2L4y0plot_fig}
\end{figure}

\begin{figure}
  \begin{center}
\includegraphics[height=10cm]{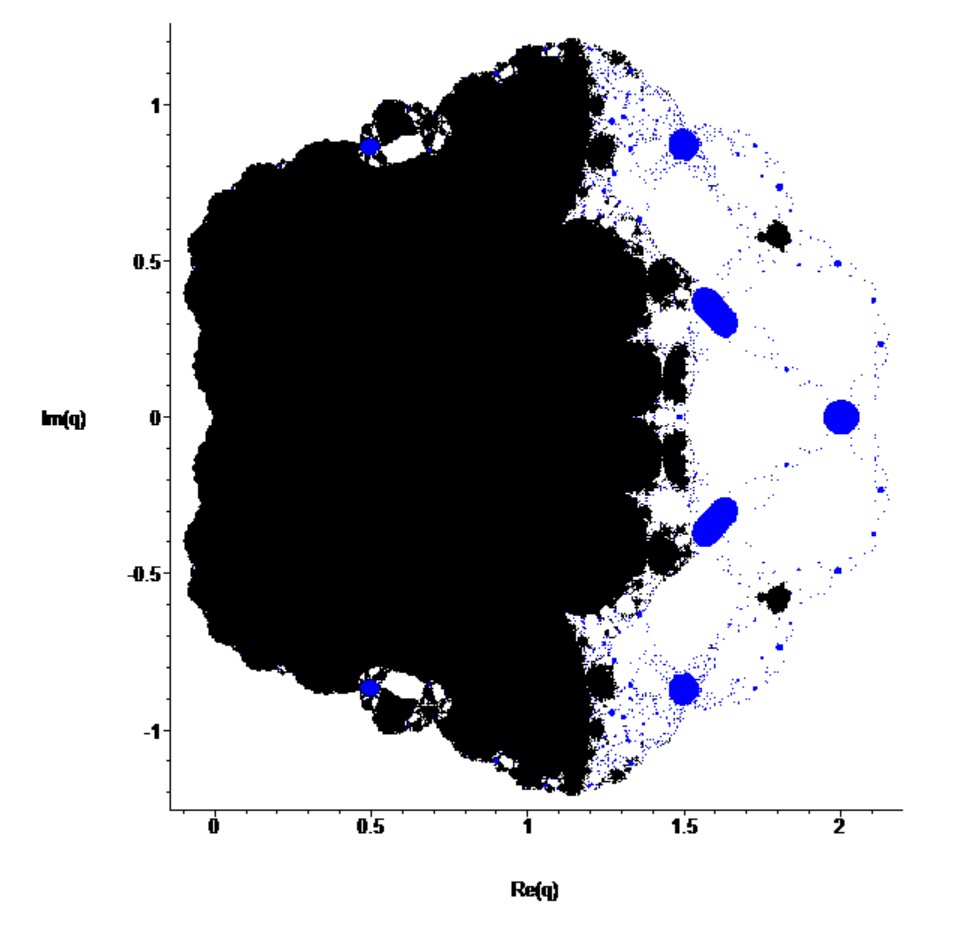}
    \end{center} 
  \caption{\footnotesize{Chromatic region diagram and locus ${\cal B}_q$ 
for $(p,\ell)=(2,6)$.}}
  \label{P2L6y0plot_fig}
\end{figure}

\begin{figure}
  \begin{center}
\includegraphics[height=10cm]{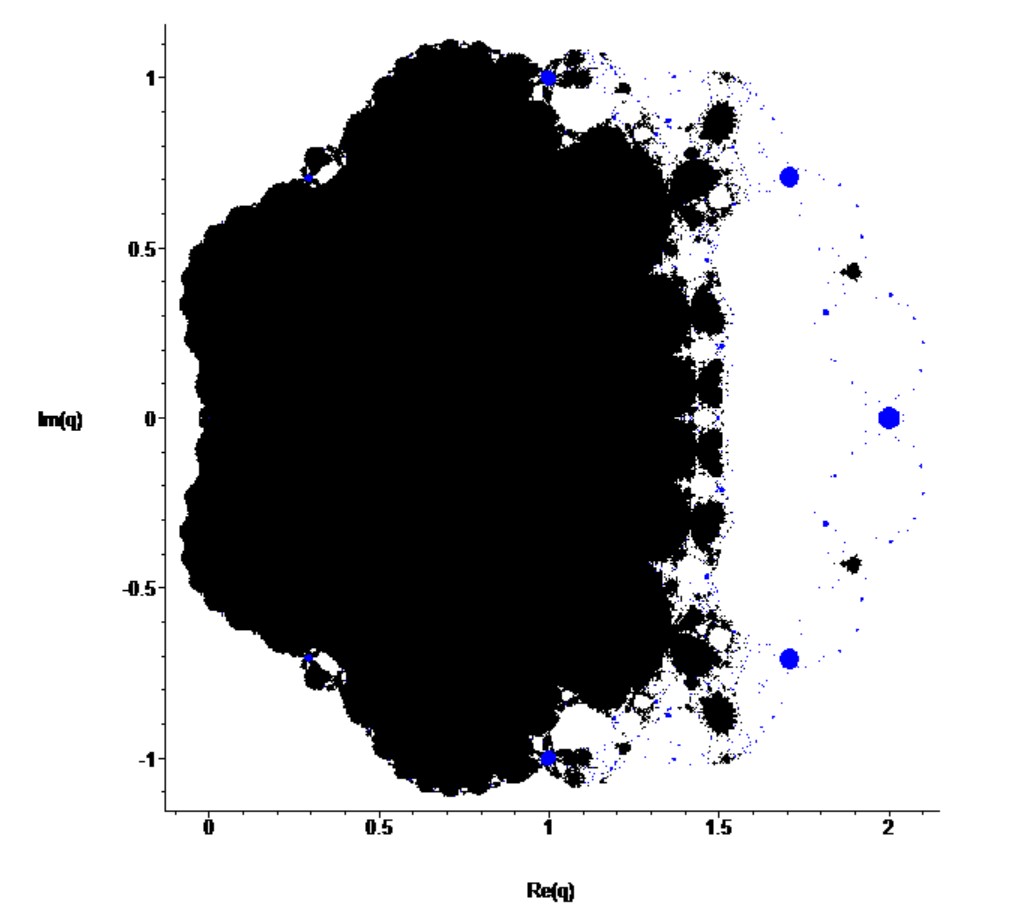}
    \end{center} 
  \caption{\footnotesize{Chromatic region diagram and locus ${\cal B}_q$ 
for $(p,\ell)=(2,8)$.}}
  \label{P2L8y0plot_fig}
\end{figure}

\begin{figure}
  \begin{center}
\includegraphics[height=10cm]{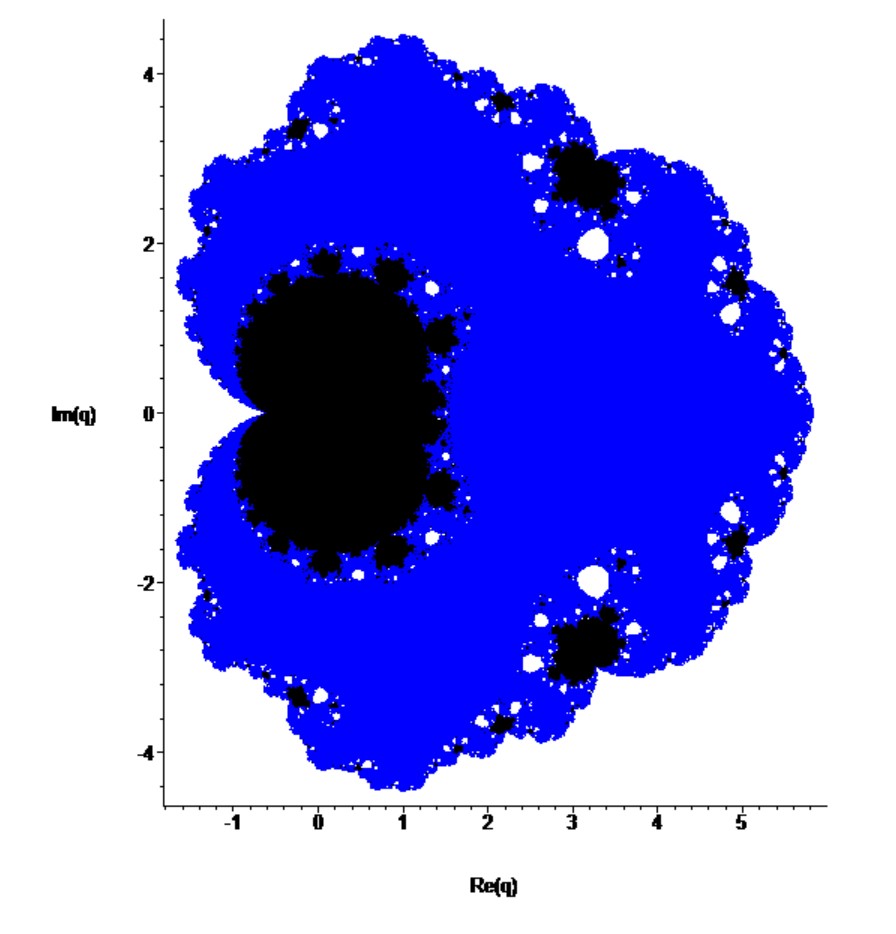}
    \end{center} 
  \caption{\footnotesize{Chromatic region diagram and locus ${\cal B}_q$ 
for $(p,\ell)=(4,2)$.}}
  \label{P4L2y0plot_fig}
\end{figure}

\begin{figure}
  \begin{center}
\includegraphics[height=10cm]{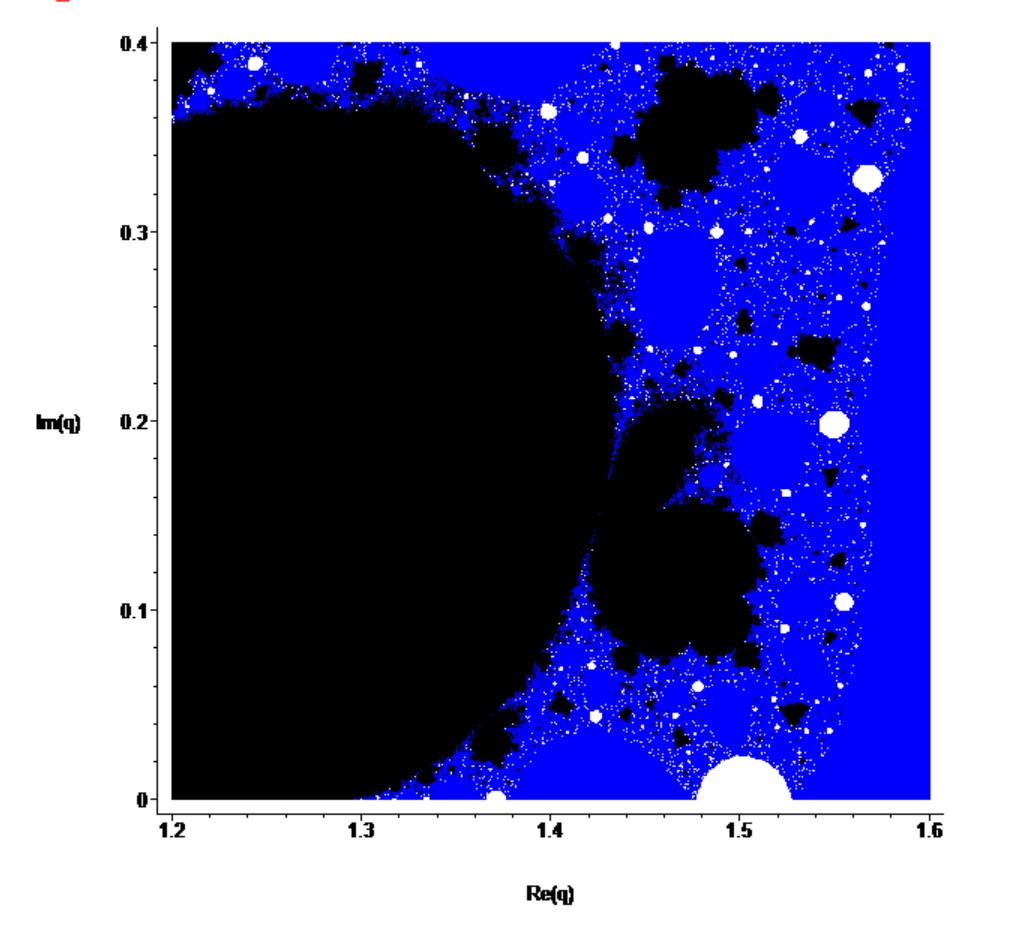}
    \end{center} 
    \caption{\footnotesize{Chromatic region diagram and locus ${\cal B}_q$ for
        $(p,\ell)=(4,2)$, showing detailed structure for the real interval $1.2
        < q < 1.6$ and associated area of the complex $q$-plane with ${\rm
          Im}(q) > 0$. This depicts part of the infinite sequence $S_\infty$ of
        crossings of the locus ${\cal B}_q$ on the real-$q$ axis.}}
  \label{P4L2y0plotq1.2_1.6_fig}
\end{figure}

\begin{figure}
  \begin{center}
\includegraphics[height=6cm]{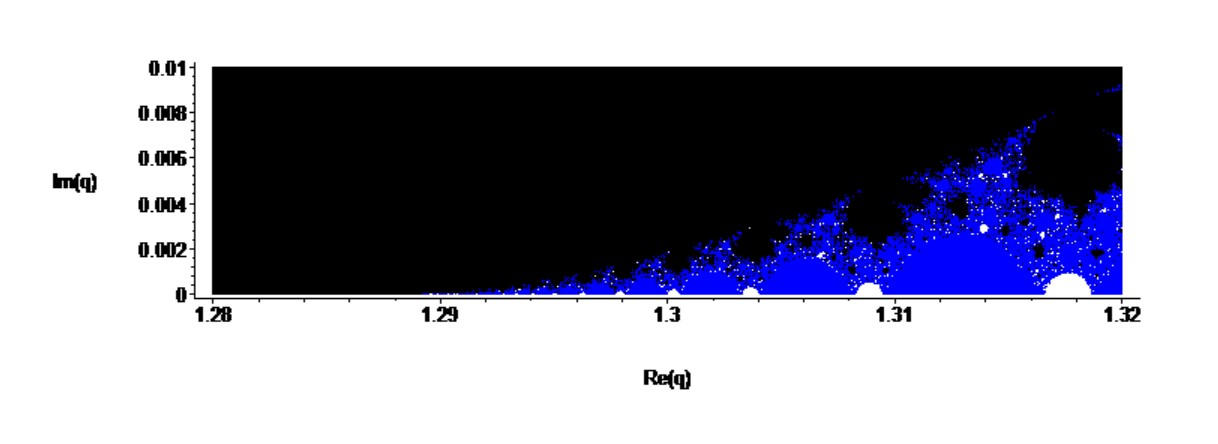}
    \end{center} 
    \caption{\footnotesize{Chromatic region diagram and locus ${\cal B}_q$ for
        $(p,\ell)=(4,2)$, showing detailed structure for the real interval
        $1.28 < q < 1.32$ and associated area of the complex $q$-plane with
        ${\rm Im}(q) > 0$. This displays part of the infinite sequence
        $S_\infty$ of crossings of the locus ${\cal B}_q$ on the real-$q$
        axis.}}
  \label{P4L2y0plotq1.28_1.32_fig}
\end{figure}

\begin{figure}
  \begin{center}
\includegraphics[height=10cm]{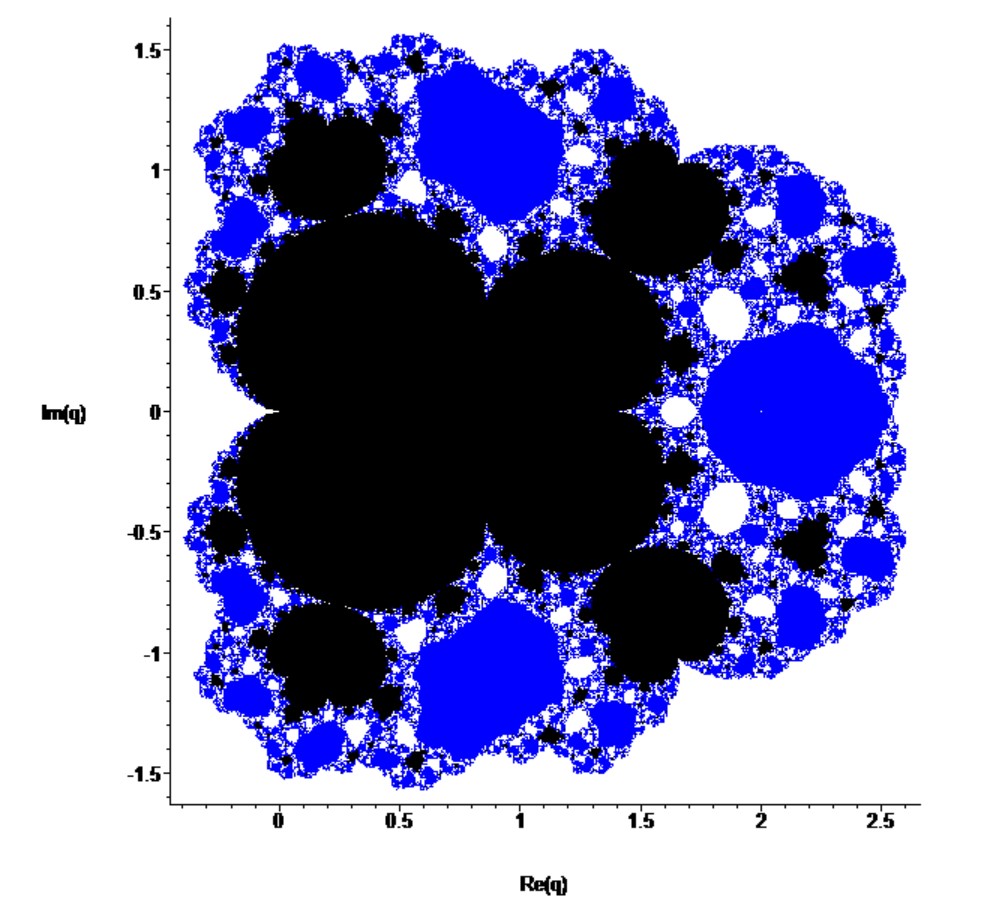}
    \end{center} 
  \caption{\footnotesize{Chromatic region diagram and locus ${\cal B}_q$ 
for $(p,\ell)=(4,4)$.}}
  \label{P4L4y0plot_fig}
\end{figure}

\begin{figure}
  \begin{center}
\includegraphics[height=10cm]{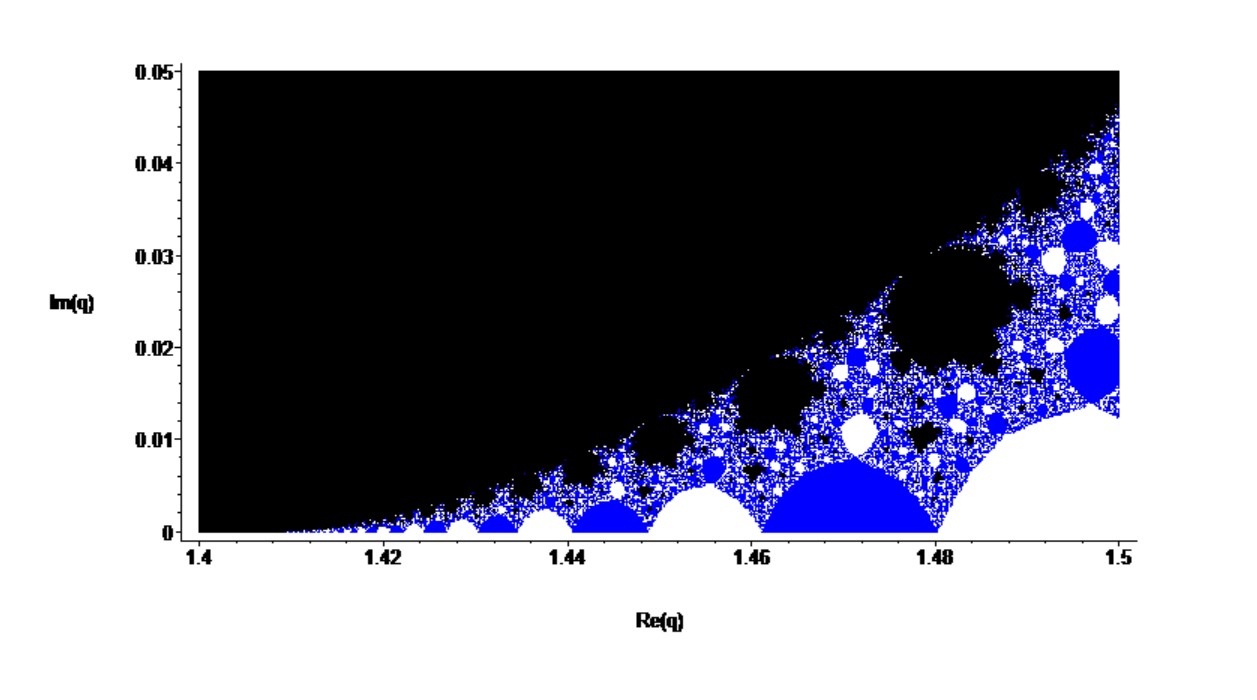}
    \end{center} 
    \caption{\footnotesize{Chromatic region diagram and locus ${\cal B}_q$ for
        $(p,\ell)=(4,4)$, showing detailed structure for the real interval $1.4
        < q < 1.5$ and associated area of the complex $q$-plane with ${\rm
          Im}(q) > 0$. This depicts part of the infinite sequence $S_\infty$ of
        crossings of the locus ${\cal B}_q$ on the real-$q$ axis. In this and
        similar detailed figures below, the corresponding area with ${\rm
          Im}(q) < 0$ is just the complex-conjugate and hence is not shown.}}
  \label{P4L4y0plotq1.4_1.5_fig}
\end{figure}

\begin{figure}
  \begin{center}
\includegraphics[height=10cm]{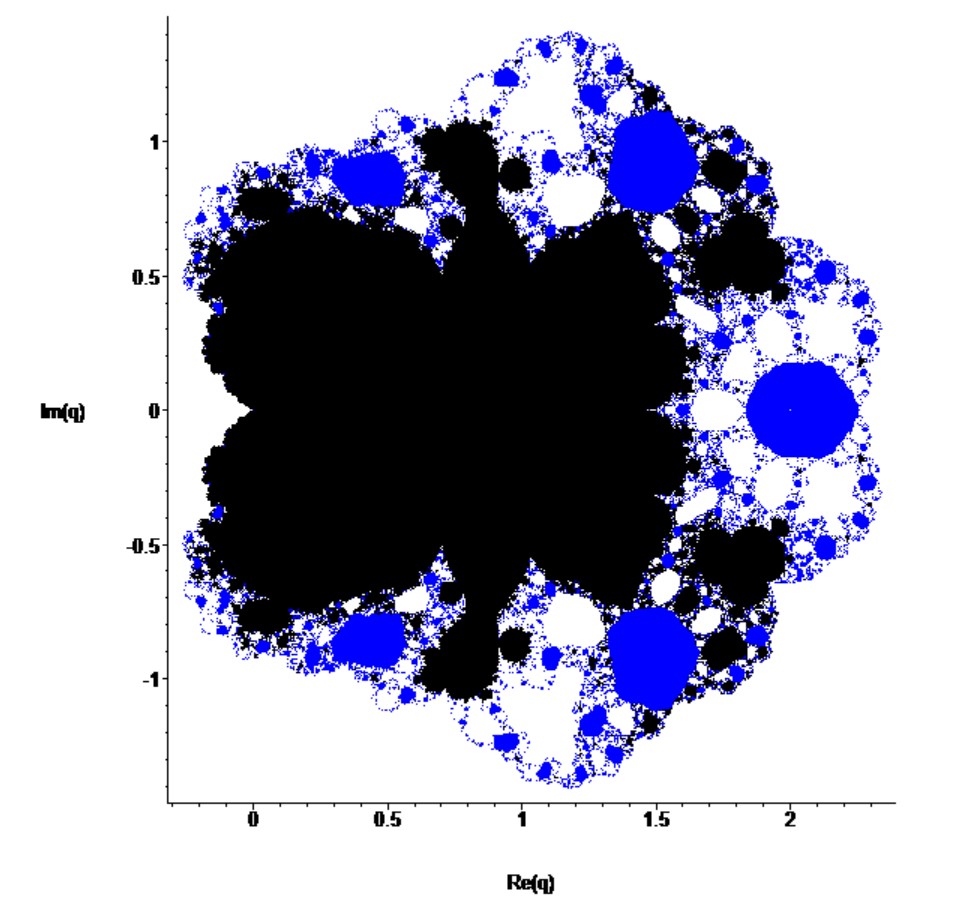}
    \end{center} 
    \caption{\footnotesize{Chromatic region diagram and locus ${\cal B}_q$ for
        $(p,\ell)=(4,6)$.}}
  \label{P4L6y0plot_fig}
\end{figure}

\begin{figure}
  \begin{center}
\includegraphics[height=10cm]{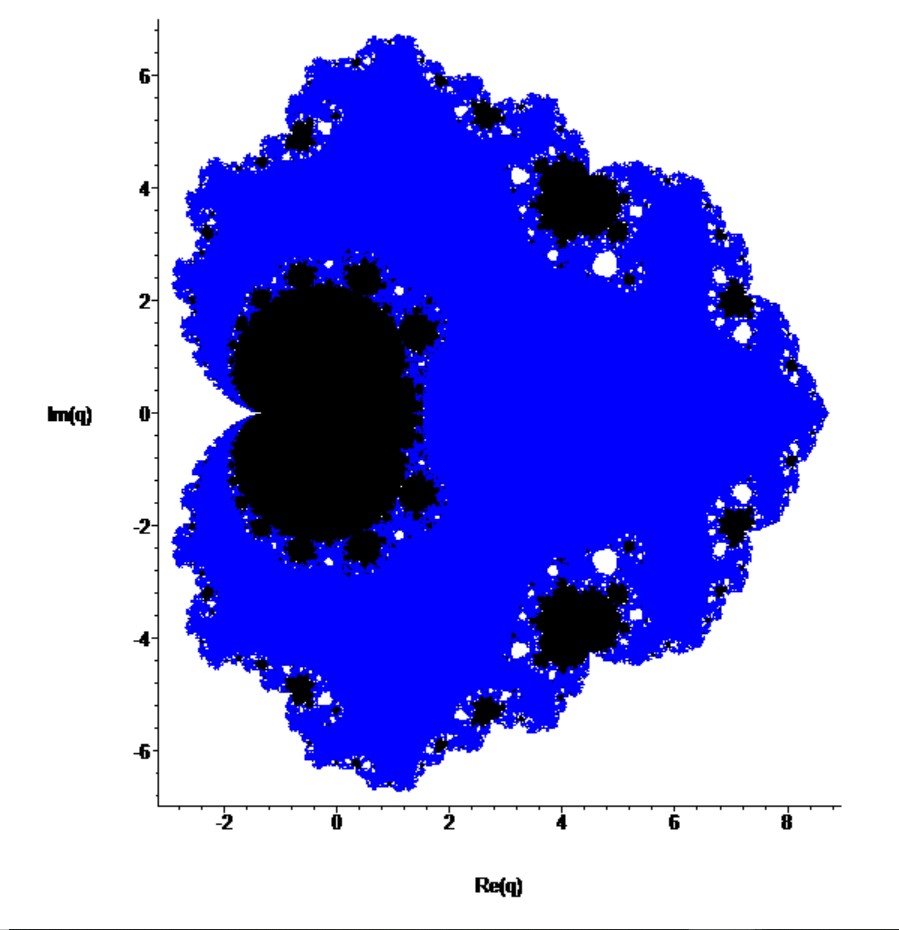}
    \end{center} 
  \caption{\footnotesize{Chromatic region diagram and locus ${\cal B}_q$ 
for $(p,\ell)=(6,2)$.}}
  \label{P6L2y0plot_fig}
\end{figure}

\begin{figure}[htbp]
  \begin{center}
\includegraphics[height=10cm]{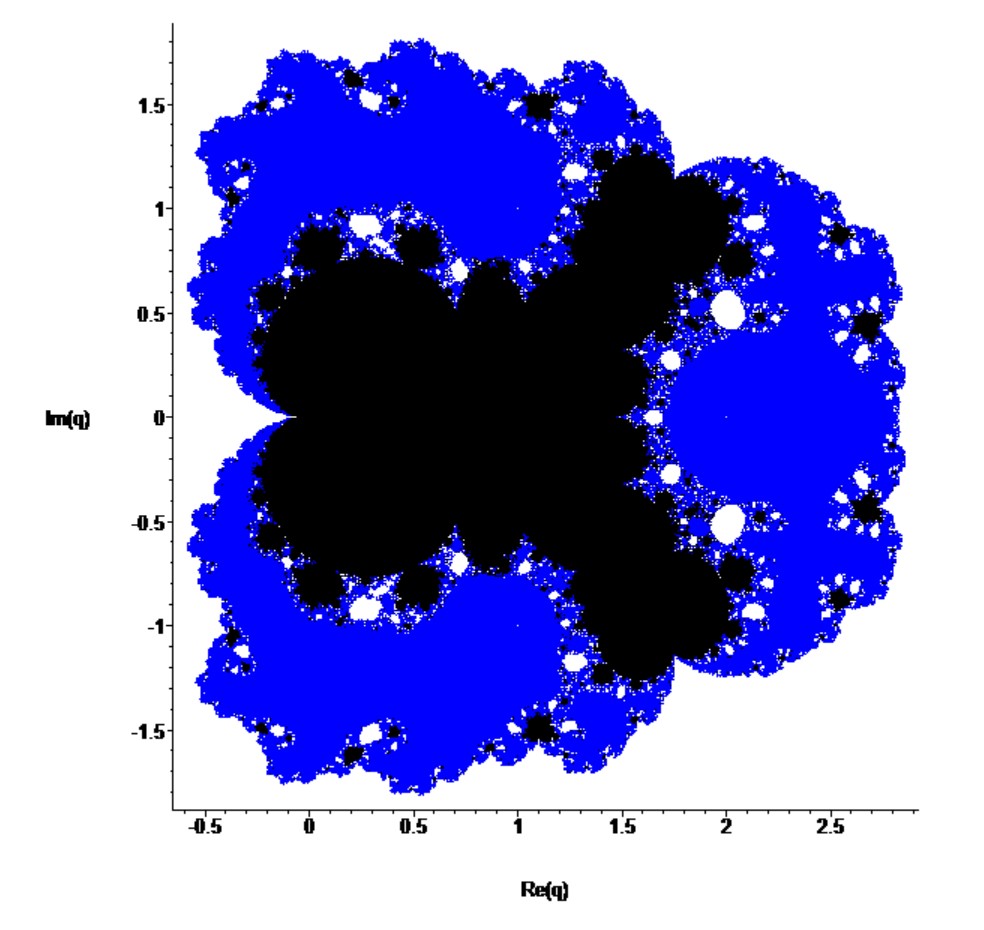}
    \end{center} 
  \caption{\footnotesize{Chromatic region diagram and locus ${\cal B}_q$ 
for $(p,\ell)=(6,4)$.}}
  \label{P6L4y0plot_fig}
\end{figure}

\begin{figure}[htbp]
  \begin{center}
\includegraphics[height=10cm]{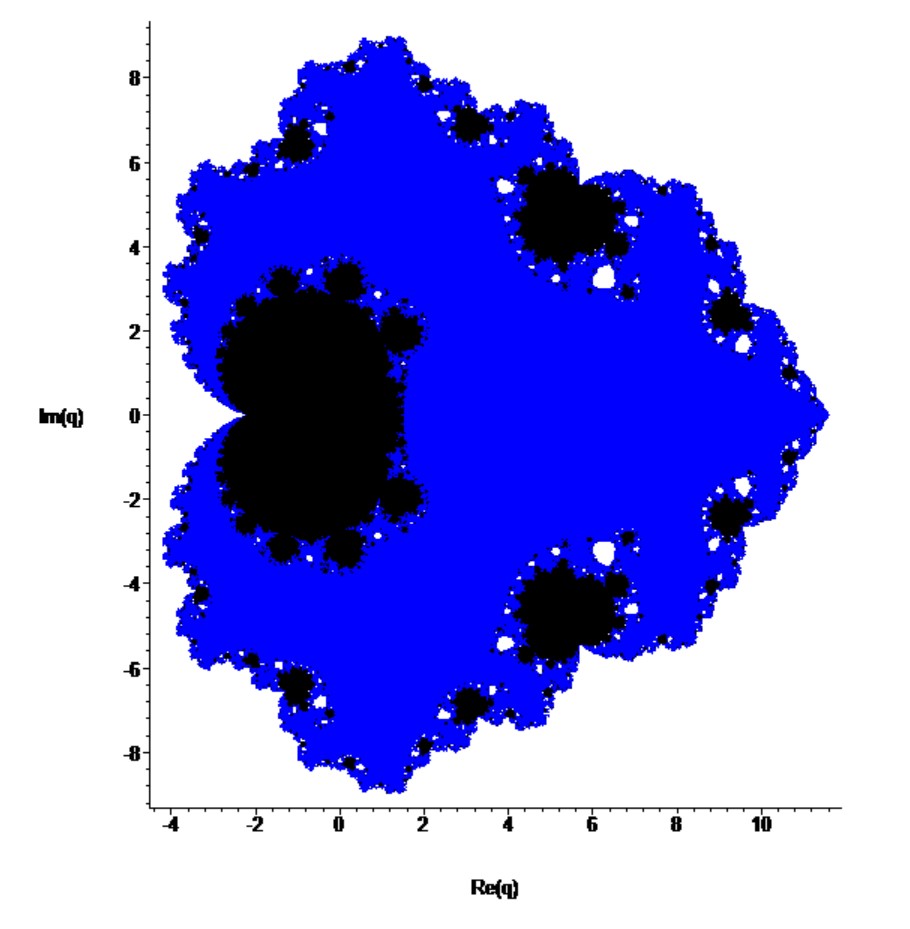}
    \end{center} 
  \caption{\footnotesize{Chromatic region diagram and locus ${\cal B}_q$ 
for $(p,\ell)=(8,2)$.}}
  \label{P8L2y0plot_fig}
\end{figure}

As illustrations of cases with higher even values of $p$ and $\ell$
for which we have calculated the regions diagrams, we show these for
$(p,\ell)=(2,4)$, (2,6), (2,8), (4,2), (4,4), (4,6), (6,2), (6,4), and
(8,2) in Figs. \ref{P2L4y0plot_fig}-\ref{P8L2y0plot_fig}.  In the
cases $(p,\ell)=(4,2)$ and (4,4), we present detailed plots of the
sequence of crossing points and associated regions in the interior
interval $0 < q < q_c(p,\ell)$. As noted, for a given case $(p_{\rm
  even}, \ell_{\rm even})$.  the $S_\infty$ sequences terminate on the
left at $q_{\infty}(p_{\rm even},\ell_{\rm even})$, with the minimal
  value being $q_\infty(2,2)$ in Eq. (\ref{qinf_P2L2}) (see Table
  \ref{qinf_table}). Although these form infinite sequences, one can
  only see the first roughly ten of the crossing points and associated
  regions with the finite grid used for these plots.  Similar
  sequences of crossing points and associated regions are evident in
  the $(p,\ell)=(2,6)$, (2,8), (4,6), and (6,4) cases.  In viewing
  these figures, the reader should recall the color coding defined in
  Eqs. (\ref{white})-(\ref{blue}).

We comment on the renormalization-group properties of the intervals
and associated bubble regions.  In each of these $(p_{\rm even},
\ell_{\rm even})$ cases, as one moves from right to left, one first
passes from the exterior white region to a blue region as one crosses
the respective $q_c(p_{\rm even},\ell_{\rm even})$ point.  Then moving
leftward, (i) as one passes a point denoted $q_1(p_{\rm
  even},\ell_{\rm even})$), one crosses from the blue region into a
white region; (ii) then, moving further left, as one passes the point
$q_2(p_{\rm even},\ell_{\rm even})$, one crosses into a blue region;
(iii) then, as one passes the point $q_3(p_{\rm even},\ell_{\rm
  even})$, one crosses into a white region, and so forth, until (iv)
as one moves leftward through the leftmost limiting point in the
infinite sequence, $q_\infty(p_{\rm even},\ell_{\rm even})$, one
crosses into a black region. Finally, as one moves leftward past the
point $q_L(p,\ell_{\rm even})$, one re-enters the external white
region.  This RG behavior can be summarized symbolically (suppressing
the dependence on $(p_{\rm even}, \ell_{\rm even})$ in the points
$q_c$, $q_1$, $q_2$..., $q_L$) as
\beq
(p_{\rm even}, \ell_{\rm even}): \quad 
{\rm white} < q_L < {\rm black} < S_\infty < {\rm blue} < q_c < 
{\rm white} \ , 
\label{P_even_L_even_intervals}
\eeq
where the symbol $S_\infty$ was defined in (\ref{Sinf}). 
In this infinite sequence $S_\infty$ of intervals, with the crossings 
$q_r$, $r=1,2,...$ enumerated going
from right to left, the RG behavior in each interval is white in the interval
$q_2 < q < q_1$, then blue in the interval $q_3 < q < q_2$, and so forth, 
{\it ad infinitem}, summarized as follows, with $r=1,2,...$:
\beqs
q \in S_\infty: \quad && q_{r+1} < q < q_r, \ r \ {\rm odd}: \ {\rm white}
 \cr\cr
                      && q_{r+1} < q < q_r, \ r \ {\rm even}: \ {\rm blue} \ .
\label{Sinf_colors}
\eeqs
As $p$ and $\ell$ increase, just as the real intervals $q_r - q_{r+1}$ become
progressively smaller, so also the bubble regions associated with these
intervals become progressively smaller and more difficult to see. For example,
in the general region diagram for $(p,\ell)=(4,2)$ shown in
Fig. \ref{P4L2y0plot_fig}, it is difficult to see this internal crossing-point
sequence, but it becomes evident when one inspects the detailed, magnified
views shown in Figs. \ref{P4L2y0plotq1.2_1.6_fig} and
\ref{P4L2y0plotq1.28_1.32_fig}.  Similar comments apply for higher values of
$p$ and $\ell$ in this $(p_{\rm even},\ell_{\rm even})$ subclass. For instance,
although the internal crossing-point sequences are also not evident in the
global plots for the (6,2) or (8,2) cases, we have checked with very magnified
views that they are present. In a later section we will discuss the calculation
of $q_\infty$ for these $(p_{\rm even},\ell_{\rm even})$ cases and, more
generally the calculation of $q_c(p,\ell)$ and $q_L(p,\ell)$.

Concerning the structure of the region diagrams away from the real axis, it is
noteworthy that for $(p_{\rm even},\ell_{\rm even})=(2,4)$, (2,6), and (2,8),
(i) there are extensive white subregions in the right-hand part of the
respective loci ${\cal B}_q$, and (ii) especially for the (2,6) and (2,8)
cases, there appear many very small ``dust''-like regions in this right-hand
area. We also find very small dust-like structures in the right-hand part of
the respective loci ${\cal B}_q$ in various other cases, such as
$(p,\ell)=(3,6)$, (3,7), (2,3), (2,5), and (2,7). To the accuracy of our
numerical calculations, we conclude that the respective continuous accumulation
sets of chromatic zeros ${\cal B}_q(p,\ell)$ with $v=-1$ for these cases are
disconnected.  There is, indeed, also a question of connectivity of ${\cal
  B}_q$ for other $(p,\ell)$ cases that do not exhibit ``dust''-like features.
Although there are known results on connectivity of the Julia set (in the $v$
plane) of $F_{(p,\ell),q}(v)$ \cite{qiao_memoir}, we are not aware of
mathematical theorems on connectivity of the locus ${\cal B}_q(p,\ell)$ for
$v=-1$ and general $(p,\ell)$.  This subject merits further study.



%

%

%

%

%

%

%

%

%

%

%

%

%

%

%

%


\clearpage

\subsection{ Chromatic Region Diagrams and Loci ${\cal B}_q$ 
for $p$ Odd and $\ell$ Even}
\label{P_odd_L_even}

\begin{figure}[htbp]
  \begin{center}
\includegraphics[height=10cm]{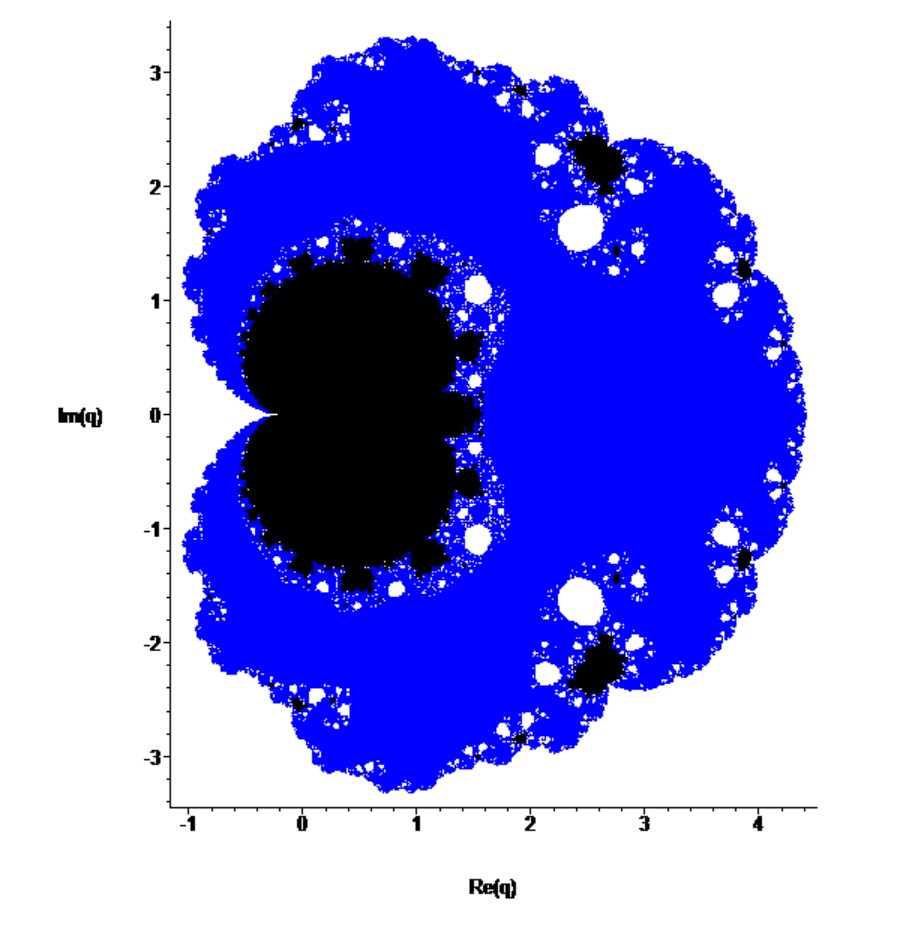}
    \end{center} 
  \caption{\footnotesize{Chromatic region diagram and locus ${\cal B}_q$ 
    for $(p,\ell)=(3,2)$.}}
  \label{P3L2y0plot_fig}
\end{figure}

\begin{figure}[htbp]
  \begin{center}
\includegraphics[height=10cm]{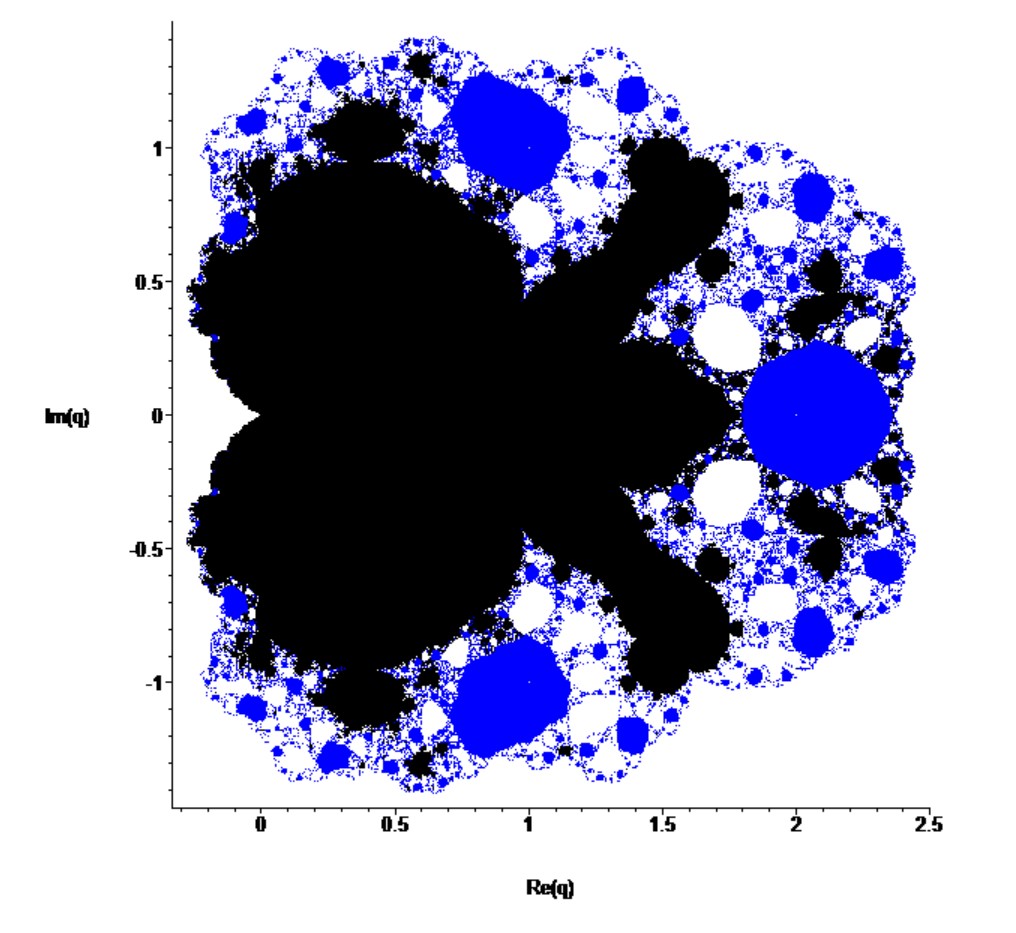}
    \end{center} 
  \caption{\footnotesize{Chromatic region diagram and locus ${\cal B}_q$ 
    for $(p,\ell)=(3,4)$.}}
  \label{P3L4y0plot_fig}
\end{figure}

\begin{figure}[htbp]
  \begin{center}
\includegraphics[height=10cm]{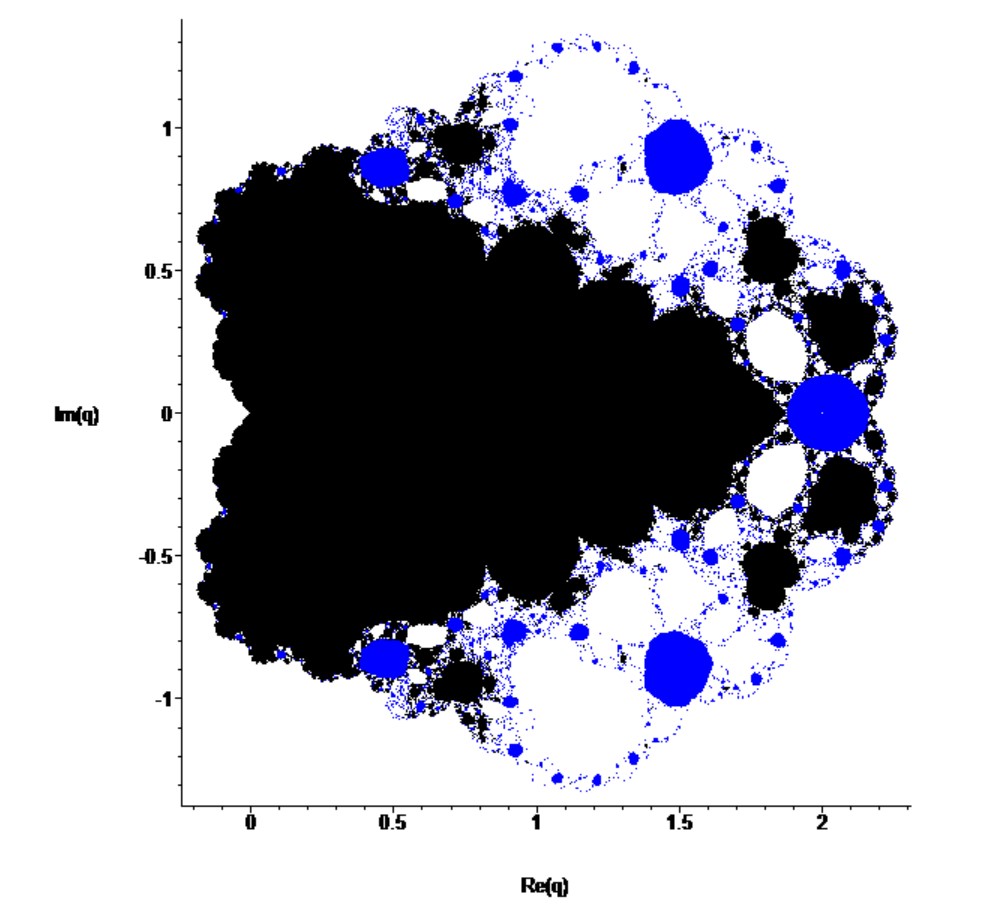}
    \end{center} 
  \caption{\footnotesize{Chromatic region diagram and locus ${\cal B}_q$ 
    for $(p,\ell)=(3,6)$.}}
  \label{P3L6y0plot_fig}
\end{figure}

\begin{figure}[htbp]
  \begin{center}
\includegraphics[height=10cm]{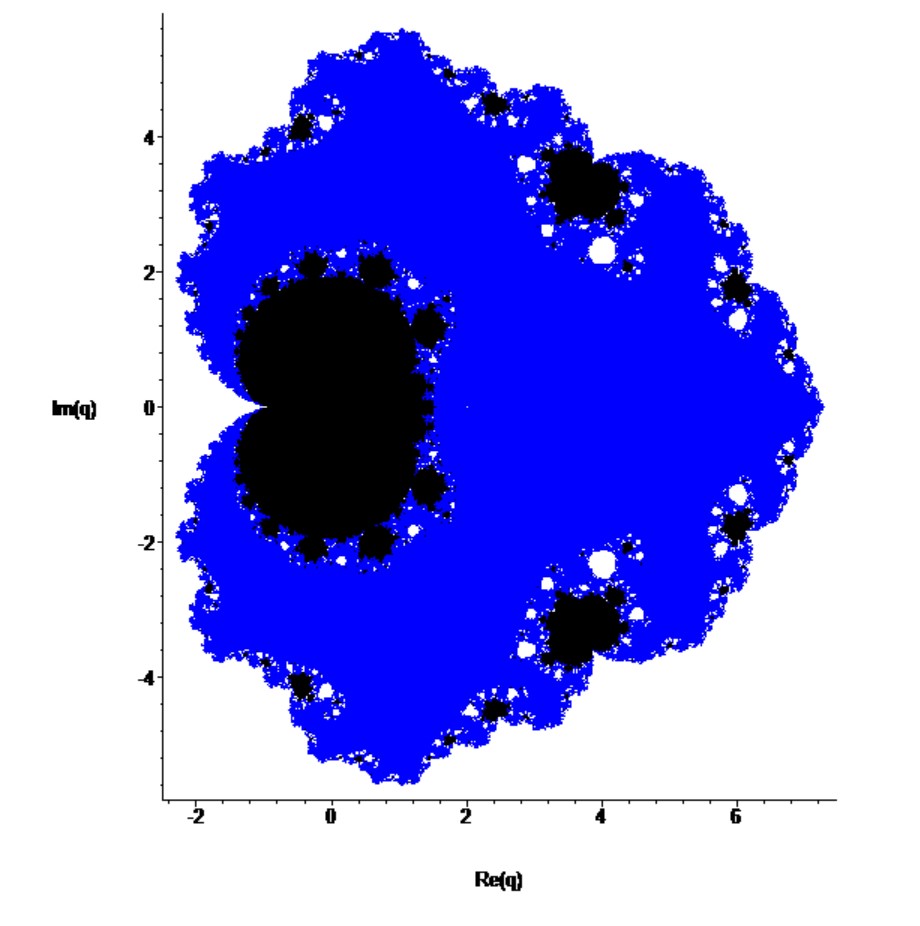}
    \end{center} 
  \caption{\footnotesize{Chromatic region diagram and locus ${\cal B}_q$ 
    for $(p,\ell)=(5,2)$.}}
  \label{P5L2y0plot_fig}
\end{figure}

\begin{figure}[htbp]
  \begin{center}
\includegraphics[height=10cm]{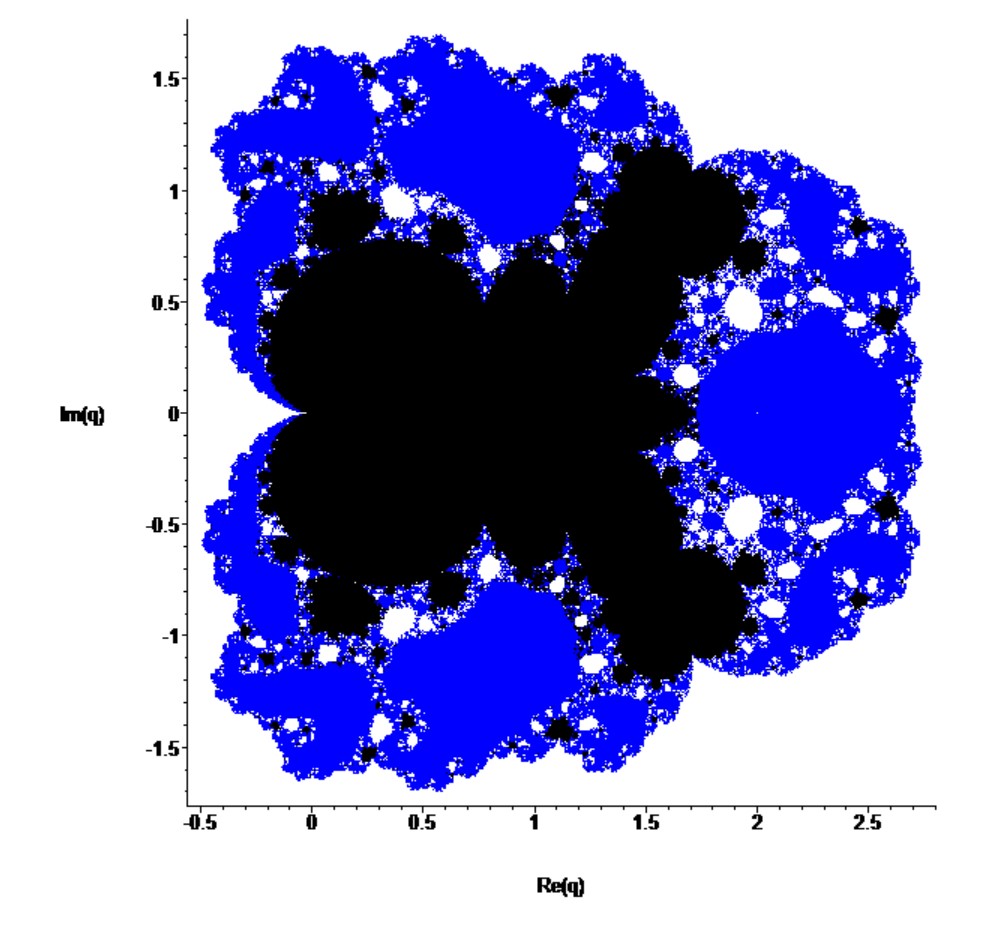}
    \end{center} 
  \caption{\footnotesize{Chromatic region diagram  and locus ${\cal B}_q$ 
    for $(p,\ell)=(5,4)$.}}
  \label{P5L4y0plot_fig}
\end{figure}

\begin{figure}[htbp]
  \begin{center}
\includegraphics[height=10cm]{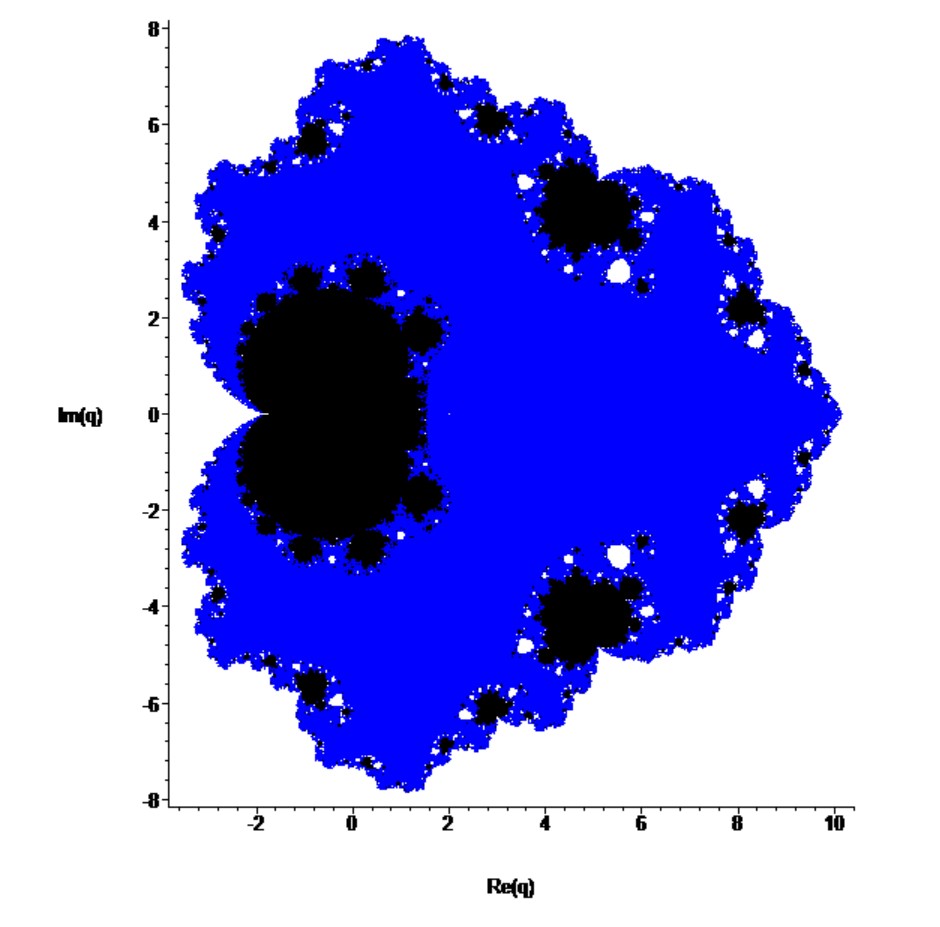}
    \end{center} 
  \caption{\footnotesize{Chromatic region diagram and locus ${\cal B}_q$ 
    for $(p,\ell)=(7,2)$.}}
  \label{P7L2y0plot_fig}
\end{figure}

We next present chromatic region diagrams and loci ${\cal B}_q$ for several
illustrative cases with odd $p$ and even $\ell$, namely
$(p,\ell)=(3,2)$, (3,4), (3,6), (5,2), (5,4), and (7,2) in Figs.
\ref{P3L2y0plot_fig}-\ref{P7L2y0plot_fig}. A general feature that we
find for these cases is that there is one and only one point at which
${\cal B}_q$ crosses the interior real interval 
$q_L(p_{\rm odd},\ell_{\rm even}) < q < q_c(p_{\rm odd},\ell_{\rm even})$,
which we denote as $q_{\rm int}(p_{\rm odd},\ell_{\rm even})$,
where the subscript ``int''
stands for ``interior''.  Calculations of 
$q_c(p_{\rm odd},\ell_{\rm even})$,
$q_L(p_{\rm odd},\ell_{\rm even})$, and 
$q_{\rm int}(p_{\rm odd},\ell_{\rm even})$ will be given below. 
Moving from right to left, as
one passes from the exterior region, which is white, through
$q_c(p,\ell)$, to the interior, one passes into a blue region; then,
as one passes through the single interior crossing point $q_{\rm
  int}(p,\ell)$, one enters a black region, and finally, as one passes
through $q_L(p,\ell)$, one re-enters the external white region. This can be
summarized symbolically as 
\beq
(p_{\rm odd}, \ell_{\rm even}): \quad
{\rm white} < q_L < {\rm black} < q_{\rm int} < {\rm blue} < q_c < 
{\rm white}  \ , 
\label{P_odd_L_even_intervals}
\eeq
where we suppress the dependence of $q_L$, $q_{\rm int}$ and $q_c$ on $(p_{\rm
  odd},\ell_{\rm even})$. 

%

%

%

%

%

%


\clearpage

\subsection{ Chromatic Region Diagrams and Loci ${\cal B}_q$ 
for $p$ Odd and $\ell$ Odd}
\label{P_odd_L_odd}

\begin{figure}[htbp]
  \begin{center}
\includegraphics[height=10cm]{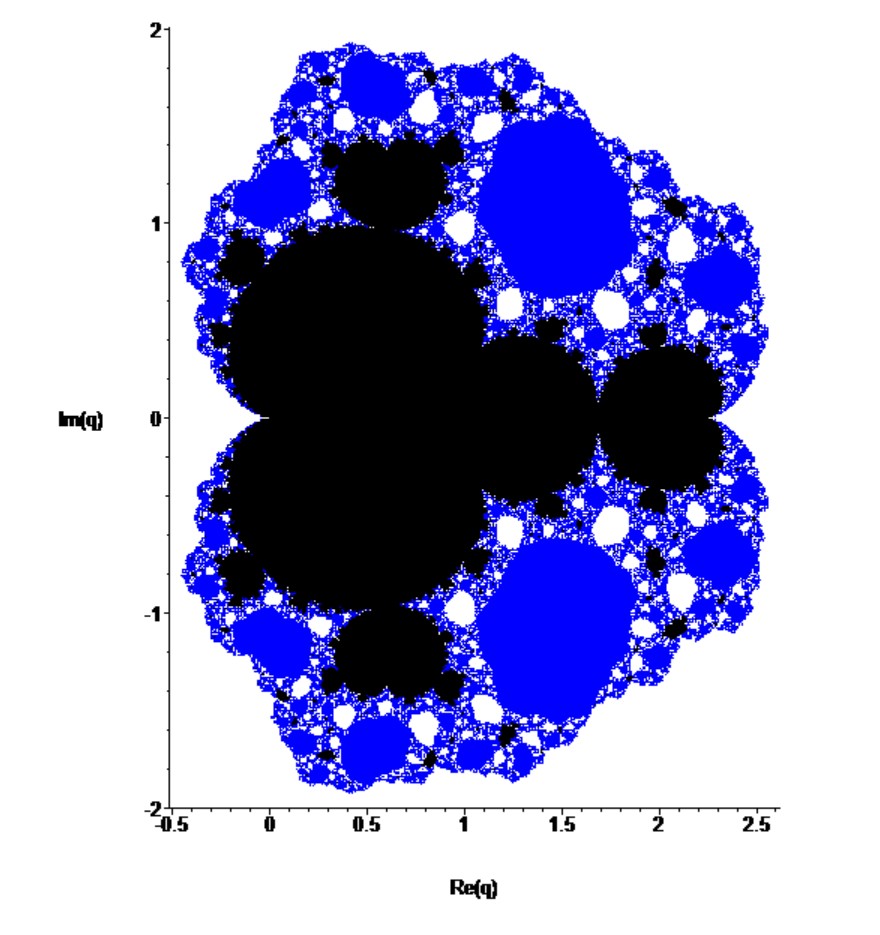}
    \end{center} 
  \caption{\footnotesize{Chromatic region diagram and locus ${\cal B}_q$ 
for $(p,\ell)=(3,3)$.}}
  \label{P3L3y0plot_fig}
\end{figure}

In this subsection we present region diagrams and loci ${\cal B}_q$ 
for illustrative cases with odd $p$
and odd $\ell$, namely $(p_{\rm odd},\ell_{\rm odd})=(3,3)$, (3,5), (3,7),
(5,3), (5,5), and (7,3), in Figs. \ref{P3L3y0plot_fig}-\ref{P7L3y0plot_fig}.
Calculations of $q_c(p,\ell)$ and $q_L(p,\ell)$ for these cases will be given
below. 

%

\begin{figure}[htbp]
  \begin{center}
\includegraphics[height=10cm]{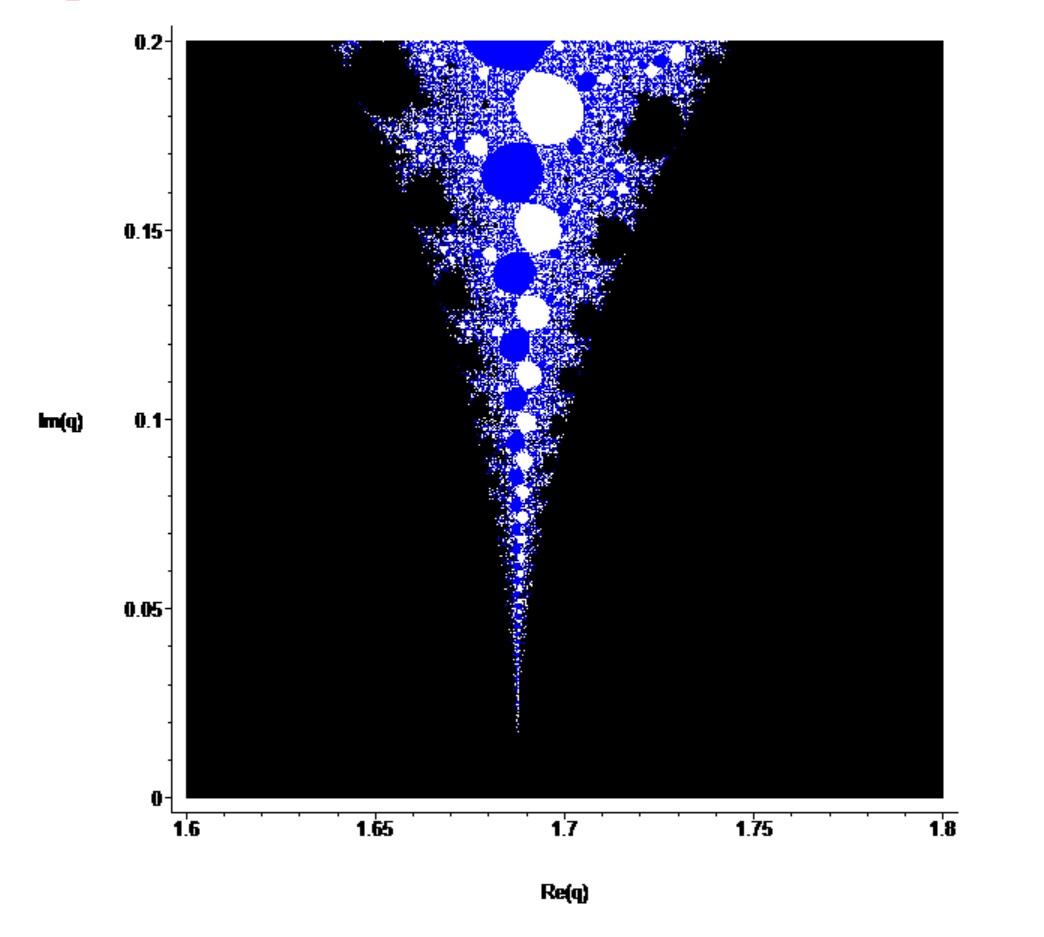}
    \end{center} 
  \caption{\footnotesize{Chromatic region diagram and locus ${\cal B}_q$ 
  for $(p,\ell)=(3,3)$ showing detail of the cusp for ${\rm Im}(q) > 0$ and
$1.6 < {\rm Re}(q) < 1.8$.}}
  \label{P3L3y0plotq1.6_1.8_fig}
\end{figure}

\begin{figure}[htbp]
  \begin{center}
\includegraphics[height=10cm]{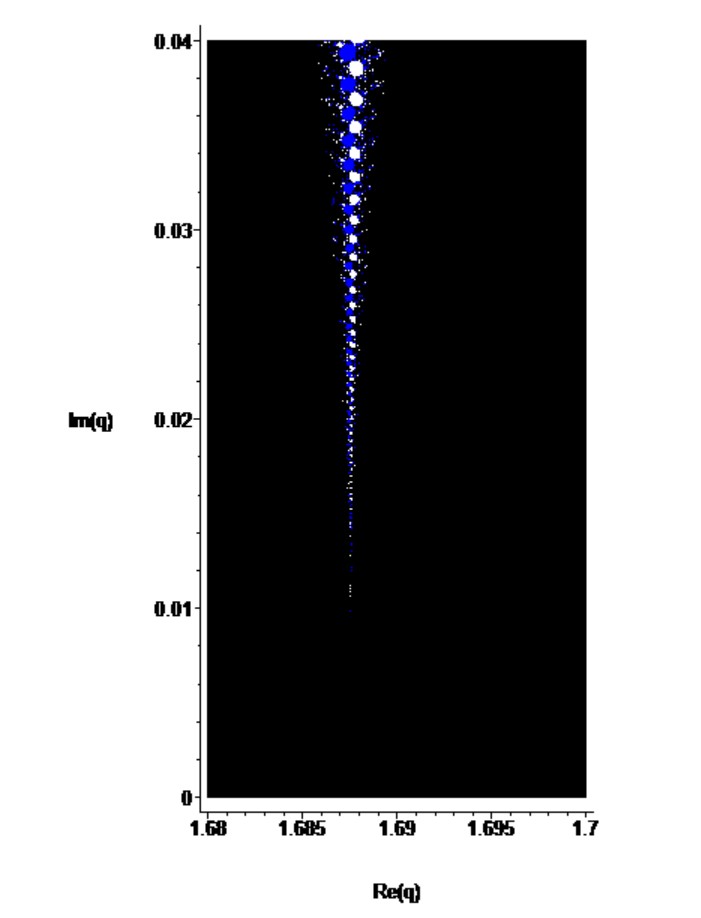}
    \end{center} 
  \caption{\footnotesize{Chromatic region diagram and locus ${\cal B}_q$ 
for $(p,\ell)=(3,3)$ showing further detail of the cusp for ${\rm Im}(q) > 0$ and
$1.68 < {\rm Re}(q) < 1.70$.}}
  \label{P3L3y0plotq1.68_1.7_fig}
\end{figure}
%

\begin{figure}[htbp]
  \begin{center}
\includegraphics[height=10cm]{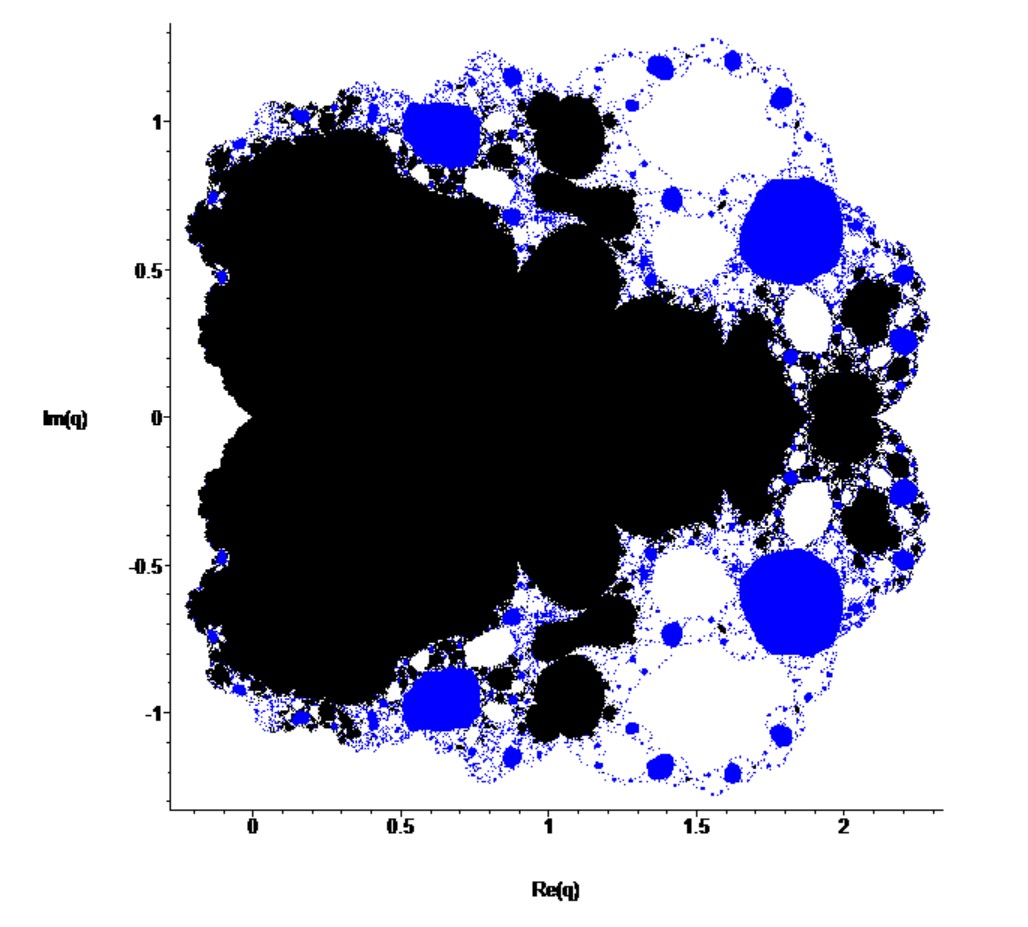}
  \end{center} 
  \caption{\footnotesize{Chromatic region diagram  and locus ${\cal B}_q$ 
for $(p,\ell)=(3,5)$.}}
  \label{P3L5y0plot_fig}
\end{figure}
%

\begin{figure}[htbp]
  \begin{center}
\includegraphics[height=10cm]{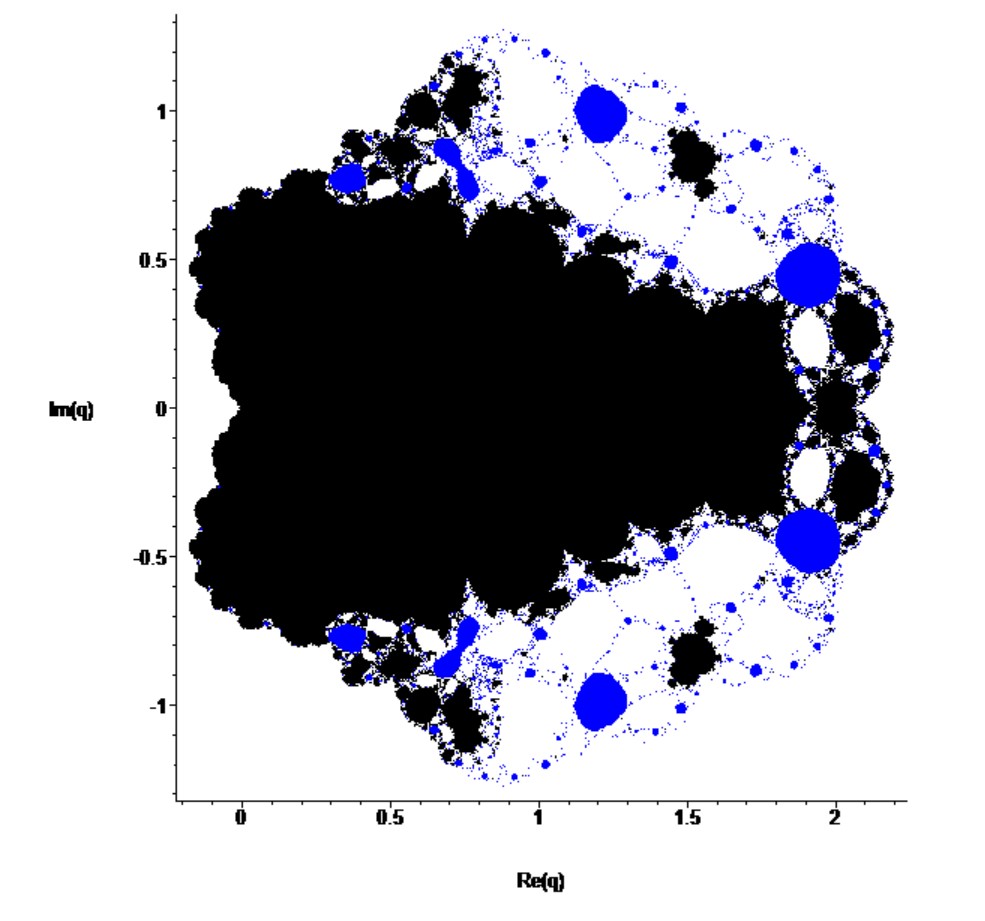}
  \end{center} 
  \caption{\footnotesize{Chromatic region diagram  and locus ${\cal B}_q$ 
for $(p,\ell)=(3,7)$.}}
  \label{P3L7y0plot_fig}
\end{figure}
%

\begin{figure}[htbp]
  \begin{center}
\includegraphics[height=10cm]{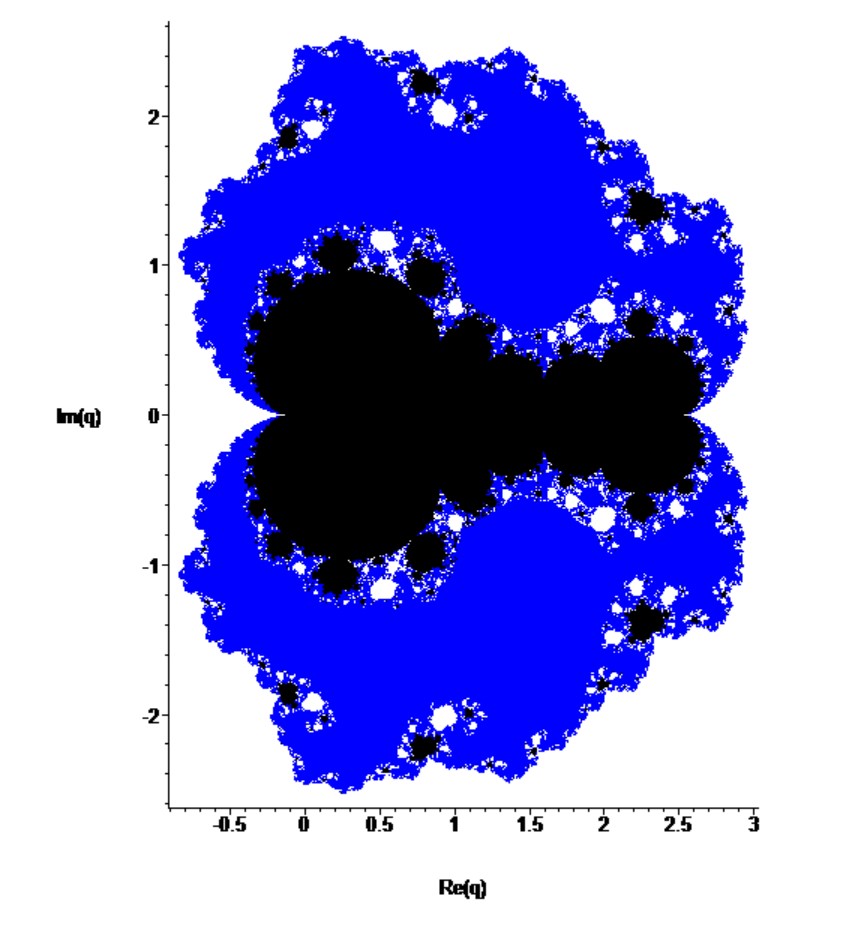}
    \end{center} 
 \caption{\footnotesize{Chromatic region diagram and locus ${\cal B}_q$ 
 for $(p,\ell)=(5,3)$.}}
  \label{P5L3y0plot_fig}
\end{figure}
%

\begin{figure}[htbp]
  \begin{center}
\includegraphics[height=10cm]{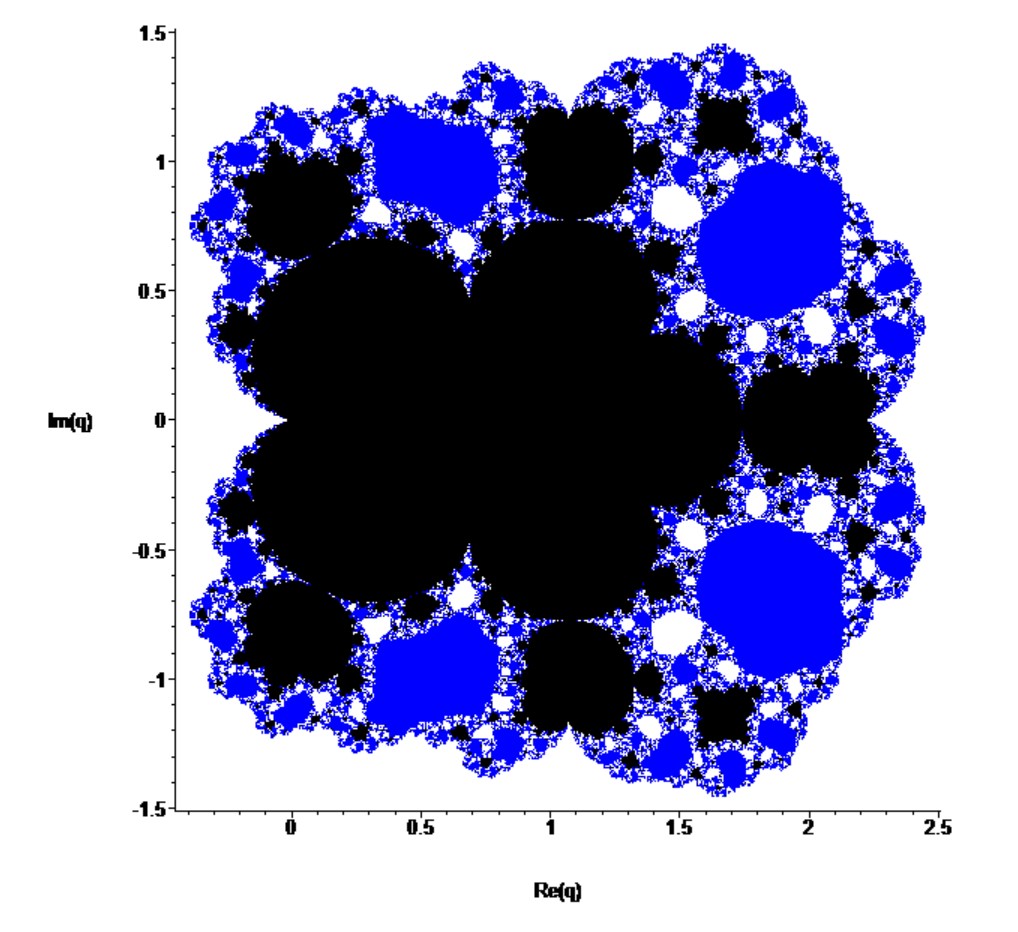}
    \end{center} 
 \caption{\footnotesize{Chromatic region diagram and locus ${\cal B}_q$ 
for $(p,\ell)=(5,5)$.}}
  \label{P5L5y0plot_fig}
\end{figure}
%

\begin{figure}[htbp]
  \begin{center}
\includegraphics[height=10cm]{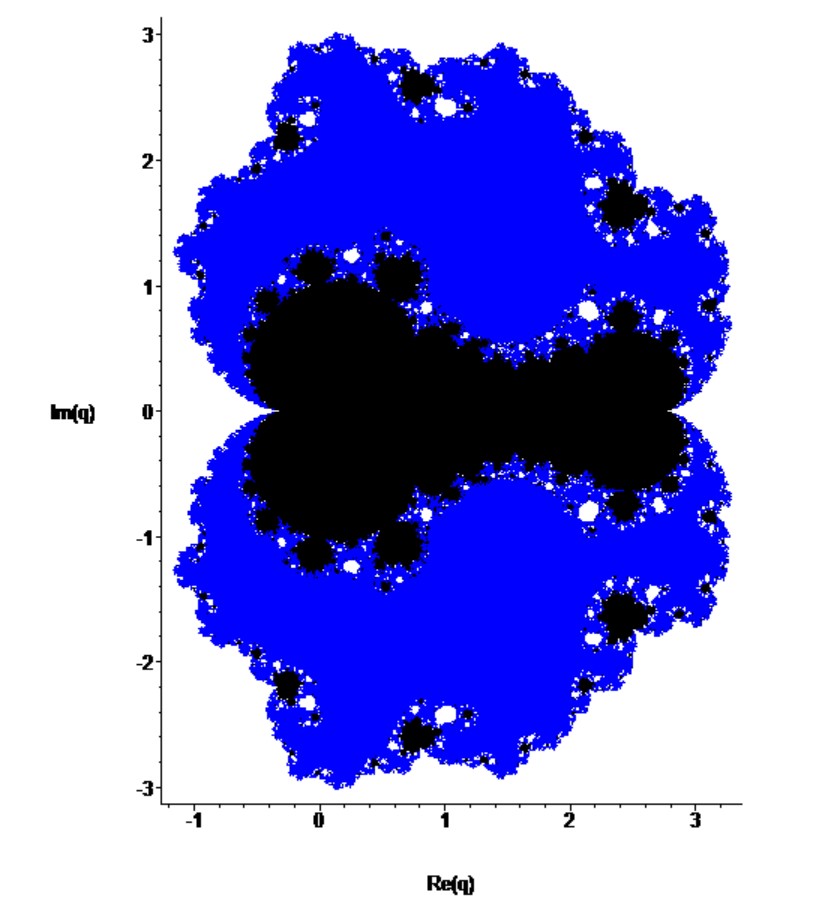}
  \end{center} 
  \caption{\footnotesize{Chromatic region diagram and locus ${\cal B}_q$
 for $(p,\ell)=(7,3)$.}}
  \label{P7L3y0plot_fig}
\end{figure}

One general feature in these cases is that the outermost part of the locus
${\cal B}_q$ intersects the real-$q$ axis on the right in a horizontally
oriented cusp at $q_c(p_{\rm odd},\ell_{\rm odd})$.  The interior real interval
$q_L(p_{\rm odd},\ell_{\rm odd}) < q < q_c(p_{\rm odd},\ell_{\rm odd})$ and the
complex-$q$ region analytically connected to this real interval are entirely
black, i.e., $F^\infty_{(p_{\rm odd},\ell_{\rm odd}),q}$ is neither 0 nor
$\infty$.  The locus ${\cal B}_q$ exhibits complex-conjugate cusp-like
structures oriented in an approximately vertical direction that extend inward
toward the real-$q$ axis.  We show two progressively more detailed views of the
upper cusp-like wedge structure for the (3,3) case in
Figs. \ref{P3L3y0plotq1.6_1.8_fig} and \ref{P3L3y0plotq1.68_1.7_fig}. The cusp
structures with ${\rm Im}(q) < 0$ are just the complex conjugates of the
structures with ${\rm Im}(q) > 0$, and hence are not shown.  As is evident
from these detailed figures, these cusp-like wedge structures become
vanishingly thin as they approach the real axis. Although our determination of
the region diagram has finite resolution, the inference that these cusps
actually extend all the way down and touch the real-$q$ axis is supported by a
calculation using discriminants of the equation for the RG fixed point equation
for this case and other $(p_{\rm odd}, \ell_{\rm odd})$ cases to be presented
below. We denote this point where the two complex-conjugate vertically oriented
wedges shrink to zero thickness and touch the real-$q$ axis as $q_x(p_{\rm
  odd},\ell_{\rm odd})$, where the subscript $x$ symbolizes the crossing
point. We calculate $q_x(3,3)=27/16 = 1.6875$ (see
Eq. (\ref{qx_P3L3}) and Table \ref{qx_table} below). We find similar
complex-conjugate cusp-like wedge regions that extend in toward the real-$q$
axis, becoming progressively narrower, and are consistent with touching this
axis at the respective $q_x(p_{\rm odd}, \ell_{\rm odd})$ points for all of the
cases $(p_{\rm odd}, \ell_{\rm odd})$ that we have studied where $p_{\rm odd}
\le \ell_{\rm odd}$. This is evident in the figures for (3,5), (3,7), (5,5),
and (5,7).  The values of $q_x$ for these cases are calculated below and listed
in Table \ref{qx_table}. From inspection of these plots, one can see that both
of the interior intervals $q_L < q < q_x$ and $q_x < q < q_c$ are black. In
contrast, for the $(p_{\rm odd}, \ell_{\rm odd})$ cases that we have studied
with $p_{\rm odd} > \ell_{\rm odd}$, such cusp-like wedge structures extending
down to the real axis do not appear. This is evident in the figures for
$(p_{\rm odd},\ell_{\rm odd})=(5,3)$ and (7,3).


\subsection{ Chromatic Region Diagrams and Loci ${\cal B}_q$ 
for $p$ Even and $\ell$ Odd}
\label{P_even_L_odd}

\begin{figure}[htbp]
  \begin{center}
\includegraphics[height=10cm]{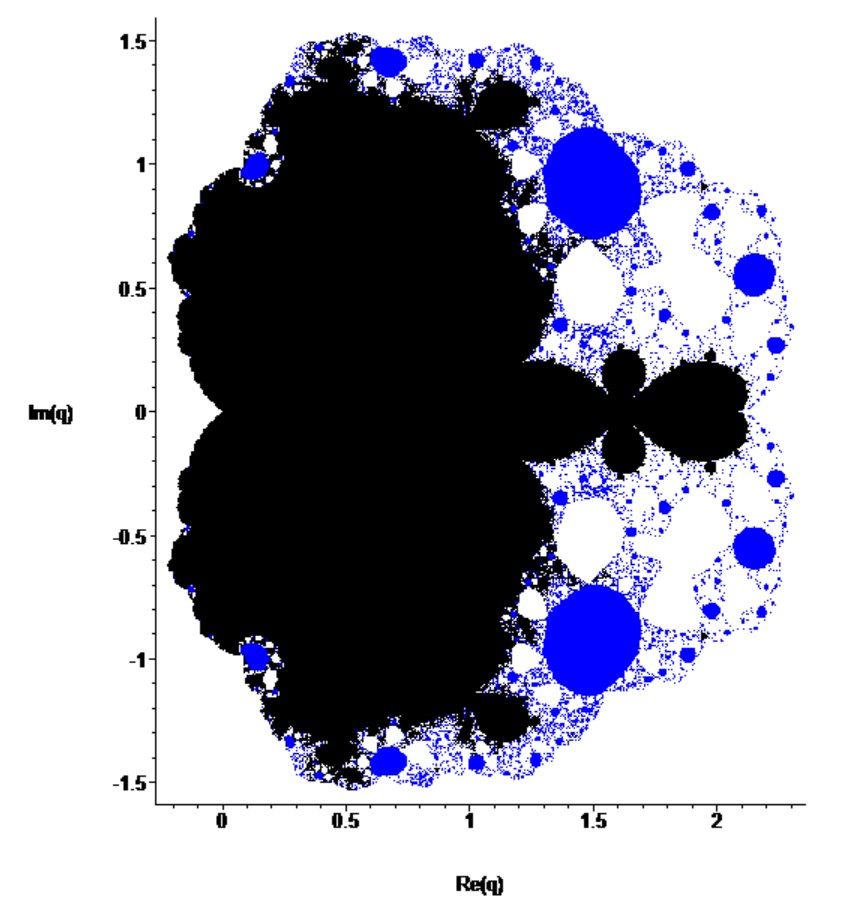}
    \end{center} 
  \caption{\footnotesize{Chromatic region diagram and locus ${\cal B}_q$ 
    for $(p,\ell)=(2,3)$.}}
  \label{P2L3y0plot_fig}
\end{figure}

\begin{figure}[htbp]
  \begin{center}
\includegraphics[height=10cm]{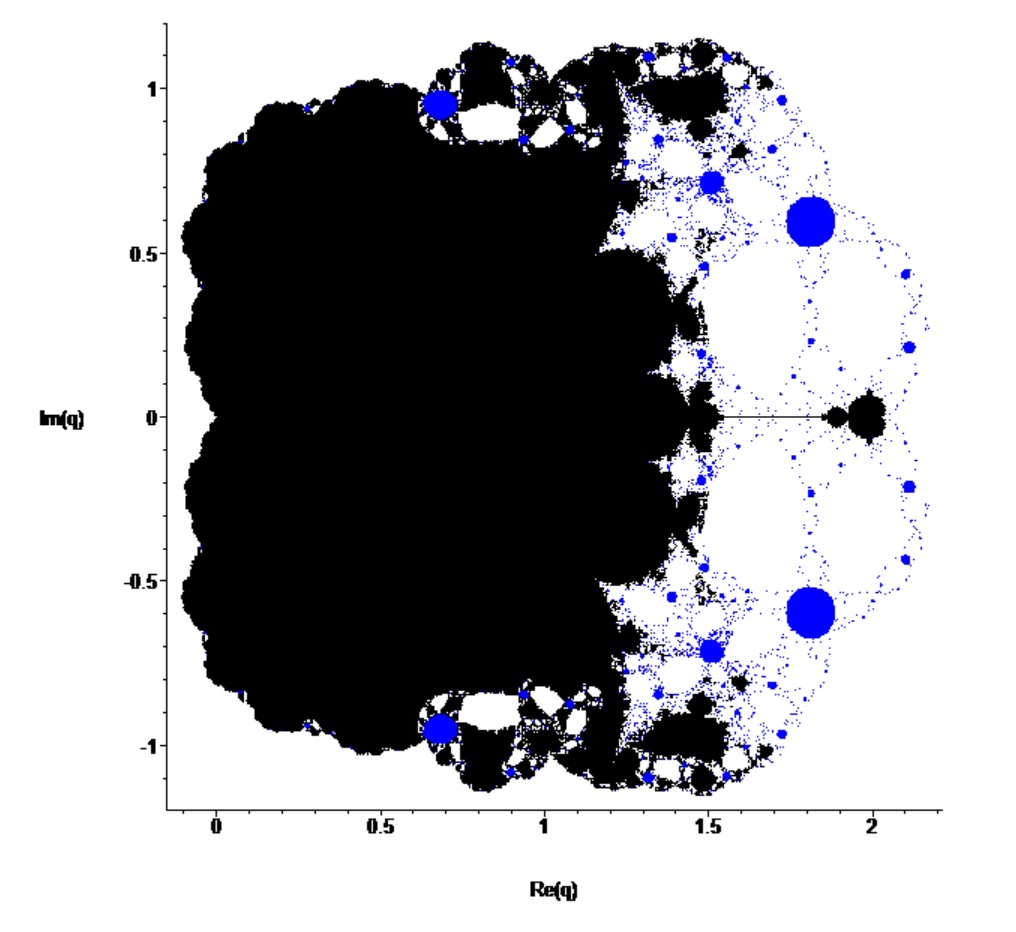}
    \end{center} 
  \caption{\footnotesize{Chromatic region diagram and locus ${\cal B}_q$ 
    for $(p,\ell)=(2,5)$.}}
  \label{P2L5y0plot_fig}
\end{figure}

\begin{figure}[htbp]
  \begin{center}
\includegraphics[height=10cm]{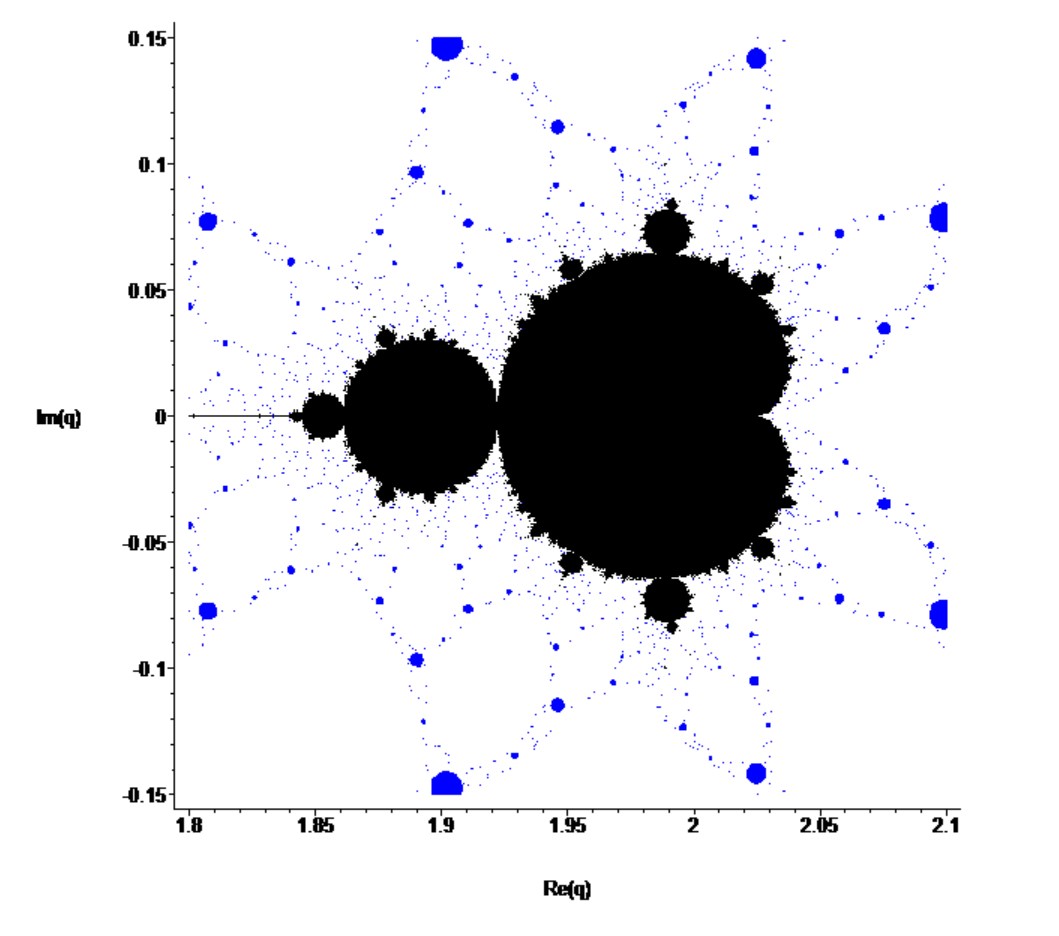}
    \end{center} 
  \caption{\footnotesize{Chromatic region diagram and locus ${\cal B}_q$ 
    for $(p,\ell)=(2,5)$  showing Mandelbrot-like sub-locus on the right-hand 
    part.}}
  \label{P2L5y0plotq1.8_2.1_fig}
\end{figure}

\begin{figure}[htbp]
  \begin{center}
\includegraphics[height=10cm]{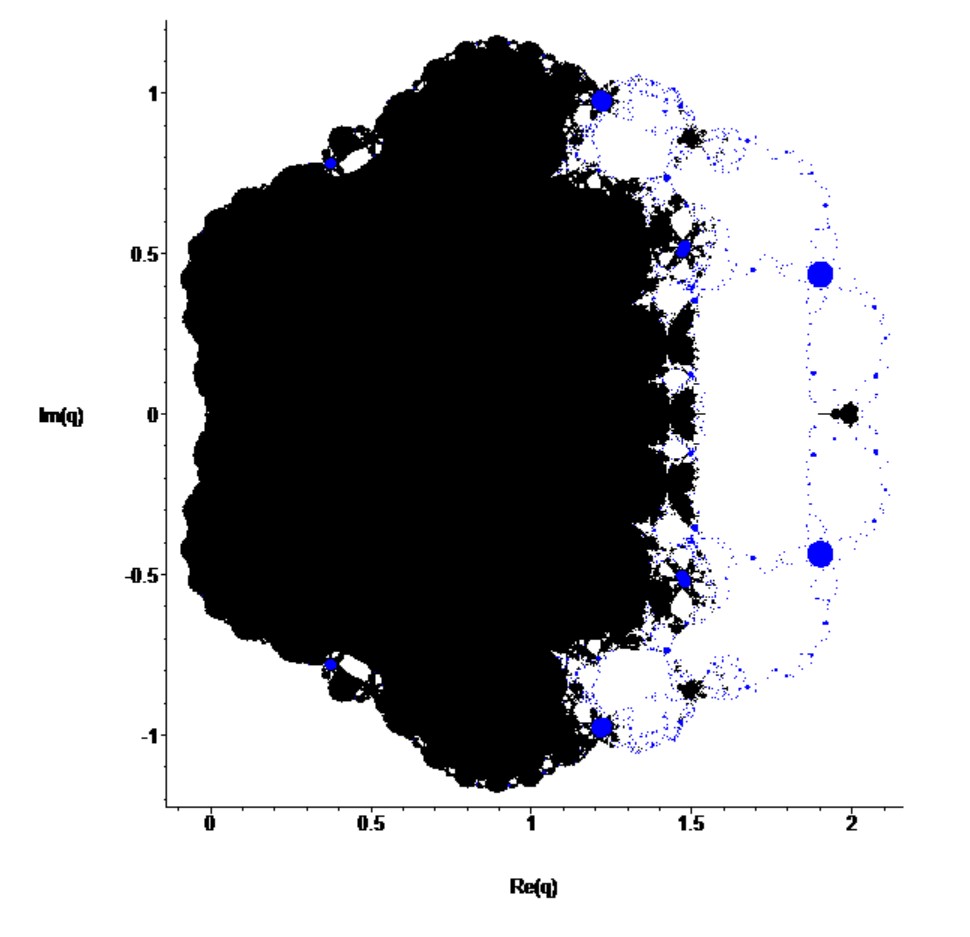}
    \end{center} 
  \caption{\footnotesize{Chromatic region diagram and locus ${\cal B}_q$ 
    for $(p,\ell)=(2,7)$.}}
  \label{P2L7y0plot_fig}
\end{figure}

\begin{figure}[htbp]
  \begin{center}   
\includegraphics[height=10cm]{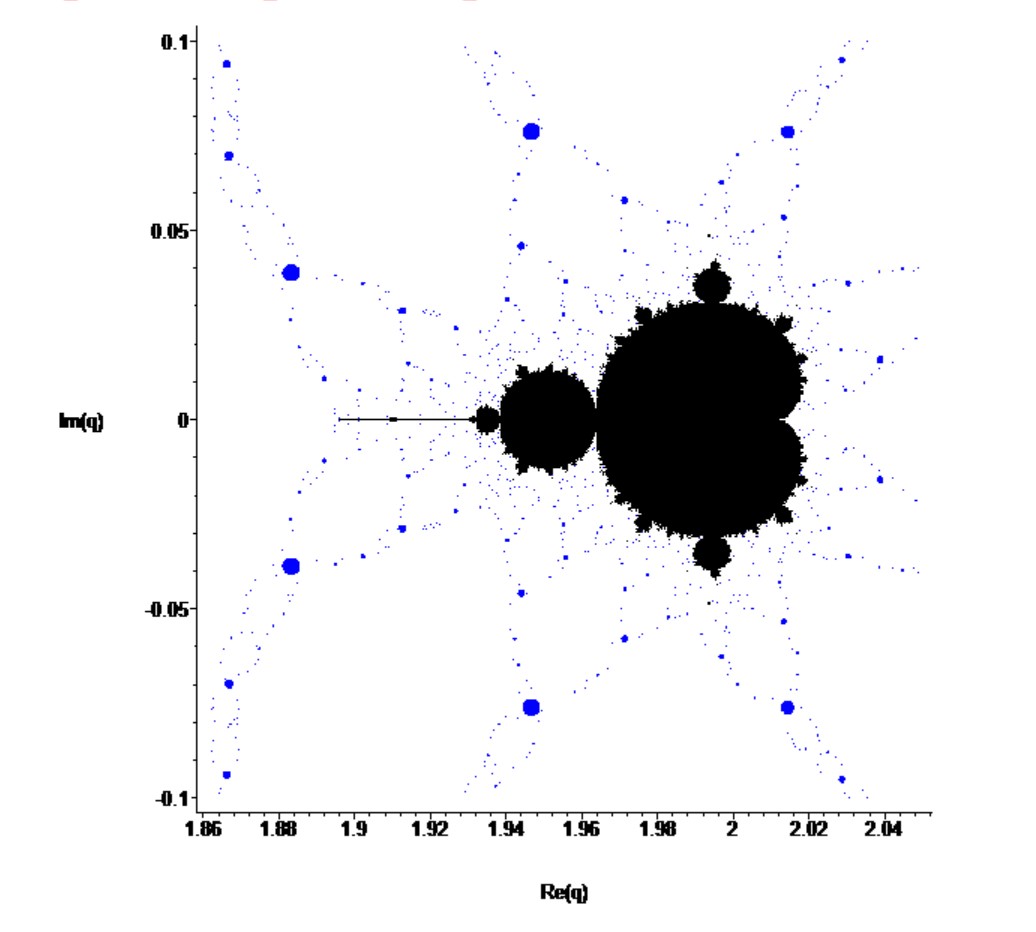}
    \end{center} 
  \caption{\footnotesize{Chromatic region diagram and locus ${\cal B}_q$ 
    for $(p,\ell)=(2,7)$ with magnified view of right-hand part 
   showing Mandelbrot-like sub-locus.}}
  \label{P2L7y0plotq1.86_2.05_fig}
\end{figure}

\begin{figure}[htbp]
  \begin{center}
\includegraphics[height=10cm]{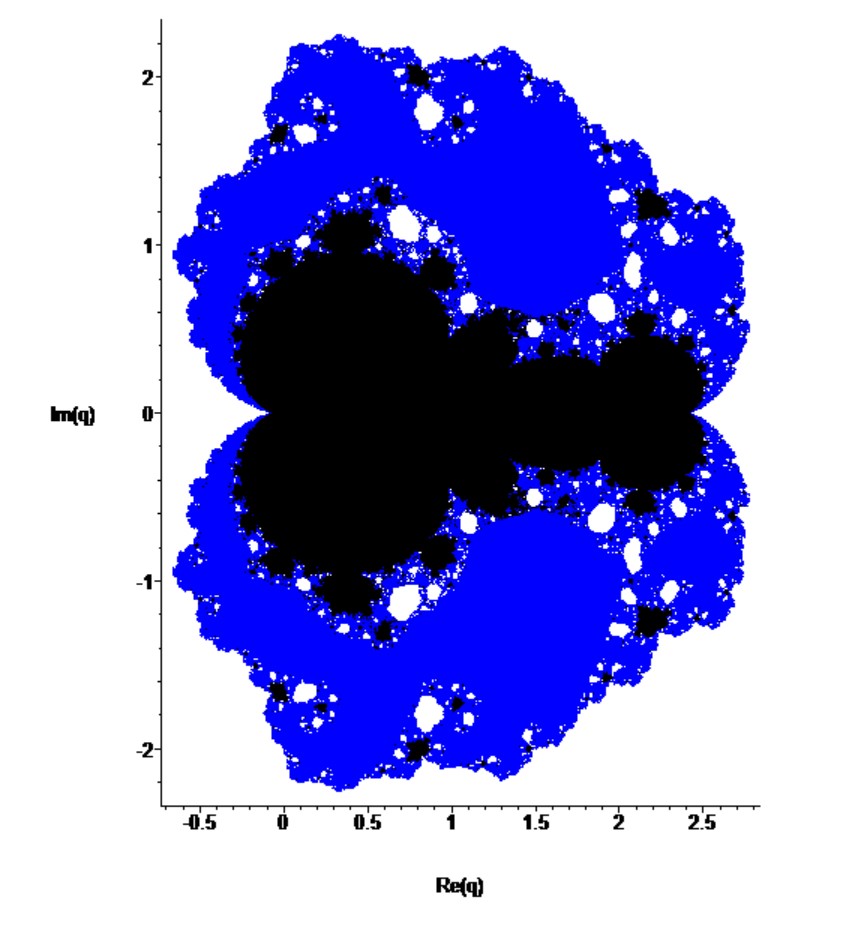}
    \end{center} 
  \caption{\footnotesize{Chromatic region diagram and locus ${\cal B}_q$ 
    for $(p,\ell)=(4,3)$.}}
  \label{P4L3y0plot_fig}
\end{figure}

\begin{figure}[htbp]
  \begin{center}
\includegraphics[height=10cm]{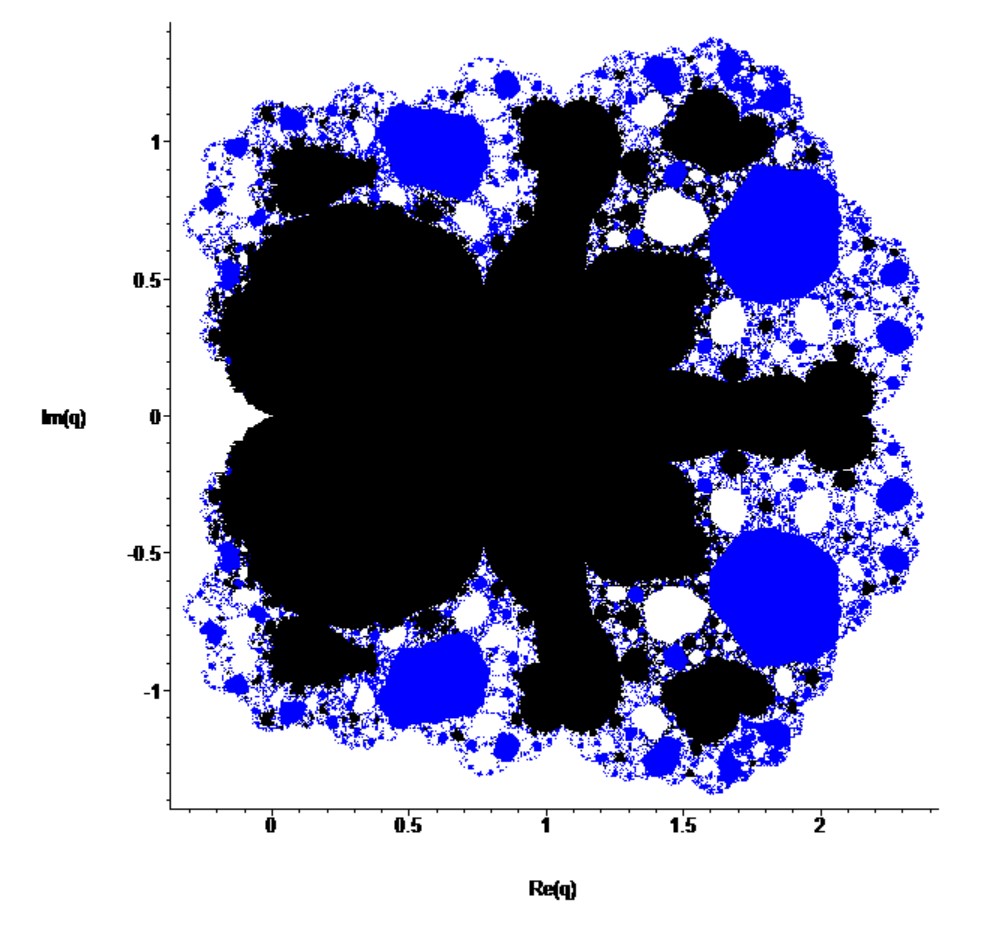}
    \end{center} 
  \caption{\footnotesize{Chromatic region diagram and locus ${\cal B}_q$ 
    for $(p,\ell)=(4,5)$.}}
  \label{P4L5y0plot_fig}
\end{figure}

\begin{figure}[htbp]
  \begin{center}
\includegraphics[height=10cm]{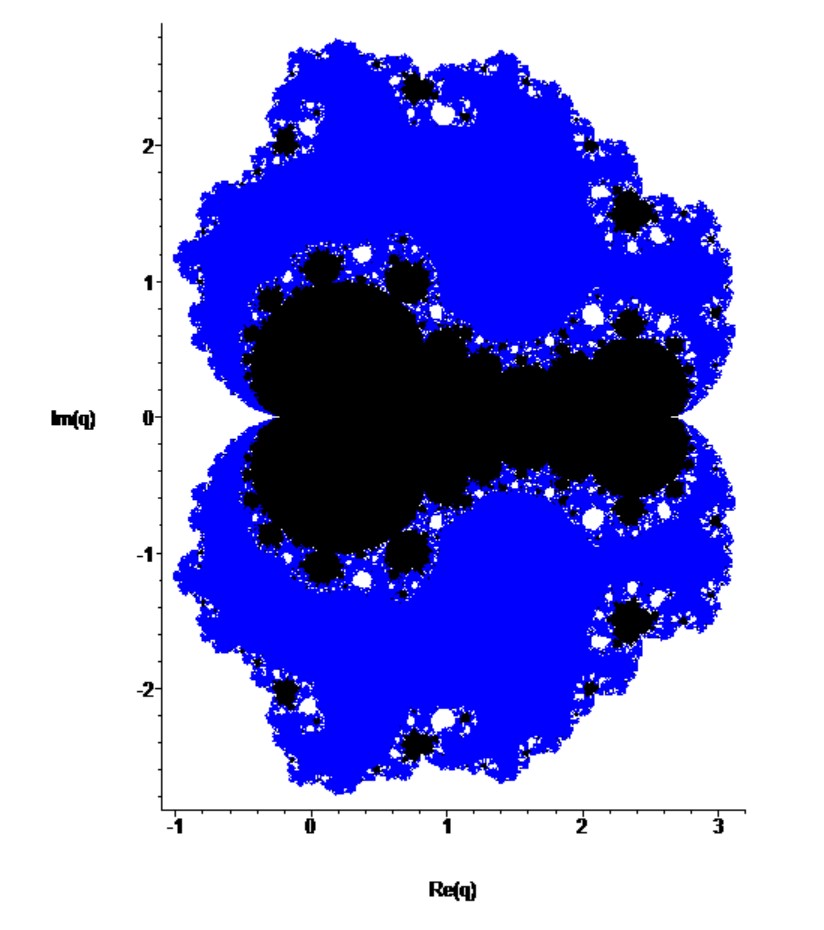}
    \end{center} 
  \caption{\footnotesize{Chromatic region diagram and locus ${\cal B}_q$ 
    for $(p,\ell)=(6,3)$.}}
  \label{P6L3y0plot_fig}
\end{figure}

In this subsection we present region diagrams and loci ${\cal B}_q$ for several
cases with even $p$ and odd $\ell$, namely $(p,\ell)=(2,3)$, (2,5), (2,7),
(4,3), (4,5), and (6,3), in Figs. \ref{P2L3y0plot_fig}-\ref{P6L3y0plot_fig}.
We remark on two general features of this class of $(p_{\rm even}, \ell_{\rm
  odd})$ cases, which are shared in common with the $(p_{\rm odd}, \ell_{\rm
  odd})=$ class. The first is that the structure of ${\cal B}_q$ in the
vicinity of $q_c(p_{\rm even},\ell_{\rm odd})$ is a horizontally oriented cusp
opening to the right. The second feature is that if $p_{\rm even} > \ell_{\rm
  odd}$, then $q_L < 0$. Combined with the analogous findings for the $(p_{\rm
  odd},\ell_{\rm odd})$ families, we infer that both of these features hold for
even and odd $p$. In particular, if $\ell$ is odd and if $p > \ell_{\rm odd}$,
where $p$ is even or odd, then $q_L < 0$.  Further combining this with our
results for even $\ell$ yields the generalization that for both even and odd
$\ell$, if $p \le \ell$, then $q_L(p,\ell)=0$, while if $p > \ell$, then
$q_L(p,\ell) < 0$, as recorded in (\ref{qLzero}) and (\ref{qLneg}).  We
calculate $q_L$ for several $(p,\ell_{\rm odd})$ values with $p > \ell_{\rm
  odd}$ below and list them in Table \ref{qleft_oddL_table}.

%

%

%

%

%

%

An interesting feature of the $(p_{\rm even}, \ell_{\rm odd})$ subclass is the
appearance for $(p,\ell)=(2,5)$ and (2,7) of a subset of the respective loci
${\cal B}_q$ that is a Mandelbrot-like set. This is evident in the magnified
views in Figs. \ref{P2L5y0plotq1.8_2.1_fig} and \ref{P2L7y0plotq1.86_2.05_fig}
for these two cases, ($p,\ell)=(2,5)$ and (2,7). Complex-conjugate
Mandelbrot-type structures on ${\cal B}_q$ oriented in an oblique manner were
also observed and analyzed for the $(p,\ell)=(2,2)$ case in 
\cite{chio_roeder} and in \cite{dhl}.

Aside from these properties, there is a large variation in the structure of the
locus ${\cal B}_q$ in this $(p_{\rm even},\ell_{\rm odd})$ class, depending on
the values of $p$ and $\ell$.  For example, in the 
$(p_{\rm even},\ell_{\rm odd})=(4,3)$, (4,5), and
(6,3) cases, the white regions in the interior are of limited extent, while, in
contrast, in the (2,5) and (2,7) cases, one sees (a) large white regions on the
right-hand interior of the plot, and (b) a ``dust''-like structure of blue and
black regions on the right-hand part of the plot.  The case (2,3) shows
features intermediate between these extremes.

Another feature for which there
is considerable variation is the RG properties of intervals and associated 
regions in the interior
$q_L(p_{\rm even},\ell_{\rm odd}) < q < q_c(p_{\rm even}, \ell_{\rm odd})$. 
In the $(p,\ell)=(2,3)$,
(4,3), (4,5), and (6,3) cases, this interior real interval is completely black,
i.e., $F^\infty_{(2,3),q}(-1)$ is neither zero nor infinite. In contrast, 
in the (2,7) case the locus ${\cal B}_q$ includes both black and white 
sub-intervals. In addition, in the (2,5) and (2,7) cases, ${\cal B}_q$ includes
black line segments in the interior on the real axis associated with the 
Mandelbrot-type sub-loci.


\section{ Calculation of $q_c$, $q_L$, $q_\infty$, $q_{\rm int}$, and 
Degeneracy $W$ at $q_c$ for Even $\ell$}
\label{qc_L_even_section} 

\subsection{Prologue}
\label{prologue_subsection}

In this section we calculate values of some special points where the locus
${\cal B}_q$ crosses the real-$q$ axis, including $q_c(p,\ell)$ and
$q_L(p,\ell)$ for various general $p$ and even $\ell$, together with other
points that occur for subclasses that depend on whether $p$ is even or odd,
namely $q_{\infty}(p_{\rm even},\ell_{\rm even})$ and $q_{\rm int}(p_{\rm
  odd},\ell_{\rm even})$.  In addition to values for these illustrative cases,
we observe certain monotonicity relations.  Note that there is a basic
difference between the effect of the RG transformation on an initial value of
$v$ for even $\ell$ and odd $\ell$, namely that the RG transformation always
maps a positive $v$ to a positive $v'$ for both even and odd $\ell$, but if $v$
is negative, then for even $\ell$, the RG transformation maps this initial
negative value of $v$ to a positive value, which remains positive under further
iterations. In contrast, if $\ell$ is odd, then the RG transformation
(\ref{vprime}) can map a negative $v$ to a negative $v'$.


\subsection{Calculation of $q_c(p,\ell_{\rm even})$ and
$q_L(p,\ell_{\rm even})$} 
\label{qc_L_even_and qint_subsection}

For even $\ell$, $q_c(p,\ell)$ can be calculated as a solution to the
equation
\beq
(q-2)\bigg [(q-1)^\ell+(q-1)\bigg ]^p=2(q-1)\bigg [(q-1)^\ell-1 \bigg ]^p \ . 
\label{qc_eq_even_L}
\eeq
A derivation of Eq. (\ref{qc_eq_even_L}) is given in Appendix
\ref{qccalc_appendix}.  Equation (\ref{qc_eq_even_L}) always has the solutions
$q=0$, $q=1$, and $q=2$ for this case of even $\ell$. Among these solutions, if
$p \le \ell_{\rm even}$, then $q_L(p,\ell_{\rm even})=0$.  If $p$ is even, then
Eq. (\ref{qc_eq_even_L}) has one additional real solution, which we determine
to be $q_c(p_{\rm even}, \ell_{\rm even})$.  If $p$ is odd, then
Eq. (\ref{qc_eq_even_L}) has, in addition to $q=0, \ 1, \ 2$, two additional
real solutions. The larger of these is $q_c(p_{\rm odd},\ell_{\rm even})$,
while the smaller one, denoted $q_{\rm int}(p_{\rm odd},\ell_{\rm even})$, is
the single point where ${\cal B}_q$ crosses the real-$q$ axis in the interior
(int) interval $q_L < q < q_c(p_{\rm odd},\ell_{\rm even})$.  In Table
\ref{qc_even_L_table} we list values of $q_c(p,\ell_{\rm even})$ for both even
and odd $p$ with $2 \le p \le 8$ and for the even values $\ell=2, \ 4, \ 6, \
8$.

In certain cases it is possible to obtain exact
analytic solutions of Eq. (\ref{qc_eq_even_L}). 
For $q_c(p,\ell_{\rm even})$, in addition to the known value
\beq
q_c(2,2)=3 \ , 
\label{qc_P2L2}
\eeq
we find
\beq
q_c(3,2) = 3 + \sqrt{2} = 4.414214 
\label{qc_P3L2}
\eeq
\beq
q_c(4,2) = 3 + 2^{2/3} + 2^{1/3} = 5.847322 
\label{qc_P4L2}
\eeq
\beq
q_c(5,2) = 3 + \sqrt{2} + \Big (4+3\sqrt{2} \, \Big )^{1/2} = 7.2852135
\label{qc_P5L2}
\eeq
\begin{widetext}
  \beq
q_c(4,4) = \frac{5}{3} +\frac{1}{3}\Big (17+3\sqrt{33} \, \Big )^{1/3} -
\frac{2}{3\Big (17+3\sqrt{33} \, \Big )^{1/3}} = 2.5436890
\label{qc_P4L4}
\eeq
and
\beq
q_c(7,2) = 3+\sqrt{2} + \Big [ 2(7+5\sqrt{2} \, ) \Big ]^{1/3} + 
\frac{4+3\sqrt{2}}{\Big [ 2(7+5\sqrt{2} \, )\Big ]^{1/3}} = 10.165795 \ , 
\label{qc_P7L2}
\eeq
\end{widetext}
where here and below, floating-point values of numbers are given to the 
indicated number of significant figures. 


\begin{table}
  \caption{\footnotesize{Values of $q_c(G^{(p,\ell)}_\infty)$ for a
      range of $p$ values and the even-$\ell$ values $\ell=2, \ 4, \ 6, \ 8$ 
      and $2 \le p \le 8$. For compact notation, in this table, we denote
      $q_c(G^{(p,\ell)}_\infty) \equiv q_c(p,\ell)$.}}
\begin{center}
\begin{tabular}{|c|c|c|c|}
  \hline\hline
$q_c(2,2)=3$         & $q_c(2,4)=2.145883$  & $q_c(2,6)=2.059518$ &
$q_c(2,8)=2.0324966$  \\ \hline 
$q_c(3,2)=4.414214$  & $q_c(3,4)=2.365550$  & $q_c(3,6)=2.168568$ &
$q_c(3,8)=2.102248$   \\ \hline
$q_c(4,2)=5.847322$  & $q_c(4,4)=2.543689$  & $q_c(4,6)=2.2563615$ &
$q_c(4,8)=2.159473$   \\ \hline
$q_c(5,2)=7.2852135$ & $q_c(5,4)=2.692763$  & $q_c(5,6)=2.327821$ &
$q_c(5,8)=2.205779$   \\ \hline
$q_c(6,2)=8.725024$  & $q_c(6,4)=2.821849$  & $q_c(6,6)=2.3880935$ &
$q_c(6,8)=2.244490$   \\ \hline
$q_c(7,2)=10.165795$ & $q_c(7,4)=2.936396$  & $q_c(7,6)=2.440359$  &
$q_c(7,8)=2.2777675$   \\ \hline
$q_c(8,2)=11.607116$ & $q_c(8,4)=3.039855$  & $q_c(8,6)=2.486625$  &
$q_c(8,8)=2.306990$   \\ \hline\hline
\end{tabular}
\end{center}
\label{qc_even_L_table}
\end{table}


 Among other entries, Table \ref{qc_even_L_table} includes 
$q_c(p,\ell)$ for the diagonal case $p=\ell$ with $\ell=2s$ even and 
$2 \le 2s \le 8$.  For later reference, it will be useful to give the next
two higher diagonal values; these are
\beq
q_c(10,10)=2.256392 \ , \quad\quad  q_c(12,12)=2.221471  \ . 
\label{qcdiagonal_higher}
\eeq

We observe several monotonicity properties describing the values of 
for $q_c(p,\ell_{\rm even})$ (for both even and odd $p$)
that we have calculated from Eq. (\ref{qc_eq_even_L}). These are evident in 
Table \ref{qc_even_L_table}:

\begin{enumerate}

\item
\label{qc_increases_with_P_for_even_L}

  $q_c(p,\ell_{\rm even})$ is a monotonically increasing function of $p$ for 
fixed even $\ell$.

\item
\label{qc_decreases_with_even_L}

  $q_c(p,\ell_{\rm even})$ is a monotonically decreasing function of even
  $\ell$ for fixed $p$.

\item
\label{qc_diagonal_decreases}
  In the diagonal case $(p_{\rm even},\ell_{\rm even})=(2s,2s)$, 
  $q_c(2s,2s)$ is a monotonically decreasing function of $s$. 

\end{enumerate}

Our results suggest the inference that as $\ell \to \infty$ through even 
values, with fixed $p$, 
\beq
\lim_{\ell_{\rm even} \to \infty} q_c(p,\ell_{\rm even}) = 2, 
\label{qc_large_even_L_lim2}
\eeq
approaching this limit from above, so that, for each $p$, $q_c(p,\ell_{\rm
  even})$ decreases monotonically from $q_c(p,2)$ to $q_c(p,\infty)=2$.  We
will combine these with results to be obtained below for $q_c(p,\ell_{\rm
  odd})$ to infer the same limit as (\ref{qc_large_even_L_lim2}) for general
$\ell$. Given the observed monotonic decrease of $q_c(2s,2s)$, our results also
suggest the inference that $\lim_{s \to \infty} q_c(2s,2s) = 2$, so that as
$2s$ increases from 2 to $\infty$, $q_c(2s,2s)$ decreases monotonically from 3
to 2. 

Concerning the leftmost point at which ${\cal B}_q$ crosses the real-$q$ axis,
for the $(p,\ell)$ cases that we have studied, we find the general properties
stated above in Eqs. (\ref{qLzero}) and (\ref{qLneg}).  For the cases with $p >
\ell$, we list our calculations of $q_L(p,\ell)$ in Table
\ref{qleft_evenL_table} with even $\ell$. The entries are listed in order of
increasing values of the ratio $p/\ell$.  Provided that $p > \ell_{\rm even}$
    so that $q_L(p,\ell_{\rm even}) < 0$, then, as is evident from Table
    \ref{qleft_evenL_table}, for the cases that we have calculated, we find
    that the magnitude $|q_L(p,\ell_{\rm even})|$ is a monotonically increasing
    function of $p$ for fixed $\ell_{\rm even}$.  Values of $q_L(p,\ell_{\rm
      odd})$ for $p > \ell_{\rm odd}$ will be discussed below, and the
    corresponding property is observed, i.e., if $p > \ell_{\rm odd}$, then 
    $|q_L(p,\ell_{\rm odd})|$ is a monotonically increasing function of $p$ 
    for fixed $\ell_{\rm odd}$. However, for both even and odd $\ell$ with
    $p > \ell$, $|q_L(p,\ell)|$ is not a monotonically increasing function of
    the ratio $p/\ell$.  For even $\ell$, an example of the non-monotonicity is
    the fact that $|q_L(7,4)|$ is less than $|q_L(3,2)|$, although 7/4
    is larger than 3/2.

\begin{table}
  \caption{\footnotesize{Values of $q_L(p,\ell_{\rm even}) < 0$ that occur for 
even $\ell$ if $p > \ell_{\rm even}$ (where $p$ can be even or odd).          
The entries are listed in order of increasing values of $p/\ell_{\rm even}$.}}
\begin{center}
\begin{tabular}{|c|c|c|} \hline\hline
$(p,\ell_{\rm even})$ & $p/\ell_{\rm even}$ & $q_L(p,\ell_{\rm even})$ 
\\ \hline
(5,4)               & 1.25            & $-0.0187563$    \\ \hline
(3,2)               & 1.5             & $-0.222373$     \\ \hline
(6,4)               & 1.5             & $-0.0554454$    \\ \hline
(7,4)               & 1.75            & $-0.0974571$    \\ \hline  
(4,2)               & 2               & $-0.5730675$    \\ \hline
(5,2)               & 2.5             & $-0.952009$     \\ \hline
(6,2)               & 3               & $-1.342060$     \\ \hline
(7,2)               & 3.5             & $-1.737693$     \\ \hline
(8,2)               & 4               & $-2.136550$     \\ \hline 
\hline
\end{tabular}
\end{center}
\label{qleft_evenL_table}
\end{table}

To explain our method of calculating $q_c(p,\ell)$ for even $\ell$ further, 
we exhibit the actual equations for several 
illustrative cases.  In the $(p_{\rm even}, \ell_{\rm even})$ cases, 
Eq. (\ref{qc_eq_even_L}) takes the form
\beq
q^p(q-1)(q-2){\cal P}_{(p_{\rm even}, \ell_{\rm even})}=0 \ , 
\label{eqqc_P_even_L_even}
\eeq
where ${\cal P}_{p_{\rm even},\ell_{\rm even}}$ is a polynomial in
$q$. In each of these cases, we find that the polynomial ${\cal
  P}_{(p_{\rm even}, \ell_{\rm even})}=0$ has one real root, which is
thus determined uniquely to be $q_c(p_{\rm even}, \ell_{\rm
  even})$. We give some explicit examples below.  For the case 
$(p,\ell)=(2,2)$ analyzed in \cite{dhl}, 
\beq
{\cal P}_{(2,2)} = q-3  \ , 
\label{polfactor_P2L2}
\eeq
leading to $q_c(2,2)=3$. For $(p,\ell)=(2,4)$, we calculate 
\beq
{\cal P}_{(2,4)} = q^5 - 5q^4 + 11q^3 - 15q^2 + 13q - 7 \ . 
\label{polfactor_P2L4}
\eeq
This polynomial has one real root, which is $q_c(2,4)$, listed in Table
\ref{qc_even_L_table}, and two pairs of complex-conjugate roots. For 
$(p,\ell)=(2,6)$ we calculate 
\beqs 
{\cal P}_{(2,6)} &=& q^9 -9q^8 + 37q^7 - 91q^6 + 148q^5 -168q^4 \cr\cr
&+&138q^3 - 84q^2 +37q - 11 \ . 
\label{polfactor_P2L6}
\eeqs
This has one real root, which is $q_c(2,6)$, listed in Table 
\ref{qc_even_L_table}, and four pairs of complex-conjugate roots. 
Increasing $p$, for $(p,\ell)=(4,2)$, we obtain 
\beq
{\cal P}_{(4,2)} = q^3 - 9q^2 + 21q - 15 \ . 
\label{polfactor_P4L2}
\eeq
This has one real root, which is $q_c(4,2)$, given in 
Eq. (\ref{qc_P4L2}) and listed in Table \ref{qc_even_L_table}, together with a
complex-conjugate pair of roots. For $(p,\ell)=(4,4)$, we find 
\beqs
{\cal P}_{(4,4)} &=& \Big ( q^3 - 5q^2 + 9q - 7 \Big ) \Big ( q^8 - 8q^7 + 
30q^6 - 72q^5 \cr\cr
&+& 125q^4 - 162q^3 + 150q^2 - 88q + 25 \Big ) \ . 
\label{polfactor_P4L4}
\eeqs
The cubic factor in this polynomial has one real root, $q_c(4,4)$, given
in Eq. (\ref{qc_P4L4}) and listed in Table \ref{qc_even_L_table},
together with a complex-conjugate pair of roots.  The factor of degree 8 has no
real roots. Increasing $p$ again to $p=6$, for $(p,\ell)=(6,2)$, we calculate 
\beq
{\cal P}_{6,2} = q^5 - 15q^4 +70q^3 - 150q^2 + 155q - 63 \ . 
\label{polfactor_P6L2}
\eeq
This has one real root, which is $q_c(6,2)$, listed in Table
\ref{qc_even_L_table}, and two pairs of complex-conjugate roots.  We find
qualitatively the same results for other $(p_{\rm even}, \ell_{\rm even})$
cases, and the resultant values of $q_c(p_{\rm even},\ell_{\rm even})$ are
listed in Table \ref{qc_even_L_table}.

As noted, the calculations of $q_c(p,\ell_{\rm even})$ in this subsection apply
for even $\ell$ and for both even and odd $p$. However, the crossing points and
associated intervals in the interior $q_L(p,\ell_{\rm even}) < q <
q_c(p,\ell_{\rm even})$ are different for even and odd $p$.  For even $p$, as
discussed above and summarized in Eq. (\ref{P_even_L_even_intervals}), the
structure in the interior real interval $q_L(p_{\rm even},\ell_{\rm even}) < q
< q_c(p_{\rm even},\ell_{\rm even})$ involves the infinite sequence of crossing
points and associated regions $S_\infty(p_{\rm even},\ell_{\rm even})$.  In
contrast, in the $(p_{\rm odd},\ell_{\rm even})$ cases, we find that ${\cal
  B}_q$ crosses the real-$q$ axis at a single point in the interior interval
$q_L(p_{\rm odd},\ell_{\rm even}) < q < q_c(p_{\rm odd},\ell_{\rm even})$,
namely $q_{\rm int}(p_{\rm odd}, \ell_{\rm even})$.  We proceed to consider
this case next.


\subsection{Calculation of $q_{\rm int}(p_{\rm odd},\ell_{\rm even})$} 
\label{qint_p_odd_and_L_even_subsection}

Here we calculate values of $q_{\rm int}(p_{\rm odd},\ell_{\rm even})$ for
a variety of cases of the type $(p_{\rm odd},\ell_{\rm even})$.
We have computed $q_{\rm int}(p_{\rm
  odd},\ell_{\rm even})$ for the same set of even values of $\ell$ as in Table
\ref{qc_even_L_table} and for several odd values of $p$. We list the results 
in Table \ref{qint_odd_P_even_L_table}.  For compact notation, in
this table, we denote $q_{\rm int}(G^{(p,\ell)}_\infty) \equiv q_{\rm
  int}(p,\ell)$, analogous to our compact notation $q_c(G^{(p,\ell)}_\infty)
\equiv q_c(p,\ell)$.  One can visually confirm that the values of
$q_c(p,\ell_{\rm even})$ in Table \ref{qc_even_L_table} and the values of
$q_{\rm int}(p_{\rm odd},\ell_{\rm even})$ agree with the results in our region
diagrams.  From our calculations, we find the following
monotonicity property for the cases that we have studied: 

\begin{enumerate}

\item 
\label{qint_decreases_with_P}
$q_{\rm int}(p_{\rm odd},\ell_{\rm even})$ is a monotonically decreasing
function of $p_{\rm odd}$ for fixed $\ell_{\rm even}$. 

\item 
\label{qint_increases_with_L}
$q_{\rm int}(p_{\rm odd},\ell_{\rm even})$ is a monotonically increasing
function of $\ell_{\rm even}$ for fixed $p_{\rm odd}$. 

\end{enumerate}

\begin{table}
  \caption{\footnotesize{Values of $q_{\rm int}(p,\ell)$ for 
      the odd-$p$ values $p=3, \ 5, \ 7$ and the even-$\ell$ values 
      $\ell=2, \ 4, \ 6, \ 8$.}}
\begin{center}
\begin{tabular}{|c|c|c|c|}
  \hline\hline
$q_{\rm int}(3,2)=1.585786$ & $q_{\rm int}(3,4)=1.793745$  & 
$q_{\rm int}(3,6)=1.875628$ & $q_{\rm int}(3,8)=1.915780$ \\ \hline
$q_{\rm int}(5,2)=1.543214$ & $q_{\rm int}(5,4)=1.727166$  &
$q_{\rm int}(5,6)=1.8118665$& $q_{\rm int}(5,8)=1.859710$  \\ \hline
$q_{\rm int}(7,2)=1.528849$ & $q_{\rm int}(7,4)=1.701986$  & 
$q_{\rm int}(7,6)=1.7846755$& $q_{\rm int}(7,8)=1.833418$  \\ \hline
\hline
\end{tabular}
\end{center}
\label{qint_odd_P_even_L_table}
\end{table}

Our method of calculation of $q_{\rm int}(p_{\rm odd},\ell_{\rm even})$ is 
as follows . For the $(p_{\rm odd}, \ell_{\rm even})$ cases,
Eq. (\ref{qc_eq_even_L}) takes the form 
\beq
q^p (q-1)(q-2){\cal P}_{(p_{\rm odd},\ell_{\rm even})}=0 \ , 
\label{eqqc_P_odd_L_even}
\eeq
where ${\cal P}_{(p_{\rm odd},\ell_{\rm even})}$ is a polynomial with two 
real roots; the larger one is $q_c$, while the smaller one is $q_{\rm int}$ 
for these cases. Thus, in these cases, this method enables us to calculate
both $q_c$ and $q_{\rm int}$. 

We proceed to give some explicit examples of these calculations for this 
$(p_{\rm odd},\ell_{\rm even})$ class. For $(p,\ell)=(3,2)$, 
\beq
{\cal P}_{(3,2)}= q^2 - 6q + 7 \ , 
\label{polfactor_P3L2}
\eeq
with roots  $q_c(3,2) = 3 + \sqrt{2}$, as listed in Eq. (\ref{qc_P3L2}), and
\beq 
q_{\rm int}(3,2) = 3 - \sqrt{2} \ . 
\label{qint_P3L2}
\eeq
For $(p,\ell)=(3,4)$, 
\beq
{\cal P}_{(3,4)} = q^8 - 9q^7 + 37q^6 - 94q^5 + 166q^4 - 214q^3 + 199q^2 
- 121q + 37 \ . 
\label{polfactor_P3L4}
\eeq
This polynomial ${\cal P}_{(3,4)}$ has two real roots, $q_c(3,4)$
listed in Table \ref{qc_even_L_table} and $q_{\rm int}(3,4)$ listed in
Table \ref{qint_odd_P_even_L_table}, together with six complex roots. 

For $(p,\ell)=(5,2)$, 
\beqs
{\cal P}_{(5,2)} &=& q^4 - 12q^3 + 42q^2 - 60q + 31 \cr\cr
&=& \Big [ q^2 -2(3+\sqrt{2})q + (7+3\sqrt{2}) \Big ] 
    \Big [ q^2 -2(3-\sqrt{2})q + (7-3\sqrt{2}) \Big ] \ . 
\label{polfactor_P5L2}
\eeqs
The first quadratic factor has the roots $q_c(5,2)$, listed in 
Eq. (\ref{qc_P5L2}), and $q_{\rm int}(5,2)$, where 
\beq
q_{\rm int}(5,2) = 3 + \sqrt{2} - \Big (4+3\sqrt{2} \Big )^{1/2} = 1.543214 \
. 
\label{qint_P5L2}
\eeq
The second quadratic factor has no real roots. 

For $(p,\ell)=(7,2)$, 
\beqs
{\cal P}_{(7,2)} &=& q^6 - 18q^5 + 105q^4 - 300q^3 + 465q^2 -378q + 127 \cr\cr
&=& \Big [ q^3 -3(3+\sqrt{2})q^2 + 3(7+3\sqrt{2})q-(15+7\sqrt{2}) \, 
\Big ] \ \times 
\cr\cr
&\times&   \Big [ q^3 -3(3-\sqrt{2})q^2 + 3(7-3\sqrt{2})q-(15-7\sqrt{2}) \,
 \Big ] \ . 
\label{polfactor_P7L2}
\eeqs
The roots of the first cubic factor are $q_c(7,2)$, given in
Eq. (\ref{qc_P7L2}), and a complex-conjugate pair. The roots of the second
cubic factor are $q_{\rm int}(7,2)$, 
\beq
q_{\rm int}(7,2) = 3-\sqrt{2} - \Big [ 2(-7 + 5\sqrt{2} \, ) \Big ]^{1/3} + 
\frac{3\sqrt{2}-4}{\Big [ 2(-7+5\sqrt{2} \, )\Big ]^{1/3}} = 1.528849 \ .  
\label{qint_P7L2}
\eeq
and another complex-conjugate pair.  We find similar results for
higher $(p_{\rm odd}, \ell_{\rm even})$ cases.


\subsection{Calculation of $q_\infty(p_{\rm even},\ell_{\rm even})$
and $q_L(p,\ell_{\rm even})$} 
\label{qinf_section} 

We have also calculated $q_\infty$ for illustrative cases in the subclass 
$(p_{\rm even},\ell_{\rm even})$ where the $S_\infty$ sequences occur. 
We list the resulting values of $q_\infty(p_{\rm even},\ell_{\rm even})$ 
in Table \ref{qinf_table}.  For the cases
that we have studied, we find that 
$q_\infty(p_{\rm even},\ell_{\rm even})$ is invariant under the interchange
of $p_{\rm even}$ and $\ell_{\rm even}$, i.e., with $p_{\rm even}=2s$ and
$\ell_{\rm even}=2t$, 
\beq
q_\infty(2s,2t) = q_\infty(2t,2s) \quad {\rm for} \ s, \ t \in 
{\mathbb Z}_+ \ . 
\label{qinf_sym}
\eeq
This is evident in Table \ref{qinf_table}. For the cases that we have 
calculated, we find the monotonicity relations

\begin{enumerate}

\item 
\label{qinf_increases_with_P}

$q_{\infty}(p_{\rm even},\ell_{\rm even})$ is a monotonically increasing
function of $p_{\rm even}$ for fixed $\ell_{\rm even}$.

\item 
\label{qinf_increases_with_L}

$q_{\infty}(p_{\rm even},\ell_{\rm even})$ is a monotonically increasing
function of $\ell_{\rm even}$ for fixed $p_{\rm even}$.

\end{enumerate}

\begin{table}
  \caption{\footnotesize{Values of $q_\infty(G^{(p,\ell)}_\infty)$ for 
      $(p_{\rm even}, \ell_{\rm even})$ with $p=2, \ 4, \ 6$ and 
      $\ell=2, \ 4, \ 6$. For compact notation, in this table, we denote
      $q_\infty(G^{(p,\ell)}_\infty) \equiv q_\infty(p,\ell)$.}}
\begin{center}
\begin{tabular}{|c|c|c|}
  \hline\hline
$q_\infty(2,2)=1.185185$ & $q_\infty(2,4)=1.289093$ & $q_\infty(2,6)=1.338702$ 
\\ \hline 
$q_\infty(4,2)=1.289093$ & $q_\infty(4,4)=1.409283$ & $q_\infty(4,6)=1.466651$ 
\\ \hline
$q_\infty(6,2)=1.338702$ & $q_\infty(6,4)=1.466651$ & $q_\infty(6,6)=1.528149$ 
\\ \hline\hline
\end{tabular}
\end{center}
\label{qinf_table}
\end{table}

We next present some details of our calculations of $q_\infty$ for the
cases where the $S_\infty$ sequences occur, namely 
$(p_{\rm even},\ell_{\rm even})$.  These generalize the analysis in \cite{dhl}.
The discussion will also elucidate the method for calculating the values of 
$q_L(p_{\rm even},\ell_{\rm even})$ for $p_{\rm even} > \ell_{\rm even}$. 
We use a similar method to calculate $q_L(p,\ell)$ for other cases
$(p,\ell)$ with $p > \ell$. 
The starting point is the RG fixed point (RGFP) equation (\ref{RGFPeq}).  
In the lowest nontrivial case $(p,\ell)=(2,2)$, Eq. (\ref{RGFPeq}) reads
\beq
 \frac{v(v^3-2qv-q^2)}{A_2^2} = 0 \ , 
\label{eqP2L2full}
\eeq
where $A_\ell$ was defined in Eq. (\ref{A_L}). Several illustrative
examples of Eq. (\ref{RGFPeq}) are given in Appendix \ref{RGFP_appendix}.  
Recall that for any graph $G$, $v=0$ leads to the trivial result
$Z(G,q,0)=q^{n(G)}$, so all zeros are at $q=0$; hence $v \ne 0$ here.
Therefore Eq. (\ref{eqP2L2full}) reduces to the equation
\beq
Eq_{(2,2)}: \ q^2+2qv-v^3=0  \ , 
\label{Eq22}
\eeq
as given in \cite{dhl} (see also Appendix \ref{RGFP_appendix}). 
The nature of the roots in $q$ of this equation is determined by the 
discriminant of the left-hand side, as a polynomial in $v$. (For a general
treatment of discriminants, see, e.g., \cite{discrim}.) For a polynomial
equation $Pol(q,v)=0$, we denote the discriminant of the equation, considered 
as an equation in $v$, as ${\rm Disc}(Pol,v)$. Now, the condition that the
discriminant of Eq. (\ref{Eq22}) vanishes is 
\beq
q^3(27q-32) = 0 \ . 
\label{discP2L2}
\eeq
Since we are not dealing with the crossing at $q_L(2,2)=0$, we take $q \ne 0$.
Then, the solution to the condition that this discriminant should vanish is 
$q=q_\infty(2,2)=32/27$, as discussed in \cite{dhl} and listed 
above in Eq. (\ref{qinf_P2L2}).

This method generalizes to the $(p_{\rm even},\ell_{\rm even})$ cases where an
infinite sequence $S_\infty$ occurs and yields the value of $q_\infty(p_{\rm
  even}, \ell_{\rm even})$.  We illustrate this with some explicit examples. 
For $(p,\ell)=(2,4)$, Eq. (\ref{RGFPeq}) reads 
\beq
v (A_4)^{-2} \, \Big [v^7-8v^6-36qv^5-60q^2v^4-54q^3v^3-28q^4v^2-8q^5v-q^6 
\Big ] = 0 \ . 
\label{eqP2L4}
\eeq
Since $v \ne 0$, this yields the equation that the expression in square
brackets is equal to 0. Calculating the discriminant of this equation as
an equation in $v$ and setting it equal to zero then gives the condition 
\beqs
&& 823543q^6 - 5054848q^5 + 12366208q^4 - 13606912q^3 + 3772416q^2 \cr\cr
&+& 4521984q - 2883584 = 0 \ .
\label{discredP2L4}
\eeqs
This equation has two real solutions, one of which is negative, namely
$q=-0.5730675$. This is not relevant to the locus ${\cal B}_q$ for this
$(p,\ell)=(2,4)$ case, which has no crossing on the negative real-$q$ axis.
However, it will be relevant to ${\cal B}_q$ for the case $(p,\ell)=(4,2)$, as
will be discussed below. The real positive solution of Eq. (\ref{discredP2L4})
is $q_\infty(2,4)$, namely
\beq
q_\infty(2,4) = 1.289093 \ . 
\label{qinf_P2L4}
\eeq
This is listed in Table \ref{qinf_table}. The other solutions of
Eq. (\ref{discredP2L4}) are complex.  

For the case $(p,\ell)=(4,2)$ the condition that
the discriminant of the equation (\ref{RGFPeq}) should vanish is an equation
that only differs from the corresponding equation for $(p,\ell)=(2,4)$ by a
different prefactor power of $q$.  Since $q_\infty \ne 0$, this means that the
conditions for the vanishing of the respective discriminants of the equation
(\ref{RGFPeq}) are the same for $(p,\ell)=(2,4)$ and (4,2), so that
$q_\infty(4,2)=q_\infty(2,4) = 1.289093$. As noted above, we have found that
this is a general result, namely that the equations (with $q \ne 0$) for the
vanishing of the respective discriminants of Eq. (\ref{RGFPeq}) for $(p_{\rm
  even},\ell_{\rm even})=(2s,2t)$ and for $(2t,2s)$, where $s, \ t \in {\mathbb
  Z}_+$, are the same, yielding the symmetry relation (\ref{qinf_sym}). 
Anticipating our study of odd-$\ell$ cases below, we generalize this to the 
finding that the discriminants of the numerators in Eq. (\ref{RGFPeq})
(divided by the prefactor $v$), as functions of $v$ for the 
cases $(p,\ell)=(a,b)$ and $(p,\ell)=(b,a)$ are the same, up to prefactors that
are different (positive) powers of $q$. Since the leftmost crossing is at
$q_L=0$ for all cases except $(p,\ell)$ with $p > \ell$, we
can assume $q \ne 0$ in solving the various discriminant equations, so these
different prefactor powers of $q$ will not be relevant.  Therefore, the
condition that the discriminant of Eq. (\ref{RGFPeq}) vanishes for
$(p,\ell)=(a,b)$ is the same as this condition for $(p,\ell)=(b,a)$. 
The negative real solution of Eq. (\ref{discredP2L4}) is the value of
$q_L(4,2)$: 
\beq
q_L(4,2) = -0.5730675 \ .
\label{qleft_P4L2}
\eeq
This is listed in Table \ref{qleft_evenL_table}. We have applied the same
method to calculate the other values of $q_L(p,\ell)$ with $p > \ell_{\rm
  even}$ in this table, and similarly for $q_L(p,\ell_{\rm odd})$ with 
$p > \ell_{\rm odd}$ below. As an example, for the case $(p,\ell_{\rm
  even})=(3,2)$,  the condition that the discriminant of Eq. (\ref{RGFPeq}) 
vanishes is 
\beq
3125q^4 -13356q^3+17244q^2-5184q-2160=0 \ . 
\label{discredP3L2}
\eeq
This has two real solutions, $q=-0.222373$ and $q=2.086026$. The negative
solution is the value $q_L$ for this case:
\beq
q_L(3,2) = -0.222373 \ .
\label{qleft_P3L2}
\eeq
The positive solution is $q_c(2,3)$, as will be discussed below.

For $(p,\ell)=(4,4)$, Eq. (\ref{RGFPeq}) is too long to list here, but the
condition that the discriminant should vanish yields the equation 
\beqs
&&\Big ( 243q^2-688q+576 \Big)^3 \, 
\Big (3125q^3-9744q^2+10432q-4096 \Big ) \ \times 
\cr\cr
&\times& \Big (3125q^4-19008q^3+49152q^2-61440q+32768 \Big )^2=0 \ . 
\label{discredP4L4}
\eeqs
The factor $(243q^2-688q+576)^3$ and the factor $(3125q^4-19008q^3+ ...)^2$ 
in this equation have no real roots. The factor 
$(3125q^3-9744q^2+10432q-4096)$ 
has one real root, which is thus $q_\infty(4,4)$, namely 
\beqs
q_\infty(4,4) &=& \frac{3248}{3125} + \frac{8}{9375}(R_{44})^{1/3} -
\frac{118936}{3125(R_{44})^{1/3}} \cr\cr
       &=& 1.409283 \ , 
\label{qinf_P4L4}
\eeqs
where
\beq
R_{44} = 69735357 + 6590625\sqrt{114} \ . 
\label{r44}
\eeq
This is listed in Table \ref{qinf_table}.  


\subsection{Calculation of Ground-State Degeneracy for the Potts
  Antiferromagnet on $G^{(p,\ell)}_\infty$ at $q_c(p,\ell)$}
\label{wc_subsection}

We define the value of the degeneracy of states per vertex of the Potts
antiferromagnet on $G^{(p,\ell)}_\infty$, evaluated at 
$q=q_c(p,\ell)$, as
\beq
W_c(p,\ell) \equiv W(G^{(p,\ell)}_\infty,q) \ {\rm at} \ q=q_c(p,\ell) \ .
\label{wc}
\eeq
Using a method from \cite{qin_yang91} applicable for even $\ell$, we 
calculate, also for even $\ell$, 
\beq
W_c(p,\ell) = [q_c(p,\ell)]^{-1/(\ell-1)} \, 
{\cal N_W}^{1/[p \ell (\ell-1)]}
\label{qcform}
\eeq
where 
\beq
{\cal N}_W = ({\cal N}_{W1})^\ell - ({\cal N}_{W2})^\ell
\label{nw}
\eeq
and (with $q_c \equiv q_c(p,\ell)$)
\beq
{\cal N}_{W1} = [(q_c-1)^\ell + (q_c-1) ]^p + (q_c-1)[(q_c-1)^\ell -1 ]^p
\label{nw1}
\eeq
and
\beq
{\cal N}_{W2} = [(q_c-1)^\ell + (q_c-1) ]^p - [(q_c-1)^\ell -1 ]^p \ . 
\label{nw2}
\eeq
We list values of $W_c(p,\ell)$ for the illustrative even values $\ell=2, \
4, \ 6, \ 8$ and a range of values of $p$, namely $2 \le p \le 8$ in 
Table \ref{wc_table}.  For all of the $(p,\ell)$ cases where we have
obtained exact analytic expressions for the respective $q_c(p,\ell)$, we have
also calculated corresponding exact analytic expressions for
$W_c(p,\ell)$. Three of these are listed below (the first was reported in 
\cite{dhl}): 
\beq
W_c(2,2) = \sqrt{3} = 1.732051
\label{wc_P2L2}
\eeq
\beq
W_c(3,2) = \frac{(1107+782\sqrt{2} \, )^{1/3}}{3+\sqrt{2}} = 2.952126 
\label{wc_P3L2}
\eeq
and
\beqs 
W_c(4,2) &=& \frac{3^{1/4} \, \Big [5612517369 + 3535664388 (2)^{2/3} + 
4454657988(2)^{1/3} \Big ]^{1/8} }{3 + 2^{2/3} + 2^{1/3}} \cr\cr
&=& 4.271773 \ . 
\label{wc_P4L2} 
\eeqs
With our analytic expressions for $q_c(p,\ell)$ for some other values of
$(p,\ell)$, we have also obtained corresponding analytic results for
$W_c(p,\ell)$ for other $(p,\ell)$ cases, but they are too lengthy to present
here.

We observe the following monotonicity relations in Table 
\ref{wc_table} for these even values of $\ell$ where our calculation applies: 
\beq
W_c(p,\ell_{\rm even}) \quad 
{\rm is \ a \ monotonically \ increasing \ function \ of} \
p \ {\rm for \ fixed} \ \ell_{\rm even} \ .
\label{w_monontonic1}
\eeq
and
\beq
W_c(p,\ell_{\rm even}) \quad 
{\rm is \ a \ monotonically \ decreasing \ function \ of} \
\ell_{\rm even} \ {\rm for \ fixed} \ p \ .
\label{w_monontonic2}
\eeq

\begin{table}
\caption{\footnotesize{Values of $W_c(p,\ell)$ for an illustrative range of 
$p$ values and the even-$\ell$ values $\ell=2, \ 4, \ 6, \ 8$.}}
\begin{center}
\begin{tabular}{|c|c|c|c|}
  \hline\hline
$W_c(2,2)=1.732051$  & $W_c(2,4)=1.089758$  & $W_c(2,6)=1.0344167$ &
$W_c(2,8)=1.018179$  \\ \hline
$W_c(3,2)=2.952126$  & $W_c(3,4)=1.249644$  & $W_c(3,6)=1.108124$ &
$W_c(3,8)=1.063256$   \\ \hline
$W_c(4,2)=4.271773$  & $W_c(4,4)=1.393347$  & $W_c(4,6)=1.174894$ &
$W_c(4,8)=1.105142$   \\ \hline
$W_c(5,2)=5.635578$  & $W_c(5,4)=1.520206$  & $W_c(5,6)=1.232912$ &
$W_c(5,8)=1.141595$   \\ \hline
$W_c(6,2)=7.023269$  & $W_c(6,4)=1.633653$  & $W_c(6,6)=1.283857$ &
$W_c(6,8)=1.173496$   \\ \hline
$W_c(7,2)=8.425396$  & $W_c(7,4)=1.736504$  & $W_c(7,6)=1.329245$  &
$W_c(7,8)=1.201783$   \\ \hline
$W_c(8,2)=9.836927$  & $W_c(8,4)=1.830828$  & $W_c(8,6)=1.370206$  &
$W_c(8,8)=1.227183$   \\ \hline\hline
\end{tabular}
\end{center}
\label{wc_table}
\end{table}

Recall that a bipartite graph $G_{bp}$ is one that can be written
formally as $G_{bp} = G_1 \oplus G_2$, where all of the vertices
adjacent to a vertex in $G_1$ are in $G_2$ and vice versa. Consider a
bipartite graph $G_{bp}$ such that $n(G_1)=n(G_2)=n(G_{bp})/2$.  A
rigorous lower bound on $P(G_{bp,n},q)$ for an $n$-vertex bipartite
($bp$) graph $G_{bp}$ with even $n$ is
\beq
P(G_{bp,n},q) \ge q(q-1)^{n/2} \ .
\label{plower}
\eeq
This is proved by assigning one of the $q$ colors to all vertices in,
say, $G_1$ and then independently assigning any of the remaining $(q-1)$
colors to each vertex in $G_2$.  Using the cluster formula with
$v=-1$, one can generalize this from positive integer $q$ to positive
real $q$, and we perform this generalization here.  In the limit
$n \to \infty$, this implies the rigorous lower bound
\beq
W(G_{bp,\infty},q) \ge \sqrt{q-1} \ .
\label{wlow}
\eeq
Since $G^{(p,\ell)}_m$ is a bipartite graph for any $m$, the lower bound
(\ref{wlow}) applies to the $W$ function on $G^{(p,\ell)}_\infty$ in the range
where the $(1/n)$'th root in Eq. (\ref{wdef}) can be chosen to be real and
positive in an unambiguous manner. This range includes $q \ge q_c(p,\ell)$, and
we have thus calculated $W(G^{(p,\ell)}_\infty,q)$ evaluated at the lower end
of this range, namely at $q=q_c(p,\ell)$ to obtain $W_c(p,\ell)$, as in
Eq. (\ref{wc}).  To measure how close the actual degeneracy per vertex
evaluated at $q_c(p,\ell)$ is to its lower bound, we define the ratio
\beq
R_{W_c}(p,\ell) \equiv \frac{W_c(p,\ell)}{\sqrt{q_c(p,\ell)-1}} \ . 
\label{wr}
\eeq
We thus have
\beq
R_{W_c}(p,\ell) \ge 1 \ . 
\label{rwbound}
\eeq
The ratios $R_{W_c}(p,\ell_{\rm even})$ are listed in Table \ref{rwc_table} for
the same set of $p$ and $\ell_{\rm even}$ values as in Table \ref{wc_table}.
We find that (i) for a given $\ell_{\rm even}$, $R_{W_c}(p,\ell_{\rm even})$
increases monotonically as a function of $p$, and (ii) for a given $p$,
$R_{W_c}(p,\ell_{\rm even})$ decreases monotonically toward its lower limit of
unity as a function of $\ell_{\rm even}$. Thus, for small $p$ and large
$\ell_{\rm even}$, $R_{W_c}(p,\ell_{\rm even})$ is quite close to this lower
bound.  For example, $R_{W_c}(2,8)-1 = 2 \times 10^{-3}$. Combining our results
for $q_c(p,\ell)$ for even and odd $\ell$ (and general $p$), we will infer the
limit (\ref{qclim_L_inf_2}) below.  Further combining this with the rigorous
lower bound (\ref{wlow}), we will infer the limit (\ref{wclim_L_inf_1}) below.

\begin{table}
  \caption{\footnotesize{Values of the ratio $R_{W_c}(p,\ell)$ for an 
illustrative range of $p$ values and the even-$\ell$ 
values $\ell=2, \ 4, \ 6, \ 8$.}}
\begin{center}
\begin{tabular}{|c|c|c|c|}
  \hline\hline
  $R_{W_c}(2,2)=1.224745$  & $R_{W_c}(2,4)=1.01802865$  &
  $R_{W_c}(2,6)=1.00494290$& $R_{W_c}(2,8)=1.00202806$  \\ \hline
  $R_{W_c}(3,2)=1.597679$  & $R_{W_c}(3,4)=1.069381$ &
  $R_{W_c}(3,6)=1.025089$  & $R_{W_c}(3,8)=1.0127406$   \\ \hline
  $R_{W_c}(4,2)=1.940248$  & $R_{W_c}(4,4)=1.121449$  &
  $R_{W_c}(4,6)=1.048193$  & $R_{W_c}(4,8)=1.026332$   \\ \hline
  $R_{W_c}(5,2)=2.247908$  & $R_{W_c}(5,4)=1.168434$  &
  $R_{W_c}(5,6)=1.069947$  & $R_{W_c}(5,8)=1.0396289$   \\ \hline
  $R_{W_c}(6,2)=2.526908$  & $R_{W_c}(6,4)=1.210329$  &
  $R_{W_c}(6,6)=1.089701$  & $R_{W_c}(6,8)=1.0519277$   \\ \hline
  $R_{W_c}(7,2)=2.782949$  & $R_{W_c}(7,4)=1.247897$  &
  $R_{W_c}(7,6)=1.107566$  & $R_{W_c}(7,8)=1.0631635$   \\ \hline
  $R_{W_c}(8,2)=3.020374$  & $R_{W_c}(8,4)=1.281881$  &
  $R_{W_c}(8,6)=1.1237905$  & $R_{W_c}(8,8)=1.0734289$   \\ \hline\hline
\end{tabular}
\end{center}
\label{rwc_table}
\end{table}


\section{Calculation of $q_c(p,\ell)$, $q_L(p,\ell)$, and $q_x(p,\ell)$
  for Odd $\ell$} 
\label{qc_L_odd_section}

In order to calculate the positions of the points where ${\cal B}_q$ crosses
the real-$q$ axis in cases with odd $\ell$, we again analyze the discriminant
associated with the RGFP equation (\ref{RGFPeq}).  As discussed above,
Eq. (\ref{RGFPeq}) sets a rational function equal to zero, so the solutions are
determined by the condition that the numerator of this rational function
vanishes.  This numerator has an overall factor of $v$, but the solution $v=0$
is not relevant here, since the RG transformation is trivial in this case,
mapping $v=0$ to $v'=0$. Hence, the relevant equation is the rest of the
numerator of Eq. (\ref{RGFPeq}) set equal to zero. The nature of the solutions
to this equation in the variable $q$ is determined by the discriminant of this
part of the numerator, as a function of $v$. The special points of interest
here correspond to the condition that this discriminant vanishes.  As is
illustrated by the examples in Appendix \ref{RGFP_appendix}, this discriminant
contains a prefactor that is a power of $q$, but since we are studying
crossings of the locus ${\cal B}_q$ on the real axis away from $q=0$, we take
$q \ne 0$.  There are several different subclasses with odd $\ell$ to
consider. We note that Ref. \cite{qin_yang91} presented numerical values for
$q_c$ for several members of this class with odd $\ell$, and where our results
overlap, they agree with those in \cite{qin_yang91}, taking into account the
requisite changes in notation and normalization \cite{spb}. We also give
exact analytic results for a number of cases.

First, we consider the diagonal cases where $p$ and $\ell$ are odd and $p_{\rm
  odd}=\ell_{\rm odd}$.  We illustrate the calculation with two examples.  For
$(p_{\rm odd},\ell_{\rm odd})=(3,3)$, given that $q \ne 0$, the condition that
the discriminant of Eq. (\ref{RGFPeq}), as a function of $v$) vanishes is the
equation
\beq
(p,\ell)=(3,3): \quad (4q-9)^2(16q-27)^3(256q^2-549q+324)=0 \ , 
\label{disceq33}
\eeq
as listed in Appendix \ref{RGFP_appendix}. 
The largest zero is $q_c(3,3)$; evidently, this is 
\beq
q_c(3,3)=\frac{9}{4} = 2.25 \ .
\label{qc_P3L3}
\eeq
The next largest zero is the point at which two approximately vertically
oriented cusp regions approach and touch the real axis.  The width of these
regions goes to zero as they approach the real axis.  As mentioned above, we
denote this point as $q_x(p_{\rm odd},\ell_{\rm odd})$. Here we calculate 
\beq
q_x(3,3) =  \frac{27}{16} = 1.6875 \ . 
\label{qx_P3L3}
\eeq
These values of $q_c$ and $q_x$ for $(p,\ell)=(3,3)$ are listed in Tables
\ref{qc_odd_L_table} and \ref{qx_table}, respectively, together with
corresponding values of $q_c$ and $q_x$ for other odd-$\ell$ cases.
The third factor in Eq. (\ref{disceq33}), namely $(256q^2-549q+324)$, 
has no real zeros; its zeros occur at
\beq
q =\frac{9}{512}\Big ( 61 \pm 5i \sqrt{15} \, \Big ) = 1.0722656 \pm 0.340399i 
\ . 
\label{q_complex_cusp}
\eeq
Although we focus here on points at which ${\cal B}_q$ crosses the real-$q$
axis, we note in passing that the points in Eq. (\ref{q_complex_cusp}) appear
to be coincident, to within the accuracy of our calculation, with cusps on
${\cal B}_q$ in the complex plane.


\begin{table}[htbp]
  \caption{\footnotesize{ Values of $q_c(G^{(p,\ell)}_\infty)$ for 
illustrative values of $(p,\ell)$ with odd $\ell$.}}
\begin{center}
\begin{tabular}{|c|c|c|} \hline\hline
$q_c(2,3)=2.086026$   & $q_c(2,5)=2.025228$   & $q_c(2,7)=2.011955$  \\ \hline
$q_c(3,3)=2.25$       & $q_c(3,5)=2.090644$   & $q_c(3,7)=2.050101$  \\ \hline
$q_c(4,3)=2.400878$   & $q_c(4,5)=2.152990$   & $q_c(4,7)=2.0883665$  \\ \hline
$q_c(5,3)=2.536216$   & $q_c(5,5)=2.207825$   & $q_c(5,7)=2.122280$  \\ \hline
$q_c(6,3)=2.659111$   & $q_c(6,5)=2.256199$   & $q_c(6,7)=2.152112$  \\ \hline
$q_c(7,3)=2.772194$   & $q_c(7,5)=2.299419$   & $q_c(7,7)=2.178605$  \\ \hline
\hline
\end{tabular}
\end{center}
\label{qc_odd_L_table}
\end{table}


\begin{table}
  \caption{\footnotesize{Values of $q_x(p_{\rm odd},\ell_{\rm odd})$, 
where complex-conjugate vertically oriented cusp 
regions come together and touch the real-$q$ axis. 
This type of crossing point occurs for 
$(p_{\rm odd},\ell_{\rm odd})$ cases with $p_{\rm odd}  \le \ell_{\rm odd}$.}}
\begin{center}
\begin{tabular}{|c|c|} \hline\hline
$(p_{\rm odd},\ell_{\rm odd})$  & $q_x(p_{\rm odd},\ell_{\rm odd})$ \\ 
\hline\hline
(3,3) & 1.687500  \\ \hline
(3,5) & 1.908264  \\ \hline
(3,7) & 1.949712  \\ \hline
(3,9) & 1.967101  \\ \hline 
(5,5) & 1.736884  \\ \hline
(5,7) & 1.872366  \\ \hline
(7,7) & 1.771918  \\ \hline
\hline
\end{tabular}
\end{center}
\label{qx_table}
\end{table}


In the next higher diagonal case, $(p_{\rm odd},\ell_{\rm odd})=(5,5)$, 
the condition that the discriminant of Eq. (\ref{RGFPeq}), as a function of
$v$, should vanish, can be written as 
\beq
(p,\ell)=(5,5): \quad 
{\cal P}_{(5,5),3a}{\cal P}_{(5,5),3b}{\cal P}_{(5,5),4}{\cal P}_{(5,5),6}=0
 \ . 
\label{disceq_P5L5}
\eeq
Because of the increased complexity here, we introduce a notation to indicate
the various polynomial factors, appending to each a label indicating the value
of $(p_{\rm odd},\ell_{\rm odd})$ and the degree of the polynomial factor in 
$q$, with a further label $a$, $b$, etc. if there are several factors of the
same degree.  Several of these polynomial factors are, themselves, raised to
various powers, but these just increase the multiplicity of the zeros,
we will not need to discuss these multiplicities. 
The cubic factors (which are each raised to further powers) are 

\beq
{\cal P}_{(5,5),3a} = (729q^3-2950q^2+4375q-3125)^2 
\label{pol_P5L5_deg3a}
\eeq
and
\beq
{\cal P}_{(5,5),3b} = (4096q^3-15925q^2+22500q-12500)^3  \ . 
\label{pol_P5L5_deg3b}
\eeq
Ignoring multiplicities of zeros, The factors ${\cal P}_{(5,5),3a}$
and ${\cal P}_{(5,5),3b}$ each have one real zero and one
complex-conjugate pair of zeros.  The real zero of ${\cal
  P}_{(5,5),3a}$ is
\beqs
q_c(5,5) &=& \frac{2950}{2187} + \frac{5}{4374}(R_{55})^{1/3} - 
\frac{346250}{2187 (R_{55})^{1/3}} \cr\cr
&=& 2.207825  \ , 
\label{qc_P5L5}
\eeqs
where
\beq
R_{55} = 100(3682297+811377\sqrt{21}) \ . 
\label{rootfun_P5L5}
\eeq
Similarly, ${\cal P}_{(5,5),3b}$ has one real root, which is
\beq
q_x(5,5) = 1.736884
\label{qx_P5L5}
\eeq
and a complex-conjugate pair. The degree-4 factor ${\cal P}_{(5,5),4}$,
\beq
{\cal P}_{(5,5),4} = 46656q^4-191275q^3+298000q^2-212500q+62500
\label{pol_P5L5_deg4}
\eeq
and the degree-6 factor ${\cal P}_{(5,5),6}$ (which, itself, is squared),
\beqs
    {\cal P}_{(5,5),6} &=& \Big ( 884736q^6-7061175q^5+24904750q^4 \cr\cr
    &-& 50590625q^3+62671875q^2-45312500q+15625000 \Big )^2 \ , 
\label{pol_P5L5_deg6}
\eeqs
have no real roots. In general, 
for this subclass $(p_{\rm odd},\ell_{\rm odd})$ with 
$p_{\rm odd}=\ell_{\rm odd}$, we find that this pattern continues; that is, 
one of the real solutions to the condition that the discriminant of Eq. 
(\ref{RGFPeq}) as a function of $v$ vanishes is the value of $q_c$, which is
larger than 2, and the other real solution is $q_x$, which is 
somewhat smaller than 2. We display these values of $q_x(p_{\rm
  odd},\ell_{\rm odd})$, together with others, in Table \ref{qx_table}. 
As in the $(p,\ell)=(3,3)$ case, one can study locations of cusps in the 
complex plane away from the real axis, but we will restrict our consideration
here to crossings of ${\cal B}_q$ on the real axis. 

We next turn to $(p_{\rm odd}, \ell_{\rm odd})$ cases where $p \ne \ell$.  We
have found that the discriminants of the numerators in Eq. (\ref{RGFPeq})
(divided by the prefactor $v$), as functions of $v$ for the cases
$(p,\ell)=(a,b)$ and $(p,\ell)=(b,a)$ are the same, up to prefactors that are
different (positive) powers of $q$. As noted above, since the leftmost crossing
is at $q_L=0$ for all cases except $(p,\ell)$ with $p > \ell$, we can assume $q
\ne 0$ in solving the various discriminant equations. Therefore, the condition
that the discriminant of Eq. (\ref{RGFPeq}) vanishes for $(p,\ell)=(a,b)$ is
the same as this condition for $(p,\ell)=(b,a)$, and hence we treat these cases
together.  Let us consider the polynomial in $q$ being equated to zero after
extraction of this prefactor power of $q$.  The degree of this polynomial in
$q$ is even and increases rapidly with $p_{\rm odd}$ and $\ell_{\rm odd}$.  For
example, for the cases $(p_{\rm odd},\ell_{\rm odd})=(3,5)$ or (5,3), its
degree is 16; for (3,7) or (7,3) its degree is 24, and so forth for higher
$p_{\rm odd}$ and $\ell_{\rm odd}$. In all of these cases, we find that this
discriminant equation has four real solutions, of which three are positive and
one is negative. Of the three real positive solutions to the above-mentioned
discriminant equation, the larger one is $q_c(p_{\rm larger},\ell_{\rm
  smaller})$, and the second largest one is $q_c(p_{\rm smaller},\ell_{\rm
  larger})$, where the notation $(p_{\rm larger},\ell_{\rm smaller})$ means
that $p > \ell$ and $q_c(p_{\rm smaller},\ell_{\rm larger})$ means that $p <
\ell$.  Further, we find that in these $(p_{\rm odd}, \ell_{\rm odd})$ cases
where $p < \ell$, the smallest real positive solution to the above-mentioned
equation is the point where two complex-conjugate approximately vertically
oriented cusp-like regions come together, with width going to zero as they
approach the real axis, and touch this real axis.  As in the diagonal $(p_{\rm
  odd}, \ell_{\rm odd})$ cases with $p_{\rm odd}=\ell_{\rm odd}$, we denote
this point as $q_x$.  In Table \ref{qx_table} we list the values of $q_x(p_{\rm
  odd}, \ell_{\rm odd})$ that we have calculated for several illustrative
$(p_{\rm odd}, \ell_{\rm odd})$ cases.

In contrast, in the cases $(p_{\rm odd},\ell_{\rm odd})$ that we have
studied with $p_{\rm odd} > \ell_{\rm odd}$, viz., $(p,\ell)=(5,3)$,
and (7,3), we do not find evidence of any point at which ${\cal B}_q$
crosses the real-$q$ axis in the interval $q_L < q < q_c$.  As stated
above, for these $(p_{\rm odd}, \ell_{\rm odd})$ cases, one of the
four real zeros of the discriminant equation is negative and is the
value of $q_L(p_{\rm odd},\ell_{\rm odd})$ if $p_{\rm odd} > \ell_{\rm
  odd}$, whereas it does not correspond to a crossing if $p_{\rm odd}
\le \ell_{\rm odd}$. We list these values in Table \ref{qleft_oddL_table}
together with other entries with even $p$, to be discussed below.

\begin{table}
  \caption{\footnotesize{Values of $q_L(p,\ell_{\rm odd}) < 0$ that occur for 
odd $\ell$ if $p > \ell_{\rm odd}$ (where $p$ can be even or odd).          
The entries are listed in order of increasing values of $p/\ell_{\rm odd}$.}}
\begin{center}
\begin{tabular}{|c|c|c|} \hline\hline
$(p,\ell_{\rm odd})$  & $p/\ell_{\rm odd}$ & $q_L(p,\ell_{\rm odd})$ \\ \hline
$(6,5)$               & 1.2                & $-0.00915523$   \\ \hline
$(4,3)$               & 1.333              & $-0.04901145$  \\ \hline
$(7,5)$               & 1.4                & $-0.02841405$   \\ \hline
$(5,3)$               & 1.667              & $-0.135481$    \\ \hline
$(6,3)$               & 2                  & $-0.229166$    \\ \hline
$(7,3)$               & 2.333              & $-0.322659$    \\ \hline
$(8,3)$               & 2.667              & $-0.413745$    \\ \hline
$(9,3)$               & 3                  & $-0.501783$    \\ \hline
\hline
\end{tabular}
\end{center}
\label{qleft_oddL_table}
\end{table}

Finally, we consider the cases $(p_{\rm even},\ell_{\rm odd})$. Here, we find a
considerable variety in the structure of ${\cal B}_q$, depending on the values
of $p_{\rm even}$ and $\ell_{\rm odd}$.  Among the cases that we have studied,
for $(p,\ell)=(2,3)$ (see Fig. \ref{P2L3y0plot_fig}), there are three
complex-conjugate pairs of cusps away from the real axis pointing toward this
axis, but, at least with the pixel resolution in our calculations, we do not
observe any wedges that actually extend down to the real axis.  This leads to
the inference that for $(p_{\rm even},\ell_{\rm odd})=(2,3)$, the locus ${\cal
  B}_q$ crosses the real-$q$ axis only at $q_L=0$ and $q_c$.  We also observe
this property for the cases $(p_{\rm even},\ell_{\rm odd})=(4,5)$. However, in
contrast, for $(p_{\rm even},\ell_{\rm odd})=(2,5)$ at the level of resolution
in our calculations, the locus ${\cal B}_q$ actually includes a line segment on
the real axis, which is part of a Mandelbrot-like set, as shown in
Fig. \ref{P2L5y0plot_fig} and, in magnified form, in
Fig. \ref{P2L5y0plotq1.8_2.1_fig}. The locus ${\cal B}_q$ also includes smaller
line segments on the real axis in the case (2,7), one of which is shown in
Fig. \ref{P2L7y0plotq1.86_2.05_fig}. 

Combining our results for $q_L$ for both even and odd $p$, we make the
following observation. If $p > \ell$, so that $q_L(p,\ell) < 0$, we
find that for the cases we have studied, the magnitude $|q_L(p,\ell)|$ 
for a fixed $\ell$ is a monotonically increasing function
of $p$.  This monotonic behavior is evident in Tables \ref{qleft_evenL_table}
and \ref{qleft_oddL_table}.  Note,
however, that $|q_L(p,\ell)|$ is not a monotonically increasing
function of $p/\ell$. This is shown, e.g., by the fact that
$|q_L(7,5)|=0.0284$, which is smaller than $|q_L(4,3)|=0.0490$, although the
ratio $p/\ell_{odd}$ has the value $7/5=1.4$ for $(p,\ell)=(7,5)$, which is
larger than the value 4/3=1.33 for $(p,\ell)=(4,3)$.  Among cases $(p,\ell)$
cases for which $p > \ell$, there are several where the ratio $p/\ell$ is the
same for even and odd $\ell$; these include (i) $p/\ell = 2$ for
$(p,\ell)=(4,2)$ and $(6,3)$; (ii) $p/\ell = 3$ for $(p,\ell)=(6,2)$ and
$(9,3)$.  In both the pair (i) and (ii) we find that $q_L$ is substantially
more negative for the even-$\ell$ case than for the odd-$\ell$ case with the
same value of $p/\ell$. .

Our discriminant method yields the values of $q_c$ for each of these 
$(p_{\rm even},\ell_{\rm odd})$ cases.  Here, the 
condition for the vanishing of the discriminant of Eq. (\ref{RGFPeq}) produces
an equation that has two real roots, one of which is negative and the other 
of which is a unique real positive solution, which is 
$q_c(p_{\rm even},\ell_{\rm odd})$. As with the other cases, the degree of 
the equation increases rapidly as $p_{\rm even}$ and $\ell_{\rm odd}$ 
increase.  As an example, we calculate that for 
$(p_{\rm even},\ell_{\rm odd})=(2,3)$, this equation is 
\beq
(p,\ell)=(2,3): \quad 3125q^4 - 13356q^3 + 17244q^2 - 5184q - 2160=0 \ . 
\label{disceq_P2L3}
\eeq
As discussed before, this is the same as the condition that the discriminant
of Eq. (\ref{RGFPeq}) vanishes  for the case 
$(p,\ell)=(3,2)$, given in Eq. (\ref{discredP3L2}). This equation has two real
solutions; the negative solution is $q_L(3,2)$ given in Eq. (\ref{qleft_P3L2})
and the positive solution is  
\beq
q_c(2,3) = 2.086026 \ , 
\label{qc_P2L3}
\eeq
listed in Table \ref{qc_odd_L_table}.  In a similar manner, we use this method
to calculate the values of $q_c$ for several other $(p_{\rm even},\ell_{\rm
  odd})$ cases; these are listed in Table \ref{qc_odd_L_table}.


\section{Further General Properties of Crossing Points on ${\cal B}_q$}
\label{further_section}

These results for odd $\ell$ are consistent with monotonicity relations
analogous to those that follow from
Eq. (\ref{qc_eq_even_L}). Combining results for even and odd $\ell$, we find
the following properties (in the nontrivial range $p \ge 2$
and $\ell \ge 2$) for both even and odd $\ell$ (and both even and odd $p$) 
for the cases we have calculated:

\begin{enumerate}

\item 
\label{qcvalue}
For $G^{(p,\ell)}_\infty$,
\beq
2 < q_c(p,\ell) \le 3 \ , 
\label{qcgt2}
\eeq
where the upper limit is realized for $q(2,2)=3$. 

\item
\label{qc_increases_with_P}
  For $G^{(p,\ell)}_\infty$, $q_c(p,\ell)$ is a monotonically increasing
  function of $p$ for fixed $\ell$. 

\item
\label{qc_decreases_with_L_even_and_L_odd}

  For $G^{(p,\ell)}_\infty$ and fixed $p$, $q_c(p,\ell)$ is a monotonically
  decreasing function of $\ell$ separately for even values of $\ell$ and for
  odd values of $\ell$. However, $q_c(p,\ell)$ is not a monotonically
  decreasing function of $\ell$ if one considers even and odd values of $\ell$
  as one set.

\item
\label{qc_large_L_lim2}

  For fixed $p$ and both even and odd $\ell$, 
\beq
\lim_{\ell \to \infty} q_c(p,\ell) = 2 \ , 
\label{qclim_L_inf_2}
\eeq
where this limit is approached from above, as is evident in Tables 
\ref{qc_even_L_table} and \ref{qc_odd_L_table}. 

\item 
\label{wclim}

The inference (\ref{qc_large_L_lim2}), in combination with the rigorous 
lower bound (\ref{wlow}), implies that, for fixed $p$,
\beq
\lim_{\ell \to \infty} W_c(p,\ell) = 1 \ ,
\label{wclim_L_inf_1}
\eeq
where this limit is approached from above, as is evident from Table 
\ref{wc_table}. 

\end{enumerate}

Concerning the second property, some examples are as follows. 
For $p=2$, $q_c$ decreases from $q_c(2,2)=3$ to
$q_c(2,3)=2.086$, but then increases to $q_c(2,4)=2.146$, as $\ell$ increases
from 2 to 3 to 4; and for $p=3$, $q_c$ decreases from $q_c(3,2)=4.114$ to
$q(3,3)=2.250$, but then increases to $q_c(3,4)=2.366$ as $\ell$ increases
from 2 to 3 to 4.

Our results for all of the $(p,\ell)$ cases we have calculated also
motivate the inference that $\lim_{\ell \to \infty} q_c(p,\ell) = 2$ for all
$p$, with this limit being approached from above.
We also observe that the $q_c(p_{\rm odd},\ell_{\rm odd})$ values that we
have calculated for the diagonal case $p=\ell=2s+1$, namely $p=\ell=3$, 5, and
7, $q_c(2s+1,2s+1)$ is a decreasing function of $2s+1$.

We can provide some heuristic insight into these results on $q_c(p,\ell)$ as
follows. The proper $q$-coloring of the vertices of a graph is easiest in the
limit of large $q$, and in the limit as the number of vertices goes to infinity
and one defines the continuous accumulation set ${\cal B}_q$ of the zeros of
$P(G,q)$, the associated $q_c$ separates a region extending from $q_c$ to
$q=\infty$ from region(s) at smaller $q$ extending, in particular, to $q_L$.
Now for a given graph $G$, the constraints on a proper $q$-coloring of its
vertices tend to increase as the maximal vertex degree increases, since the
larger the maximal vertex degree in a graph that contains circuits (as the
graphs $G^{(p,\ell)}_m$ do), the more paths involving adjacent vertices there
generically are, and these yield constraints on a proper $q$-coloring of $G$.
Although $G^{(p,\ell)}_m$ is not a $\Delta$-regular graph except for the
initial $m=0$ graph and $G^{(2,\ell)}_1$, this is not a complication, since we
are only interested in the behavior in the $m \to \infty$ limit, where we can
focus on the effective vertex degree, $\Delta_{\rm eff}(p,\ell)$.

Related to the above, the region of large integer $q$ values is the region in
which the number of proper $q$-colorings of the graph $G^{(p,\ell)}_m$ grow
exponentially and hence where there is nonzero degeneracy per site in the $m
\to \infty$ limit, $W(G^{(p,\ell)}_\infty,q)$. Let us denote this region (with
$q$ generalized from positive integers to positive real numbers) as $R_1$.  The
lower end of this semi-infinite interval occurs at $q_c(p,\ell)$.  Since an
increase in $\Delta_{\rm eff}(G^{(p,\ell)}_m)$ generically increases the
constraints on a proper $q$-coloring of $G^{(p,\ell)}_m$, this suggests that an
increase in $\Delta_{\rm eff}(G^{(p,\ell)}_m)$ has the effect of reducing the
interval $R_1$. Taking the limit $m \to \infty$, this suggests that an increase
in $\Delta_{\rm eff}(G^{(p,\ell)}_\infty)$ and the resultant reduction in the
interval $R_1$ involves an increase in $q_c(p,\ell)$.  For the same reason, a
decrease in $\Delta_{\rm eff}(G^{(p,\ell)}_\infty)$ is expected to decrease
$q_c(p,\ell)$.  Given that (in the nontrivial range $p \ge 2$ and $\ell \ge
2$), $\Delta_{\rm eff}(G^{(p,\ell)}_\infty)$ is a monotonically increasing
function of $p$ for fixed $\ell$ and a monotonically decreasing function of
$\ell$ for fixed $p$, this motivates the monotonicity relation stated above,
that for fixed $\ell$, $q_c(p,\ell)$ increases monotonically with
$p$. Moreover, in view of the differences between the operation of the RG
transformation $F_{(p,\ell),q}(v)$ on $v$ for even and odd $\ell$, one expects
that the monotonicity property of $q_c(p,\ell)$ as a function of $\ell$ for
fixed $p$ would apply separately for even and odd $\ell$. This motivates the
two separate monotonicity relations stated above for even and odd $\ell$.
$q_c(p,\ell)$ is a monotonically decreasing function (i) of even $\ell$ for
fixed $p$ and, separately, (ii) of odd $\ell$ for fixed $p$.  

The limit (\ref{qclim_L_inf_2}) can be understood as follows.  As background,
we recall the a result from graph theory that the chromatic polynomial of the
$n$-vertex circuit graph $C_n$ is $P(C_n,q) = (q-1)^n + (q-1)(-1)^n$.  Hence,
in the limit $n \to \infty$, $q_c(C_\infty)=2$ \cite{w}. Now, applying this to
the present study, as $\ell \to \infty$ for fixed $p$, the proper $q$-coloring
of $G^{(p,\ell)}_m$ as $m \to \infty$ is dominantly determined by the proper
$q$-coloring of the circuits, each of which is approaching infinite length,
which leads to the inference (\ref{qclim_L_inf_2}). Combining this with the
rigorous lower bound (\ref{wlow}) then leads to the inference
(\ref{wclim_L_inf_1}).


\section{Conclusions }
\label{conclusions_section}

In conclusion, in this work we have calculated the continuous
accumulation set ${\cal B}_q(p,\ell)$ of zeros of the chromatic
polynomial $P(G^{(p,\ell)}_m,q)$ in the limit $m \to \infty$, on a
family of graphs $G^{(p,\ell)}_m$ defined such that $G^{(p,\ell)}_m$
is obtained from $G^{(p,\ell)}_{m-1}$ by replacing each edge (i.e.,
bond) on $G^{(p,\ell)}_m$ by $p$ paths each of length $\ell$ edges,
starting with the tree graph $T_2$.  This work extends a previous
study with R. Roeder of the $(p,\ell)=(2,2)$ case to higher $p$ and
$\ell$ values. Our method uses the property that the chromatic
polynomial $P(G,q)$ of a graph $G$ is equal to the $v=-1$ evaluation
of the partition function of the $q$-state Potts model, together with
(i) the property that $Z(G^{(p,\ell)}_m,q,v)$ can be expressed via an
exact closed-form real-space renormalization group transformation in
terms of $Z(G^{(p,\ell)}_{m-1},q,v')$, where $v'=F_{(p,\ell),q}(v)$ is
a rational function of $v$ and $q$ and (ii) ${\cal B}_q(p,\ell)$ is
the locus in the complex $q$-plane that separates regions of different
asymptotic behavior of the $m$-fold iterated RG transformation
$F^m_{(p,\ell),q}(v)$ in the $m \to \infty$ limit, starting from the
initial value $v=v_0=-1$.  Our results involve calculations of region
diagrams in the complex $q$-plane showing the types of behavior in the
$m \to \infty$ limit of the iterated mapping $F^m_{(p,\ell),q}(v)$
with initial value $v=v_0=-1$.  We find a number of differences in
this region, and the nature of the crossings of ${\cal B}_q$ on the
real axis, depending on whether $p$ and/or $\ell$ is even or odd, and
thus study the four different types of classes $(p_{\rm
  even},\ell_{\rm even})$, $(p_{\rm odd},\ell_{\rm even})$, $(p_{\rm
  odd},\ell_{\rm odd})$, and $(p_{\rm even},\ell_{\rm
  odd})$. Calculations are presented of the maximal (rightmost) point
$q_c(G^{(p,\ell)}_\infty)$ at which the locus ${\cal B}_q$ crosses the
real-$q$ axis.  For the cases we have studied, we find that the
leftmost point $q_L(p,\ell)$ where ${\cal B}_q$ crosses the real axis
is $q_L(p,\ell)=0$ if $p \le \ell$ and $q_L(p,\ell) < 0$ if $p >
\ell$.  For cases $(p_{\rm even},\ell_{\rm even})$, we observe the
occurrence of a sequence $S_\infty$ of crossings and calculate the
values of the left-endpoint of this sequence, $q_\infty(p_{\rm
  even},\ell_{\rm even})$.  In the cases $(p_{\rm odd},\ell_{\rm
  even})$ we find that ${\cal B}_q(p_{\rm odd},\ell_{\rm even})$
crosses the real axis at an interior point $q_{\rm int}(p_{\rm
  odd},\ell_{\rm even})$ and calculate the value of this point.  For
cases $(p_{\rm even},\ell_{\rm odd})$ we find that there are
complex-conjugate cusps that extend down and touch the real axis at a
point $q_x(p_{\rm even},\ell_{\rm odd})$, and we calculate this point
for illustrative cases.  The characteristics of these various points
as functions of $p$ and $\ell$ are further described.  In general, our
study reveals a wealth of structural features that will be interesting
for further analysis.


\bigskip
\bigskip

{\bf Acknowledgments}

    We are grateful to Prof. Roland Roeder for collaboration on Ref. \cite{dhl}
    and subsequent valuable discussions.  
    This research was partly supported by the Taiwan
    National Science and Technology Council (NSTC) grant No. 
    113-2115-M-006-006-MY2 (S.-C.C.) and by the U.S. National Science
    Foundation grant No. NSF-PHY-22-10533 (R.S.).

\bigskip
\bigskip

{\bf Declarations}: (1) The authors have no financial or
non-financial conflicts of interest relevant to this article; (2)
concerning data accessibility, this article is theoretical, and
relevant data are included herein.

\bigskip
\bigskip


\begin{appendix}

\section{Examples of Renormalization-Group Fixed-Point and 
Discriminant Equations}
\label{RGFP_appendix}

Here we display some examples of RGFP equations for illustrative
$(p,\ell)$ cases. These have the general form 
\beq
(p,\ell): \quad \frac{v {\cal P}_{p,\ell}}{(A_\ell)^p} = 0 \ , 
\label{eqcrit}
\eeq
where ${\cal P}_{p,\ell}$ is a polynomial in $q$ and $v$, and $A_\ell$ is
a polynomial in $q$ and $v$ depending on $\ell$, given in Eq. (\ref{A_L}),
with
\beq
A_2= q+2v 
\label{AL2}
\eeq
\beq
A_3 = q^2 + 3qv + 3v^2 
\label{AL3}
\eeq
and
\beq
A_4 = (q+2v)(q^2+2qv+2v^2) \ . 
\label{AL4}
\eeq

Some examples of Eq. (\ref{eqcrit}) for various $(p,\ell)$ are:
\beq
(p,\ell)=(2,2): \quad v A_2^{-2} (v^3-2qv-q^2) = 0
\label{eqcrit_P2L2}
\eeq
\beq
(p,\ell)=(3,2): \quad v A_2^{-3} \Big [v^5+6v^4+(3q+4)v^3-3q^2v-q^3\Big ]=0 
\label{eqcrit_P3L2}
\eeq
\beq
(p,\ell)=(2,3): \quad vA_3^{-2}\Big ( v^5-3v^4-12qv^3-13q^2v^2
-6q^3v-q^4 \Big )=0
\label{eqcrit_P2L3}
\eeq
and
\beqs
(p,\ell)=(3,3): \quad && v A_3^{_-3}
         \Big (v^4 + 9v^3 +12qv^2 +6q^2v + q^3\Big ) \times \cr\cr
&\times& \Big (v^4 - 3qv^2 - 3q^2v -q^3 \Big ) = 0  \ . 
\label{eqcrit_P3L3}
\eeqs
Since $v=0$ yields a trivial RG transformation, these RGFP equations are
equivalent to the corresponding conditions that the rest of the numerator
for a given $(p,\ell)$ is zero.  The discriminants of these latter equations,
as functions of $v$, are denoted $D_{(p,\ell)}$. We denote the conditions
that these discriminants vanish as $Deq_{(p,\ell)}$.
For our illustrative cases, these are 

\beq
Deq_{(2,2)}: \quad q^3(27q-32)=0 
\label{disc_P2L2}
\eeq
\beq
Deq_{(3,2)}: \quad  q^8(3125q^4-13356q^3+17244q^2-5184q-2160)=0
\label{disc_P3L2}
\eeq
\beq
Deq_{(2,3)}: \quad q^{12}(3125q^4-13356q^3+17244q^2-5184q-2160)=0
\label{disc_P2L3}
\eeq
and
\beq
Deq_{(3,3)}: \quad q^{35}(4q-9)^2(16q-27)^3(256q^2-549q+324)=0 \ . 
\label{disc_P3L3}
\eeq
As these examples illustrate, the discriminant equations for the
$(p,\ell)=(a,b)$ and $(b,a)$ cases are the same except for different
prefactors of (positive) powers of $q$.

  
\section{ Details of Calculation of $q_c(p,\ell)$ for Even $\ell$.}
\label{qccalc_appendix}

In this appendix we derive Eq. (\ref{qc_eq_even_L}). The analysis begins with
the observation that if $\ell$ is even, then the RG transformation
(\ref{vprime}), or equivalently, (\ref{yprime}), maps a model with the
antiferromagnetic sign of the spin-spin coupling, $J < 0$, to a model with $J'
> 0$, i.e., the ferromagnetic sign of the spin-spin coupling.  In terms of $v$,
if $\ell$ is even and initially, $v$ is in the AFM range, $v \in [-1,0]$, then
the RG transformation (\ref{vprime}) yields $v' \ge 0$, in the FM range.  One
may infer that the RG fixed-point (i.e., criticality) relation
$F_{(p,\ell_{even}),q}(v)=v$, should also yield a criticality condition for the
antiferromagnetic model \cite{qin_yang91}. The RG fixed point for the
ferromagnet is given by Eq. (\ref{vprime}) with $v'=v = v_{c,PM-FM}$, or
equivalently, Eq. (\ref{yprime}) with $y'=y = y_{c,PM-FM}$. For compact
notation, we denote $y_{c,PM-FM} \equiv y_c$ and keep the
dependence of $y_c$ on $(p,\ell)$ implicit, so Eq. (\ref{yprime}) reads
\beq
y_c = \Bigg [ \frac{(q+y_c-1)^\ell+(q-1)(y_c-1)^\ell}
  {(q+y_c-1)^\ell-(y_c-1)^\ell} \ \Bigg ]^p  \ .
\label{yc}
\eeq
Choosing the real positive root among the $p$ $(1/p)$'th roots of both sides
of this equation yields a resultant formula for $(y_c)^{1/p}$:
\beq
y_c^{1/p} = \frac{(q+y_c-1)^\ell+(q-1)(y_c-1)^\ell}
                 {(q+y_c-1)^\ell-(y_c-1)^\ell} \ . 
\label{ycroot}
\eeq

We recall that in the context of the $q$-state Potts antiferromagnet on the $n
\to \infty$ limit of a regular lattice or family of graphs, as $q$ increases,
the temperature $T_{c,PM-AFM}$ at which the Potts model makes a phase
transition from the $S_q$-symmetric high-temperature phase to a low-temperature
phase with spontaneously broken $S_q$ symmetry decreases, or equivalently, the
critical value of $y$, $y_{c,PM-AFM} =\exp(K_{c,PM-AFM})$, decreases
(recall that $K_{c,PM-AFM} < 0$). As $q$
approaches $q_c$ from below, $y_{c,PM-AFM}$ approaches 0 from above.  These
properties may be formally generalized from positive integer $q$ to real
positive $q$ by use of the cluster representation for $Z(G,q,v)$ in
Eq. (\ref{cluster}).  We now apply this to the Potts model on the limit of
hierarchical graphs $G^{(p,\ell)}_\infty$ under consideration.  To avoid
cumbersome notation, we set $y_{c,PM-AFM} \equiv y_{ac}$ and $v_{ac} =
y_{ac}-1$. Then the above-mentioned relation linking the FM critical point and
the AFM critical point is \cite{qin_yang91}
\beq
y_c = \Bigg [ \frac{(q+y_{ac}-1)^\ell+(q-1)(y_{ac}-1)^\ell}
  {(q+y_{ac}-1)^\ell-(y_{ac}-1)^\ell} \ \Bigg ]^p \ .
\label{yc_yac}
\eeq
Next, we combine this with the above-mentioned correspondence that
$q_c$ is the value of $q$ such that the antiferromagnetic $q$-state
Potts model has a zero-temperature critical point, so that $q_c$
corresponds to setting $y_{ac}=0$.  This means that $q_c$ is a solution
to the equation 
\beq
y_c = \Bigg [ \frac{ (q-1)^\ell+(q-1)(-1)^\ell}
  {(q-1)^\ell-(-1)^\ell} \ \Bigg ]^p 
    = \Bigg [ \frac{ q_d^\ell + q_d}{q_d^\ell-1} \Bigg ]^p \ ,
\label{ycrel}
\eeq
where we have introduced the compact notation
\beq
q_d \equiv q-1 
\label{qd}
\eeq
and have
used the fact that $\ell$ is even here, so each
factor $(-1)^\ell=1$ in Eq. (\ref{ycrel}). Next, we take the $(1/p)$'th power
of both sides of Eq. (\ref{ycrel}) and substitute the right-hand side of
this equation for the value of $y_c^{1/p}$ in Eq. (\ref{ycroot}), thereby
obtaining
\beqs
&& \bigg ( \frac{ q_d^\ell + q_d }{ q_d^\ell - 1 } \Bigg )  
\Bigg [\bigg [
 \bigg (\frac{ q_d^\ell+q_d }{q_d^\ell -1} \bigg )^p +
 q_d \bigg ]^\ell 
- \bigg [ \bigg ( \frac{ q_d^\ell + q_d }{ q_d^\ell - 1 } \bigg )^p 
- 1 \bigg ]^\ell \Bigg ] \cr\cr
&=& \bigg [ \bigg ( \frac{ q_d^\ell + q_d }{ q_d^\ell - 1}\bigg )^p 
+ q_d \bigg ]^\ell 
+ q_d\bigg [ \bigg (\frac{q_d^\ell+q_d}{q_d^\ell-1}\bigg )^p 
- 1 \bigg ]^\ell \ .
\label{yceqlong1}
\eeqs
Regrouping terms in Eq. (\ref{yceqlong1}), we have
\beq
\Bigg ( \frac{ q_d^\ell + q_d } { q_d^\ell - 1 } - 1 \Bigg ) \, 
\Bigg [ \bigg ( \frac{ q_d^\ell + q_d } { q_d^\ell - 1 } \bigg )^p 
+ q_d \Bigg ]^\ell = 
\Bigg ( \frac{ q_d^\ell + q_d }{q_d^\ell - 1 } + q_d \Bigg ) 
\, 
\Bigg [\bigg (\frac{q_d^\ell +q_d }{q_d^\ell-1 }\bigg )^p-1 \Bigg ]^\ell  \ . 
\label{yceqlong2}
\eeq
The prefactors in parentheses on the left-hand and right-hand sides of Eq. 
(\ref{yceqlong2}) can be simplified as 
\beq
\frac{ q_d^\ell + q_d}{q_d^\ell-1 }-1=\frac{ q_d + 1}{q_d^\ell - 1 } 
\label{simp1}
\eeq
and
\beq
\frac{ q_d^\ell + q_d } { q_d^\ell - 1 } + q_d = 
\frac{ q_d^\ell + q_d^{\ell+1}} { q_d^\ell -1 } = 
\frac{ q_d^\ell ( q_d + 1) } { q_d^\ell - 1 } \ .
\label{simp2}
\eeq
Substituting these expressions into Eq. (\ref{yceqlong2}) and rearranging
terms, we have 
\beq
\Bigg ( \frac{ q_d^\ell + q_d } { q_d^\ell - 1 } \Bigg )^p + q_d
 = q_d \Bigg [ \bigg (\frac{ q_d^\ell + q_d } { q_d^\ell - 1 } \bigg )^p - 1 
\Bigg ] \ ,
\label{simp3}
\eeq
that is 
\beq
(q_d^\ell + q_d)^p + q_d (q_d^\ell - 1)^p = q_d (q_d^\ell + q_d)^p - 
q_d (q_d^\ell - 1)^p  .
\label{simp4}
\eeq
Again, regrouping terms, we get 
\beq
(q_d - 1) (q_d^\ell + q_d)^p = 2q_d (q_d^\ell - 1)^p \ .
\label{simp5}
\eeq
Inserting $q_d \equiv q_c-1$, we finally obtain 
\beq
(q-2)\bigg [(q-1)^\ell+(q-1)\bigg ]^p=2(q-1)\bigg [(q-1)^\ell-1 \bigg ]^p \ ,
\label{qc_eq_even_L_appendix}
\eeq
which is Eq. (\ref{qc_eq_even_L}) in the text. 

\end{appendix}



\begin{thebibliography}{99}
\bibitem{julia}
  Some relevant definitions are noted here
  \cite{beardon}-\cite{devaney}. Let $F_q(v)$ be an iterated transformation on
  $v \in {\mathbb C}$ depending on a parameter $q \in {\mathbb C}$,
  i.e. $v_{n+1} = F_q(v_n)$.  The Julia set of $F_q(v)$, denoted $J(F_q(v))$,
  is, roughly speaking, the closure of the points in the $v$-plane that are
  unstable under the action of $F_q(v)$ upon infinitesimally small changes.
  For almost all parameters $q$, $J(F_q(v))$ is a (self-similar) fractal.
  Another characterization of $J(F_q(v))$ is that it comprises the boundary of
  the Fatou set of $F_q(v)$, where this Fatou set is the set of points in the
  complex $v$-plane that, under the iterated action of $F_q(v)$ map to
  attractors, such as fixed points (e.g., 0 or $\infty$) or periodic orbits.
\bibitem{spb} 
Ref. \cite{qin_yang91} gave results for the ground-state
entropy per bond (edge), labelled with nonstandard notation as $W$.
In standard notation, the ground-state entropy
is related to the ground-state degeneracy according to Eq. (\ref{s0}).
Note that in the $m \to \infty$ limit on $G^{(p,\ell)}_m$, the ground state
entropy per site (vertex), $S_0$, as defined in Eq. (\ref{s0}), is related to
the ground-state entropy per bond for $G^{(p,\ell)}_\infty$ as
$S_0 = \Big ((p\ell - 1)/[p(\ell-1)]\Big ) S_{pb}$.
\end{thebibliography}
\end{document}